\documentclass[onecolumn,showpacs,11pt,nofootinbib]{revtex4-2}
\usepackage{amsmath, graphicx}
\usepackage{dcolumn}
\usepackage{bm}
\usepackage{color}
\usepackage{epstopdf}
\usepackage{eurosym}
\usepackage{multirow}
\usepackage{amsfonts}
\usepackage{amsmath}
\usepackage{hyperref}
\usepackage{amssymb}
\usepackage[english]{babel}
\usepackage{graphicx}
\usepackage{epsfig}
\usepackage{subfigure}
\usepackage{unicode}
\begin{document}
\newcommand{\hs}{\hspace*{0.2cm}}
\newcommand{\hsp}{\hspace*{0.5cm}}
\newcommand{\vs}{\vspace*{0.5cm}}
\newcommand{\be}{\begin{equation}}
\newcommand{\ee}{\end{equation}}
\newcommand{\bea}{\begin{eqnarray}}
\newcommand{\eea}{\end{eqnarray}}
\newcommand{\ben}{\begin{enumerate}}
\newcommand{\een}{\end{enumerate}}
\newcommand{\bde}{\begin{widetext}}
\newcommand{\ede}{\end{widetext}}
\newcommand{\nn}{\nonumber}
\newcommand{\crn}{\nonumber \\}
\newcommand{\Tr}{\mathrm{Tr}}
\newcommand{\non}{\nonumber}
\newcommand{\noi}{\noindent}
\newcommand{\al}{\alpha}
\newcommand{\la}{\lambda}
\newcommand{\bet}{\beta}
\newcommand{\ga}{\gamma}
\newcommand{\va}{\varphi}
\newcommand{\om}{\omega}
\newcommand{\pa}{\partial}
\newcommand{\+}{\dagger}
\newcommand{\fr}{\frac}
\newcommand{\sq}{\sqrt}
\newcommand{\bc}{\begin{center}}
\newcommand{\ec}{\end{center}}
\newcommand{\Ga}{\Gamma}
\newcommand{\de}{\delta}
\newcommand{\De}{\Delta}
\newcommand{\ep}{\epsilon}
\newcommand{\varep}{\varepsilon}
\newcommand{\ka}{\kappa}
\newcommand{\La}{\Lambda}
\newcommand{\si}{\sigma}
\newcommand{\Si}{\Sigma}
\newcommand{\ta}{\tau}
\newcommand{\up}{\upsilon}
\newcommand{\Up}{\Upsilon}
\newcommand{\ze}{\zeta}
\newcommand{\ps}{\psi}
\newcommand{\Ps}{\Psi}
\newcommand{\ph}{\phi}
\newcommand{\vph}{\varphi}
\newcommand{\Ph}{\Phi}
\newcommand{\Om}{\Omega}

\newcommand{\Vien}[1]{{\color{red}#1}}
\newcommand{\lh}[1]{{\color{blue}#1}}
\newcommand{\ac}[1]{{\color{cyan}#1}}
\newcommand{\Green}[1]{{\color{green}#1}}
\newcommand{\JM}[1]{{\color{teal}#1}}
\newcommand{\Red}[1]{{\color{red}#1}}
\newcommand{\Revised}[1]{{\color{red}#1}}
\newcommand{\Blue}[1]{{\color{blue}#1}}
\newcommand{\Long}[1]{{\color{cyan}#1}}
\title{Fermion masses and mixings in an extended SM based on $A_4$ flavor symmetry with the linear seesaw for majorana neutrino}

\author{V. V. Vien}
\email{vovanvien@tdmu.edu.vn}
\affiliation{Institute of Applied Technology, Thu Dau Mot University, Binh Duong Province, Vietnam,}
\date{\today}

\begin{abstract}
We propose a $U(1)_L$ model with $A_4$ symmetry in light of the linear seesaw for majorana neutrino that capable of generating the current lepton and quark mass and mixing patterns. The smallness of Majorana neutrino mass is reproduced through the linear seesaw mechanism. The model can accommodate the current observed patterns of lepton and quark mixing in which the solar neutrino mixing angle and the Dirac CP violating phase are in $2\sigma$ range for both NO and IO, the Majorana violating phases are predicted to be $\eta_{1} \in (2.29, 10.31)^\circ$ and $\eta_{2} \in (57.30, 302.70)^\circ $ for NO while $\eta_1\in (3.44, 10.31)^\circ$ and $\eta_{2} \in(79.07, 87.09)^\circ$ for  IO. The obtained sum of neutrino mass and the effective Majorana neutrino mass are in good consistent with the recent upper limits. For quark sector, all the quark masses can get the best-fit values and all the elements of the quark mixing matrix are in agreement with the experimental constraints except one element, $(V_{\mathrm{CKM}})_{21}$, with a deviation about $0.25\,\%$ deviation.
\end{abstract}
\maketitle
\section{\label{intro}Introduction}
In the elementary particle physics, the Standard Model (SM) is one of the most successful theories. However, it leaves some unresolved problems that have been experimentally confirmed, such as the observed lepton and quark mass and mixing patterns.
The canonical Type I, II, and III seesaw mechanisms (see, for example, Refs.\cite{seesaw,S3VLAEPJC20, VienQ620} and the references therein) are the most simplest ways to explain the smallness of neutrino mass. However, the mass scale of the right-handed neutrinos, in the canonical seesaw mechanisms, is too high  that cannot  be achieved by the near future experiments.
In the linear seesaw mechanism, the smallness of neutrino masses generate as a result of physics at $\mathrm{TeV}$ scale which can be verified by LHC \cite{linearseesaw1,linearseesaw2, linearseesaw3, linearseesaw4,linearseesaw5, linearseesaw6}.

Experimentally, lepton mixing angles and the Dirac CP violating phase as well as the neutrino mass squared differences are given in Ref. \cite{Salas2021} as shown in Table \ref{Salas2021T}.
\begin{table}[ht]
\begin{center}
\caption{\label{Salas2021T}Neutrino oscillation parameters \cite{Salas2021}}
\vspace{0.25cm}
\begin{tabular}{|c|c|c|c|c|c|c|c|c|c|c|}\hline
\multirow{2}{2.3cm}{\hfill Parameter  \hfill }&\hspace{0.1 cm} $\mathrm{Normal\hspace{0.1cm} ordering\, (NO)}$ \hspace{0.15 cm}&\hspace{0.15 cm}$\mathrm{Inverted \hspace{0.1cm}ordering\, (IO)}$\hspace{0.15 cm} \\
\cline{2-3}   & $\mathrm{Best\,\,fit\,\, point}\pm 1\sigma \, (3\sigma \,\, \mathrm{range})$  & $\mathrm{Best\,\,fit\,\, point}\pm 1\sigma \, (3\sigma \,\, \mathrm{range})$ \\  \hline
$10^{2} \sin^2\theta_{13}$&  $2.200^{+0.069}_{-0.062}\, (2.00-2.405)$& $2.225^{+0.064}_{-0.070}\, (2.018-2.424)$\\
$\sin^2\theta_{12}$&\hspace{0.1cm}$0.318\pm 0.016\, (0.271-0.369)$ \hspace{0.1cm}&\hspace{0.1cm} $0.318\pm 0.016\, (0.271-0.369)$ \hspace{0.1cm}\\
$\sin^2\theta_{23}$&\hspace{0.1cm}$0.574\pm 0.014 \, (0.434-0.610)$ \hspace{0.1cm}&\hspace{0.1cm} $0.578^{+0.010}_{-0.017}\, (0.433-0.608)$ \hspace{0.1cm}\\
$\delta/\pi$&  $1.08^{+0.13}_{-0.12}\, (0.71-1.99)$ & $1.58^{+0.15}_{-0.16}\, (1.11-1.96)$ \\
$\Delta m^2_{21} \big[\mathrm{meV}^2\big]$&$75.0^{+2.2}_{-2.0}\, \big(69.4-81.4\big)$& $75.0^{+2.2}_{-2.0}\, \big(69.4-81.4\big)$\\
$10^{-3} |\Delta m^2_{31}| \,\big[\mathrm{meV}^2\big]$&$2.55^{+0.02}_{-0.03} \, (2.47-2.63)$& $2.45^{+0.02}_{-0.03}$ \, (2.37-2.53)\\
 \hline
\end{tabular}
\end{center}
\vspace{-0.25cm}
\end{table}
Besides that, the magnitude of the entries of the lepton mixing matrix 
are determined in Ref.\cite{Esteban2020}:
\begin{eqnarray}
\left|U_{3\sigma}\right| =\left(
\begin{array}{ccc}
0.801 \to 0.845 &\hs\hs  0.513 \to 0.579 &\hs\hs 0.143 \to 0.156 \\
0.233 \to 0.507 &\hs\hs 0.461 \to 0.694 &\hs\hs 0.631 \to 0.778 \\
 0.261 \to 0.526 &\hs\hs 0.471 \to 0.701 &\hs\hs  0.611 \to 0.761 \\
\end{array}\right). \label{2019lepmix}
\end{eqnarray}
Furthermore, the quark mixing angles, the CP violating phase in the quark sector and quark masses have been measured with high accuracy \cite{PDG2022},
\begin{eqnarray}
&&s^q_{12}= 0.22500\pm 0.00067,\hs s^q_{23}= 0.04182^{+0.00085}_{-0.00074}, \crn
&&s^q_{13}= 0.00369\pm 0.00011, \hs \delta^q = 1.144\pm 0.027,\label{PDG22qangles}\\
&&m_u =2.16^{+0.49}_{-0.26} \,\textrm{MeV},\hs m_{c}=1.27\pm 0.02 \, \textrm{GeV} ,\hs  m_{t}=172.9\pm 0.30 \,\textrm{GeV}, \crn
&&m_d = 4.67^{+0.48}_{-0.17} \, \textrm{MeV},\hs m_{s}= 93.4^{+8.6}_{-3.4}\,\textrm{MeV},\hs\hs\hs\, m_{b}=4.18^{+0.03}_{-0.02}\, \textrm{GeV}, \label{PDG22qmass} \end{eqnarray}
where the quark mixing angles are very small in comparison to those of lepton and the masses of up and down quarks 
 are very small in comparison to the others.
The standard parameterization 
\cite{PDG2022} with the quark mixing angles and the CP violating phase at the best-fit points in Eq. (\ref{PDG22qangles}) corresponding to the following quark mixing matrix:
\begin{eqnarray}
&&V^{\mathrm{exp}}_{\mathrm{CKM}}=\left(
\begin{array}{ccc}
 0.97435 & 0.22500 & 0.0015275-0.003359 i \\
 -0.22487 & 0.97349 & 0.04182 \\
 0.0079225 -0.00327 i & -0.041091 & 0.99912 \\
\end{array}
\right),\label{PDG22qmix}\end{eqnarray}
which has the following approximate pattern
\begin{eqnarray}
V^{\mathrm{appro}}_{\mathrm{CKM}} \simeq \left(
\begin{array}{ccc}
0.97435 & 0.22500 & 0 \\
 -0.22487 & 0.97349 & 0\\
 0 & -0.0411 & 0.99912 \\
\end{array}\right).\label{PDG22qmixapp}
\end{eqnarray}
Discrete symmetry plays a crucial role in explaining the observed masses and mixing patterns of leptons and quarks (see, for example, Refs.\cite{S3VLAEPJC20, VienQ620} and the references therein), among them, $A_4$ has been used in different works \cite{A41, A42, A43, A44, A46, A47, A48, A49, A410, A411, A412, A413, A414, A415, A416, A417, A418, A419, A420Ishimori10, A421VL2015, A422, A423, A424, A425}. The linear seesaw mechanism for Dirac neutrino in which the heavy neutral singlet leptons $N$ and $S$ are Dirac fermions possessing the lepton number $L=\pm 1$ 
in Refs. 
\cite{LseesawS32, LseesawA43, LseesawAD274, LseesawA45, LseesawA46, LseesawD277, LseesawS48, LseesawS49,LseesawZ2Z2, LseesawZ2, LseesawU1H, A4DiracVL}.  
In Refs. \cite{LseesawMajo1,LseesawMajo2} the linear seesaw for Majorana neutrino has been considered with $SO(10)$ symmetry \cite{LseesawMajo1} and $G_{\mathrm{SM}}\times A_4\times Z_3$ symmetry \cite{LseesawMajo2} with 
do not refer to the quark sector and thus differ crucially from the current work.
We construct a SM extension with the non-Abelian discrete symmetry $A_4$ supplemented by three Abelian discrete symmetries $Z_4, Z_3$ and $Z_2$ and one 
lepton number symmetry $U(1)_L$ with nine singlet scalars and one doublet which can accommodate the currently observed lepton and quark mass and mixing patterns.

This work is organized as follows. In section \ref{model} we present a SM extension with $A_4$ symmetry supplemented by three discrete Abelian symmetries $Z_4, Z_3$ and $Z_2$ and one global lepton number symmetry $U(1)_L$. The stability condition of the scalar potential is presented in Section \ref{Higgstable}. Sections \ref{lepton} and \ref{quark} are devoted to the lepton and quark sectors, respectively. Finally, some conclusions are drawn in Section \ref{conclusion}.

\section{The model\label{model}}
The SM is extended by adding one non-Abelian discrete symmetry $A_4$ supplemented by three Abelian discrete symmetries $Z_4, Z_3, Z_2$ and one global lepton number symmetry $U(1)_L$ so as to forbid some unwanted terms otherwise allowed by $A_4$ which are listed in Table \ref{preventedtermsT}. The total symmetry of the model is $\mathbf{\Gamma}=G_{\mathrm{SM}}\times U(1)_L\times A_4\times Z_4\times Z_2 \times Z_2$ with $G_{\mathrm{SM}}$ being the gauge symmetry of the SM. The SM lepton doublet $\psi_L$ are assigned to the triplet representation of $A_4$, three right-handed charged lepton $l_{1R}, l_{2R}$ and $l_{3R}$ are assigned to the $A_4$ singlets $\underline{1}, \underline{1}^{''}$ and $\underline{1}^'$, respectively, while the SM quark doublet $Q_{1L}, Q_{2L}, Q_{3L}$ and right-handed up-and down quark $u_{1,2,3R}, d_{1,2,3R}$ are assigned to $\underline{1}$ of $A_4$.
For the lepton fields,  three right-handed neutrino $\nu_R$ and two types of neutral 
leptons 
$N_{L,R}, S_{L,R}$, which are all put in $\underline{3}$ of $A_4$, are additionally introduced. For the scalar sector, five extra scalars $\phi, \varphi, \chi, \rho$ and $\eta$ are added to the SM where $\phi$ and $\varphi$ are aligned in $\underline{3}$ of $A_4$ while $\chi, \rho$ and $\eta$ are put in $\underline{1}$ of $A_4$.
The transformation properties of the leptons, quarks and scalars 
are sumarized in Table \ref{lepcont}.

\begin{table}[ht]
\caption{\label{lepcont} The assignment 
for fermions and scalars.}
\vspace{0.25cm}
\centerline{\begin{tabular}{|c|ccccc|ccc|cccccc|c|c|c|c|c|c|c|c|c|c|}
\hline
Fields& $\psi_{L}$ & $l_{1, 2, 3R}$ &$\nu_{R}$& $N_L, S_L$ &$N_{R}, S_R$ &$Q_{1L}, Q_{2 L}, Q_{3 L}$ &$u_{1,2,3R}$& $d_{1,2,3R}$&$H$& $\phi$ & $\varphi$&$\chi$ &$\rho$&$\eta$  \\ \hline
$SU(2)_L$ & $2$ & 1 &1 &1 &1 &$2$&$1$&$1$&$2$& 1& 1& 1& 1 & 1 \\ 
$\mathrm{U}(1)_Y$ & $-\frac{1}{2}$ & $-1$ &0 &0 &0 &$\frac{1}{6}$&$\frac{2}{3}$&$-\frac{1}{3}$&$\frac{1}{2}$& 0& 0  & 0 & 0 & 0 \\ 
$\mathrm{U}(1)_{L}$ & $x$ &$x$&$x$&$0$&$0$&$-\frac{x}{3}$ &$-\frac{x}{3}$&$-\frac{x}{3}$&$0$& $0$ & $0$ & $0$ & $-x$ & 0 \\ 
$A_4$ & $\underline{3}$ & $\underline{1}, \underline{1}^{''}, \underline{1}^'$& $\underline{3}$  & $\underline{3}$ &$\underline{3} $ &$\underline{1}$  & $\underline{1}$  &  $\underline{1}$& $\underline{1}$ &$\underline{3}$&$\underline{3}$&$\underline{1}$ &$\underline{1}$ & $\underline{1} $ \\ 
$Z_4$ & $i$&$i$&$1$ &$i$& $i$  &$1$&$1$ &$1$& $1$ &$1$& $-1$ & $-1$&$-i$ & $1$ \\
$Z_3$ &$\om$ &$\om$ & $1$  & $1$&$\om^2$ &$1, 1, \om$ &$1, 1, \om$ & $1, 1, \om$ &$1$&$1$&$\om^2$&$\om^2$&$\om$ & $\om^2$  \\ 
$Z_2$ &$ +$ &$-$ & $-$ & $+$&$+$ &$+, +,-$& $+, +,-$ \hs & $+, +,-$&$+$&$-$&$+$&$+$&$-$ &$-$\\ \hline
\end{tabular}}
\end{table}

The fermion Yukawa terms, up to five-dimensions, read\footnote{ Two Majorana neutrino mass terms $(\bar{N}^c_R N_R)_{\underline{3}_a} \varphi$ and $(\bar{S}^c_R S_R)_{\underline{3}_a}  \varphi$ are vanished due to the antisymmetry in $\bar{N}^c_{kR}\, (\bar{S}^c_{kR})$ and $N_{l R}\, (S_{l R})$ with $k, l=1\div 3$ and $k\neq l$ so they are not included in Eq. (\ref{LYlep}).}:
\bea
-\mathcal{L}^{l}_Y&=&\fr{\lambda_{e}}{\La }(\bar{\psi}_L\phi)_{\underline{1}} (H l_{1R})_{\underline{1}}
+\fr{\lambda_{\mu}}{\La }(\bar{\psi}_L \phi)_{\underline{1}^{'}}(H l_{2R})_{\underline{1}^{''}}
+\fr{\lambda_{\tau}}{\La }(\bar{\psi}_L \phi)_{\underline{1}^{''}}(H l_{3R})_{\underline{1}^{'}}\crn
 &+& x_{1\nu} (\bar{\nu}^c_R N_R)_{\underline{1}} \rho +x_{2\nu} (\bar{\nu}^c_R S_R)_{\underline{1}} \rho
 +x_{3\nu} (\bar{N}^c_R \nu_R)_{\underline{1}} \rho + x_{4\nu} (\bar{S}^c_R \nu_R)_{\underline{1}} \rho \crn
 &+&y_{1\nu}(\bar{N}^c_R N_R)_{\underline{1}} \chi +y_{2\nu}(\bar{N}^c_R N_R)_{\underline{3}_s} \varphi  + y_{3\nu}(\bar{N}^c_R N_R)_{\underline{3}_a} \varphi \crn
 &+&z_{1\nu}(\bar{N}^c_R S_R)_{\underline{1}} \chi +z_{2\nu}(\bar{N}^c_R S_R)_{\underline{3}_s} \varphi  +z_{3\nu}(\bar{N}^c_R S_R)_{\underline{3}_a} \varphi\crn
 &+&t_{1\nu}(\bar{S}^c_R N_R)_{\underline{1}} \chi +t_{2\nu}(\bar{S}^c_R N_R)_{\underline{3}_s} \varphi  +t_{3\nu}(\bar{S}^c_R N_R)_{\underline{3}_a} \varphi\crn
 &+&w_{1\nu}(\bar{S}^c_R S_R)_{\underline{1}} \chi +w_{2\nu}(\bar{S}^c_R S_R)_{\underline{3}_s} \varphi  + w_{3\nu}(\bar{S}^c_R S_R)_{\underline{3}_a}  \varphi + \mathrm{h.c}., \label{LYlep}\\
-\mathcal{L}^q_Y&=&x^u_{11}(\bar{Q}_{1L} u_{1R})_{\underline{1}} \widetilde{H}
+ x^u_{22}(\bar{Q}_{2L} u_{2R})_{\underline{1}} \widetilde{H}
+ x^u_{33}(\bar{Q}_{3L} u_{3R})_{\underline{1}} \widetilde{H}\crn
&+&x^u_{12}(\bar{Q}_{1L} u_{2R})_{\underline{1}} \widetilde{H}
+ x^u_{21}(\bar{Q}_{2L} u_{1R})_{\underline{1}} \widetilde{H}
+\fr{x^u_{13}}{\La }(\bar{Q}_{1L} u_{3R})_{\underline{1}} (\widetilde{H}\eta)_{\underline{1}}\crn
&+&\fr{x^u_{23}}{\La }(\bar{Q}_{2L} u_{3R})_{\underline{1}} (\widetilde{H}\eta)_{\underline{1}}
+\fr{x^u_{31}}{\La }(\bar{Q}_{1L} u_{3R})_{\underline{1}} (\widetilde{H}\eta^*)_{\underline{1}}
+\fr{x^u_{32}}{\La }(\bar{Q}_{2L} u_{3R})_{\underline{1}} (\widetilde{H}\eta^*)_{\underline{1}}\crn
&+&x^d_{11}(\bar{Q}_{1L} d_{1R})_{\underline{1}} H
+ x^d_{22}(\bar{Q}_{2L} d_{2R})_{\underline{1}} H
+ x^d_{33}(\bar{Q}_{3L} d_{3R})_{\underline{1}} H \crn
&+&x^d_{12}(\bar{Q}_{1L} d_{2R})_{\underline{1}} H
+ x^d_{21}(\bar{Q}_{2L} d_{1R})_{\underline{1}} H
+\fr{x^d_{13}}{\La }(\bar{Q}_{1L} d_{3R})_{\underline{1}} (H\eta)_{\underline{1}}\crn
&+&\fr{x^d_{23}}{\La }(\bar{Q}_{2L} d_{3R})_{\underline{1}} (H\eta)_{\underline{1}}
+\fr{x^d_{31}}{\La }(\bar{Q}_{1L} d_{3R})_{\underline{1}} (H\eta^*)_{\underline{1}}
+\fr{x^d_{32}}{\La }(\bar{Q}_{2L} d_{3R})_{\underline{1}} (H\eta^*)_{\underline{1}} + \mathrm{h.c}. \label{Yquark}\eea
The vacuum expectation value (VEV) structure of the scalars, which comes from the minimum condition of the model scalar
 potential (see section \ref{Higgstable}), reads:
 \bea
&&\langle H \rangle = \left( 0  \hspace{0.35cm} v\right)^T, \hspace{0.42cm} \langle \phi \rangle = (v_{\phi}, \hspace{0.15cm} v_{\phi},\hspace{0.15cm} v_{\phi}),  \crn
&& \langle \varphi \rangle = (0,\hspace{0.15cm} v_{\varphi},\hspace{0.15cm} 0), \hspace{0.15cm} \langle \Psi \rangle= v_{\Psi} \,\, (\Psi=\chi, \rho, \eta). \label{scalarvev}
\eea
The cut-off scale and VEV of singlets are assumed to be at a very high scale,
\bea
\Lambda \sim 10^{13}\, \mathrm{GeV}, \hs v_\phi \sim  v_\varphi  \sim v_\chi \sim v_\rho \sim v_\eta \simeq 10^{11} \, \mathrm{GeV}. \label{vevscales}
\eea
\section{\label{Higgstable} The stability condition}
\begin{figure}[h]
\begin{center}
\vspace{0.25 cm}
\hspace{-1.25 cm}\includegraphics[width=0.475\textwidth]{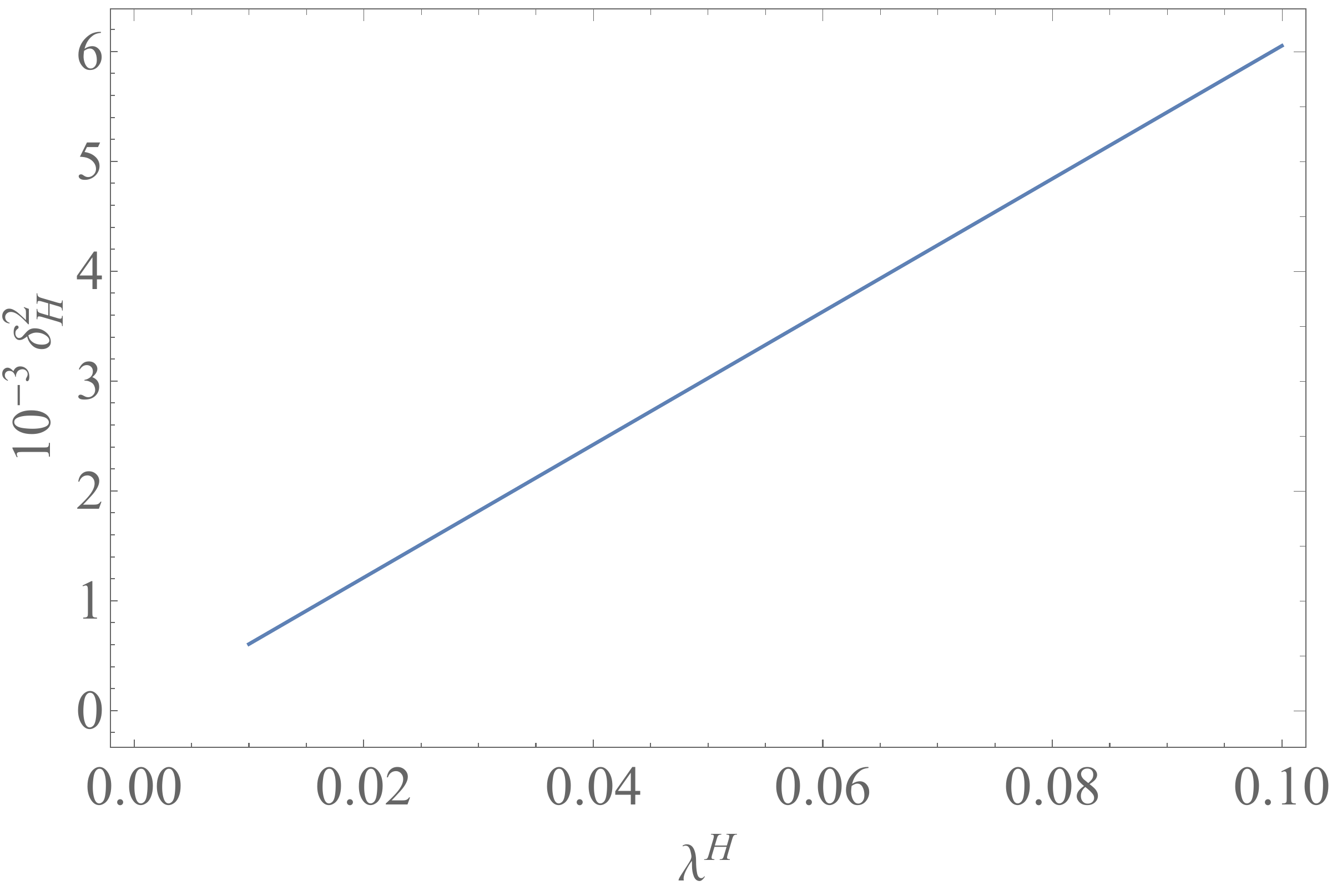}\hspace{0.1 cm}
\includegraphics[width=0.475\textwidth]{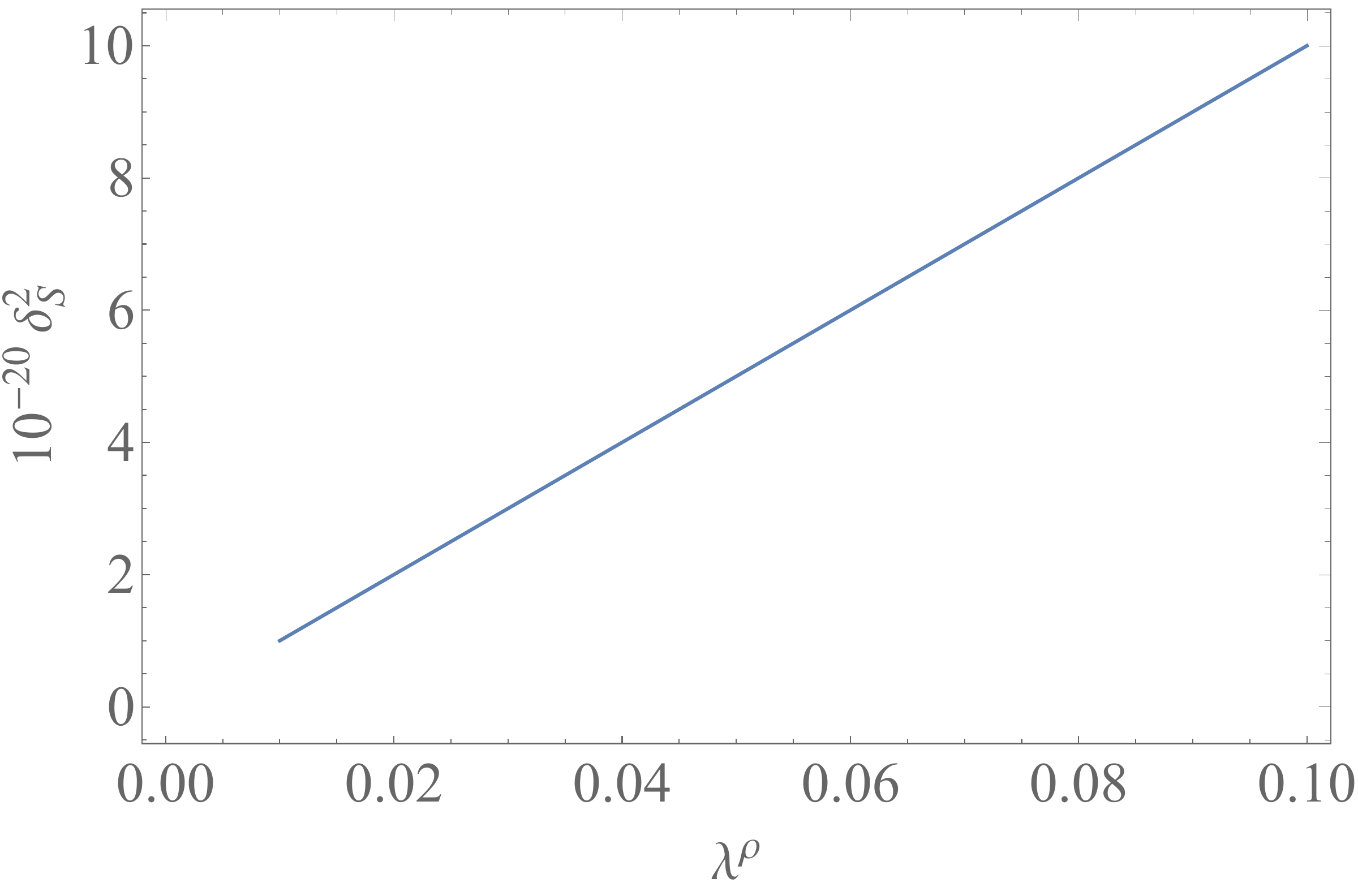}\hspace*{-1.25 cm}\hspace{-15.5 cm}
\end{center}
\vspace{-0.65 cm}
\caption{$\delta^2_{H}$ versus $\lambda^{H}$ (left panel) and $\delta^2_{\rho}$ versus $\lambda^{\rho}$ (right panel) with $\lambda^{H, \rho} \in (10^{-2}, 10^{-1})$.}
\label{deltaPhi}
\end{figure}

\begin{figure}[h]
\begin{center}
\vspace{-0.5 cm}
\hspace{-1.25 cm}\includegraphics[width=0.675\textwidth]{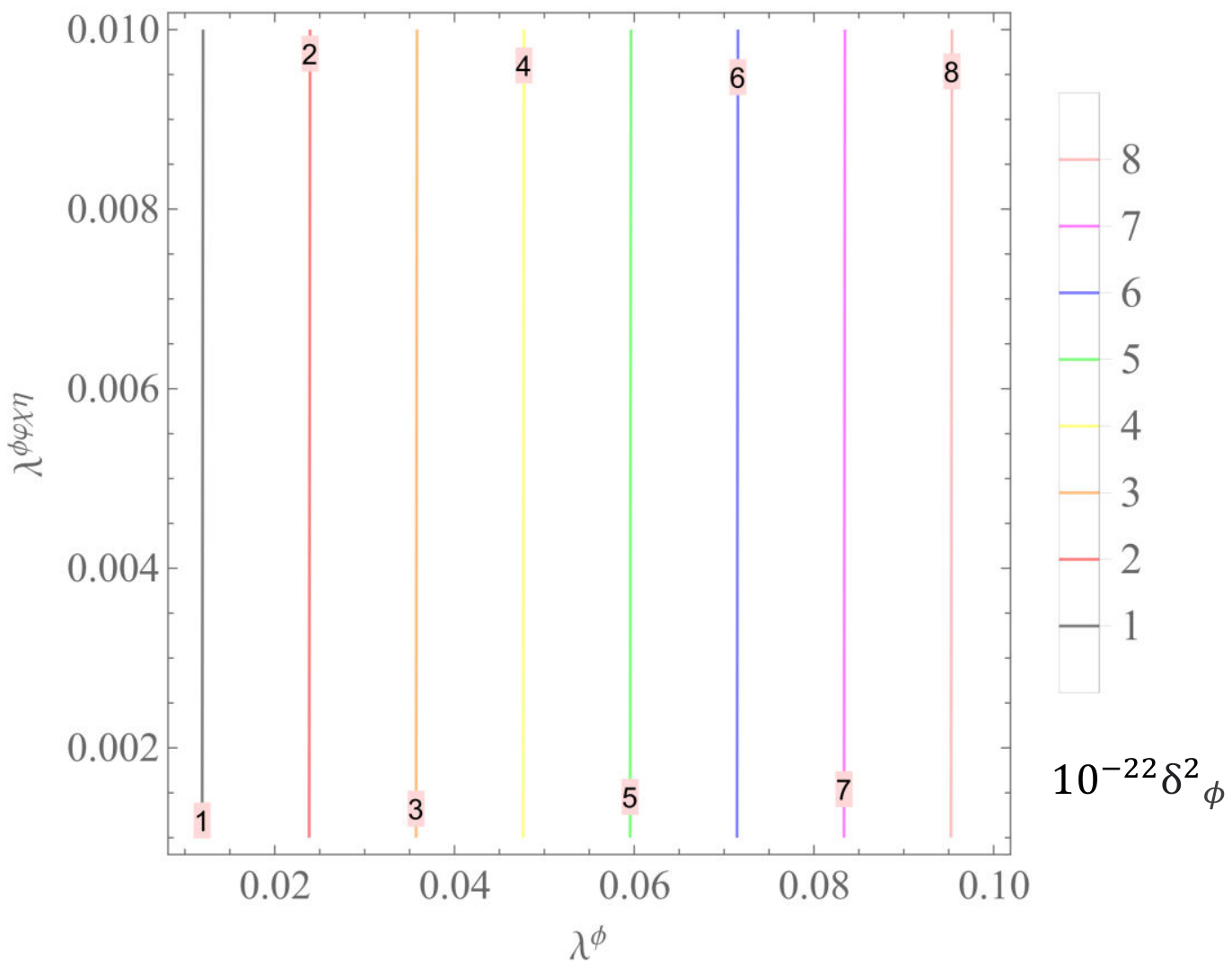}\hspace{-3.5 cm}
\includegraphics[width=0.675\textwidth]{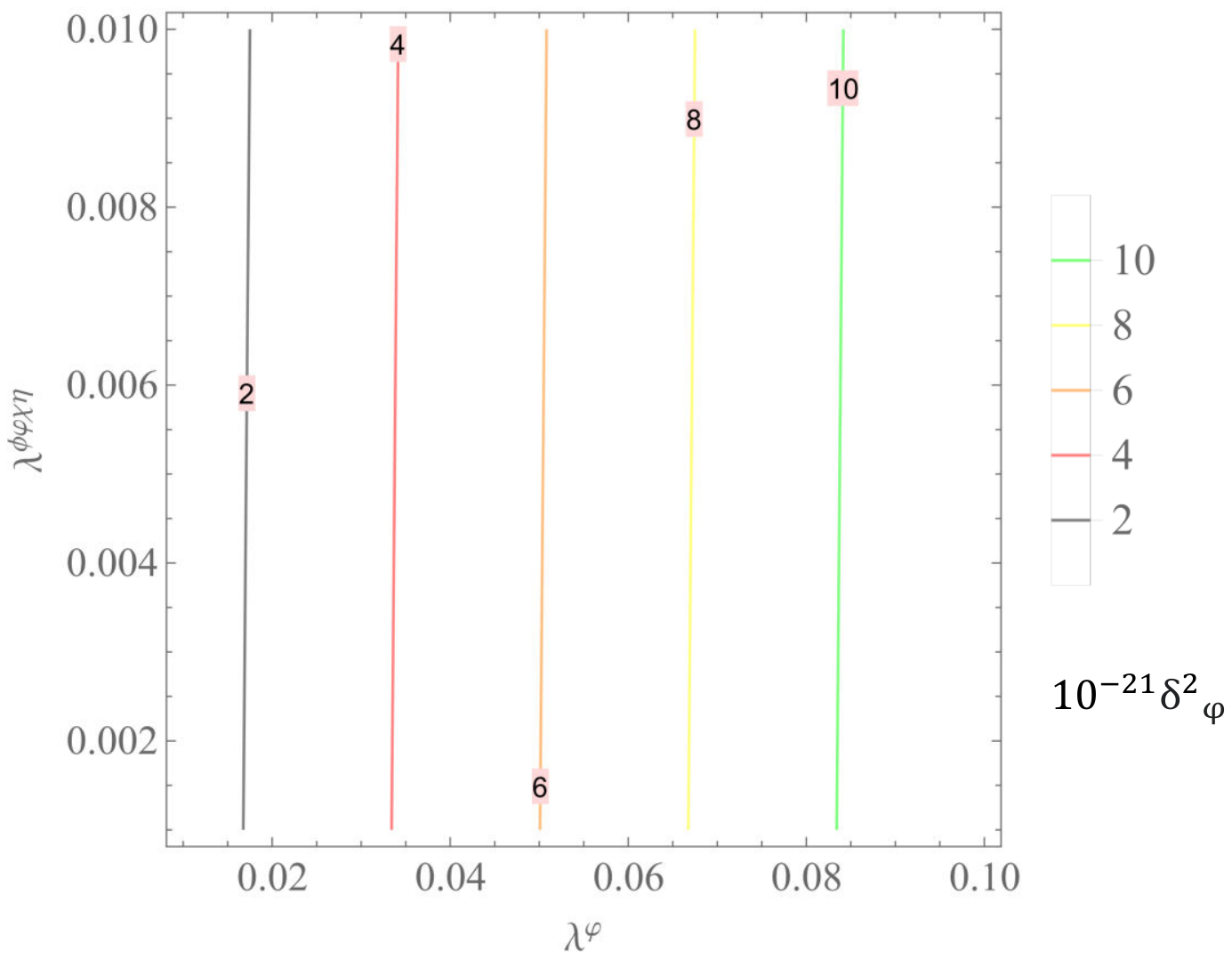}\hspace*{-1.75 cm} \\
\vspace{-9.55 cm}
\hspace{-1.25 cm}\includegraphics[width=0.675\textwidth]{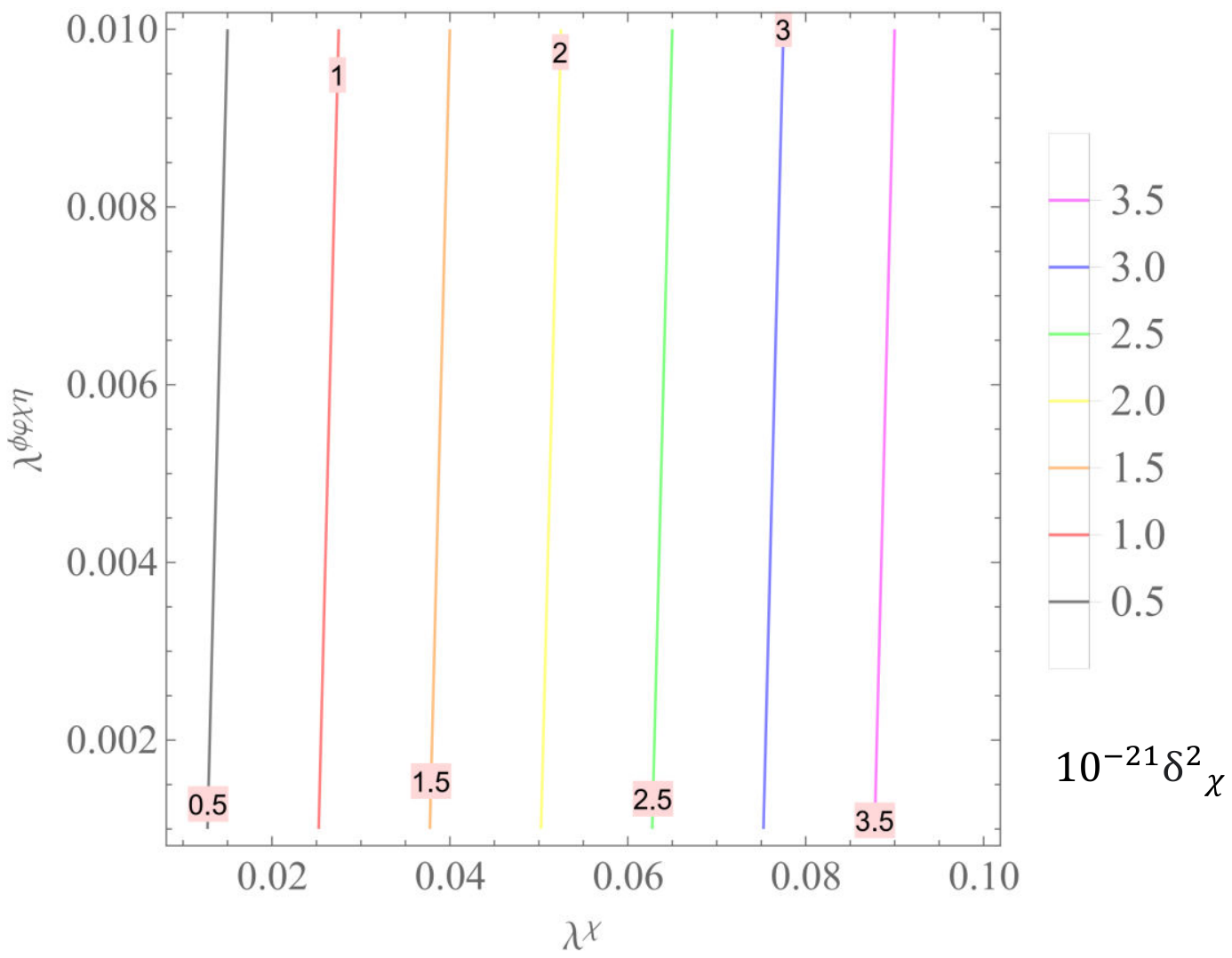}\hspace{-3.5 cm}
\includegraphics[width=0.675\textwidth]{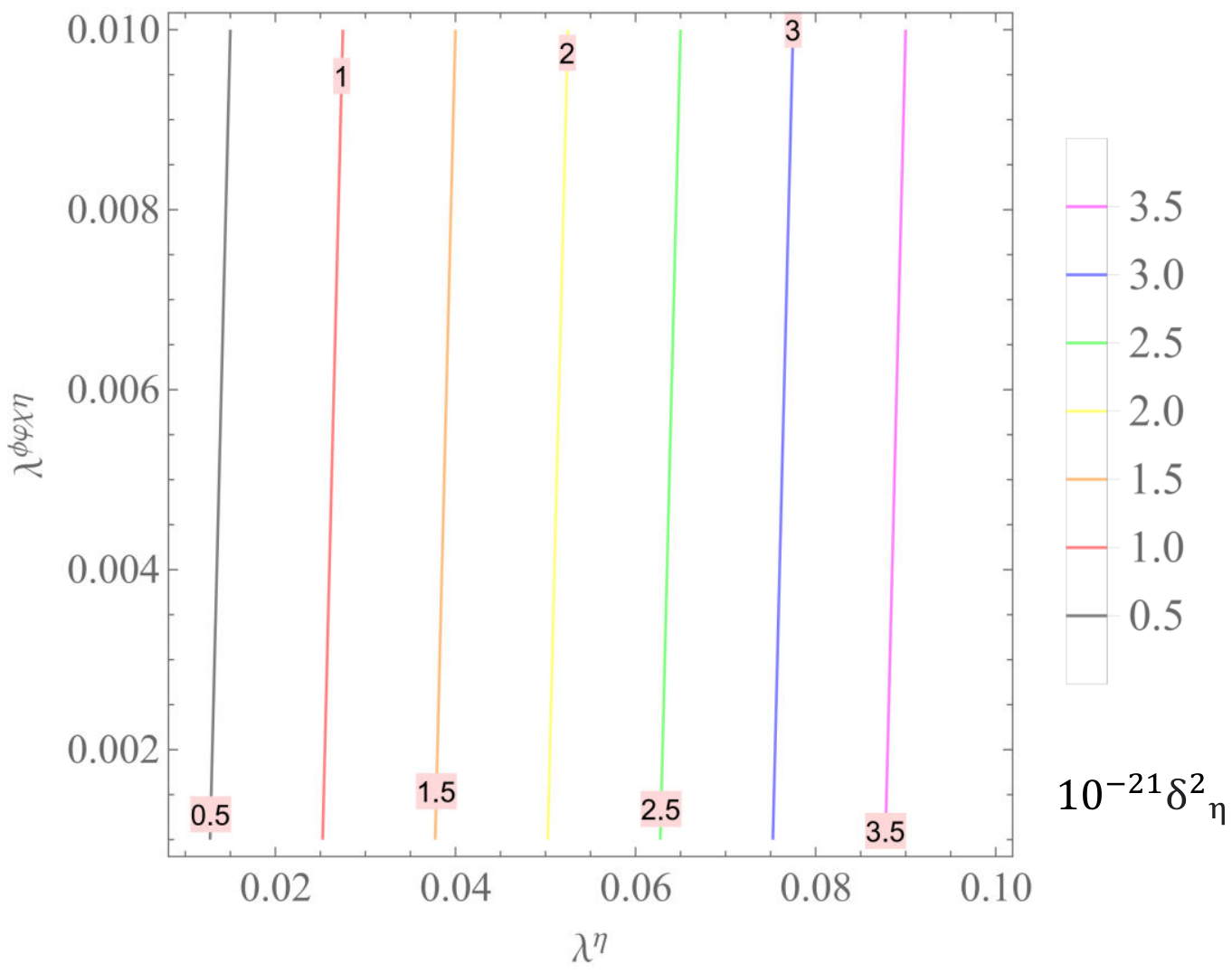}\hspace*{-1.75 cm}
\end{center}
\vspace{-9.5 cm}
\caption{$\delta^2_{\phi, \varphi, \chi, \eta}$ versus $\lambda^{\phi, \varphi, \chi, \eta}$ and $\lambda^{\phi\varphi\chi\eta}$  with $\lambda^{\phi, \varphi, \chi, \eta} \in (10^{-2}, 10^{-1})$ and $\lambda^{\phi\varphi\chi\eta} \in (10^{-3}, 10^{-2})$.}
\label{deltaphivarchieta}
\vspace{-0.25 cm}
\end{figure}

The potential invariant under $\mathbf{\Gamma}$ is given by\footnote{The following Yukawa terms $(\phi^*\phi)_{\underline{3}_s}(\phi^*\phi)_{\underline{3}_a},\,(\phi^*\phi)_{\underline{3}_a}(\phi^*\phi)_{\underline{3}_s},\,
(\phi^*\phi)_{\underline{3}_a}(\phi^*\phi)_{\underline{3}_a},\,
(\varphi^*\varphi)_{\underline{3}_s}(\varphi^*\varphi)_{\underline{3}_s},\,(\varphi^*\varphi)_{\underline{3}_s}(\varphi^*\varphi)_{\underline{3}_a}$,\,
$(\varphi^*\varphi)_{\underline{3}_a}(\varphi^*\varphi)_{\underline{3}_s},
(\varphi^*\varphi)_{\underline{3}_a}(\varphi^*\varphi)_{\underline{3}_a},
(\phi^* \phi)_{\underline{3}_s}(\varphi^* \varphi)_{\underline{3}_s},
(\phi^* \phi)_{\underline{3}_s}(\varphi^* \varphi)_{\underline{3}_a},
(\phi^* \phi)_{\underline{3}_a}(\varphi^* \varphi)_{\underline{3}_s},(\phi^* \phi)_{\underline{3}_a}(\varphi^* \varphi)_{\underline{3}_a}, (\phi^* \phi)_{\underline{1}^{'}}(\varphi^* \varphi)_{\underline{1}^{''}},$
$(\phi^* \phi)_{\underline{1}^{''}}(\varphi^* \varphi)_{\underline{1}^{'}},\,(\phi^* \varphi)_{\underline{3}_s}(\varphi^* \phi)_{\underline{3}_a}
,\,(\phi^* \varphi)_{\underline{3}_a}(\varphi^* \phi)_{\underline{3}_s}$ are vanished due to the tensor product of $A_4$ group and the VEV alignment of $\phi$ and $\varphi$ in Eq. (\ref{scalarvev}). Besides, other terms with three and four different scalars are not invariant under one or some of symmetries of $\mathbf{\Gamma}$ thus don't contribute to $\mathcal{V}_{\mathrm{scalar}}$. Further, for simplicity, we assume that the couplings in the same type of interactions are in the same scale. 
}:
\bea  \mathcal{V}_{\mathrm{tot}}&=& \mathcal{V}(H)+\mathcal{V}(\phi)+\mathcal{V}(\varphi)+ \mathcal{V}(\chi)+ \mathcal{V}(\rho)+ \mathcal{V}(\eta)+\mathcal{V}(H,\phi)+\mathcal{V}(H,\varphi)\crn
&+& \mathcal{V}(H,\chi)+ \mathcal{V}(H,\rho)+ \mathcal{V}(H,\eta)+ \mathcal{V}(\phi,\varphi)
+ \mathcal{V}(\phi,\chi)+ \mathcal{V}(\phi,\rho)+ \mathcal{V}(\phi,\eta) \crn
&+& \mathcal{V}(\varphi,\chi)+ \mathcal{V}(\varphi,\rho)+ \mathcal{V}(\varphi,\eta)
+ \mathcal{V}(\chi,\rho)+ \mathcal{V}(\chi,\eta)+\mathcal{V}(\rho,\eta)+\mathcal{V}_{\mathrm{multi}},
\label{Higgspoten}\eea
where
\bea
&&\mathcal{V}(H)=\mu^2_H H^\+H +\lambda^H (H^\+H)^2, \, V(\chi)=V(H\rightarrow \chi), \, V(\rho)=V(\chi\rightarrow \rho),\, V(\eta)=V(\chi\rightarrow \eta), \\
&&\mathcal{V}(\phi)=\mu^2_\phi (\phi^*\phi)_{\underline{1}}
+\lambda^{\phi} \big[(\phi^*\phi)_{\underline{1}}(\phi^*\phi)_{\underline{1}}
+(\phi^*\phi)_{\underline{1}^{'}}(\phi^*\phi)_{\underline{1}^{''}}
+(\phi^*\phi)_{\underline{1}^{''}}(\phi^*\phi)_{\underline{1}^{'}}
+(\phi^*\phi)_{\underline{3}_s}(\phi^*\phi)_{\underline{3}_s}\big], \hspace{0.65 cm}\\
&&\mathcal{V}(\varphi)=\mu^2_\varphi (\varphi^*\varphi)_{\underline{1}}
+\lambda^{\varphi} \big[(\varphi^*\varphi)_{\underline{1}}(\varphi^*\varphi)_{\underline{1}}
+(\varphi^*\varphi)_{\underline{1}^{'}}(\varphi^*\varphi)_{\underline{1}^{''}}
+(\varphi^*\varphi)_{\underline{1}^{''}}(\varphi^*\varphi)_{\underline{1}^{'}}\big], \\
&&\mathcal{V}(H,\phi)=\lambda^{H\phi} \big[(H^\+ H)_{\underline{1}}(\phi^* \phi)_{\underline{1}} +(H^\+ \phi)_{\underline{3}}(\phi^* H)_{\underline{3}}\big], \hs \mathcal{V}(H,\varphi)=\mathcal{V}(H,\phi\rightarrow \varphi),  \\
&& \mathcal{V}(H,\chi)=\lambda^{H\chi}\big[ (H^\+ H)(\chi^* \chi)+(H^\+ \chi)(\chi^* H) \big], \mathcal{V}(H,\rho)=\mathcal{V}(H,\chi\hspace{-0.1 cm}\rightarrow \hspace{-0.1 cm}\rho), \mathcal{V}(H,\eta)=\mathcal{V}(H,\chi\hspace{-0.1 cm}\rightarrow \hspace{-0.1 cm}\eta), \hspace{0.75 cm} \\
&&\mathcal{V}(\phi,\varphi)=\lambda^{\phi\varphi}\big[ (\phi^* \phi)_{\underline{1}}(\varphi^* \varphi)_{\underline{1}}
+(\phi^* \varphi)_{\underline{1}}(\varphi^* \phi)_{\underline{1}}
+(\phi^* \varphi)_{\underline{1}^{'}}(\varphi^* \phi)_{\underline{1}^{''}}
+ (\phi^* \varphi)_{\underline{1}^{''}}(\varphi^* \phi)_{\underline{1}^{'}}\crn
&&\hspace{1.35 cm} +(\phi^* \varphi)_{\underline{3}_s}(\varphi^* \phi)_{\underline{3}_s}
+(\phi^* \varphi)_{\underline{3}_a}(\varphi^* \phi)_{\underline{3}_a}\big], \, \mathcal{V}(\phi,\chi)=\lambda^{\phi\chi}\big[ (\phi^* \phi)_{\underline{1}}(\chi^* \chi)_{\underline{1}}
+(\phi^* \chi)_{\underline{3}}(\chi^* \phi)_{\underline{3}}\big], \crn
&&\mathcal{V}(\phi,\rho)=\mathcal{V}(\phi,\chi\rightarrow \rho),
\mathcal{V}(\phi,\eta)=\mathcal{V}(\phi,\chi\rightarrow\eta),
\mathcal{V}(\varphi,\chi)=\mathcal{V}(\phi\rightarrow\varphi,\chi),
 \mathcal{V}(\varphi,\rho)= \mathcal{V}(\varphi,\chi\hspace{-0.1 cm}\rightarrow\hspace{-0.1 cm}\rho), \\
&&\mathcal{V}(\varphi,\eta)=\mathcal{V}(\varphi,\chi\rightarrow\eta),
\mathcal{V}(\chi,\rho)=\lambda^{\chi\rho}\big[(\chi^* \chi)_{\underline{1}}(\rho^* \rho)_{\underline{1}}
+(\chi^* \rho)_{\underline{3}}(\rho^* \chi)_{\underline{3}}\big],
\mathcal{V}(\chi,\eta)=\mathcal{V}(\chi,\rho\rightarrow\eta), \hspace{0.75cm} \\
&&\mathcal{V}(\rho,\eta)=\lambda^{\rho\eta}\big[ (\rho^* \rho)_{\underline{1}}(\eta^*\eta)_{\underline{1}}
+(\rho^* \eta)_{\underline{3}}(\eta^* \rho)_{\underline{3}}\big], \hs
\mathcal{V}_{\mathrm{multi}}=\lambda^{\phi\varphi\chi\eta} \big[(\phi\varphi)_{\underline{1}} (\chi\eta)_{\underline{1}}
+(\phi\varphi^*)_{\underline{1}} (\chi^*\eta^*)_{\underline{1}}\big].
\eea
We will demonstrate that the VEV alignment in Eq. (\ref{scalarvev}) satisfies the minimum condition of the scalar potential with the assumption that all the VEVs are real, i.e., $\fr{\partial \mathcal{V}_{\mathrm{tot}}}{\partial v^*_\beta} =\fr{\partial \mathcal{V}_{\mathrm{tot}}}{\partial v_\beta},\, \fr{\partial^2 \mathcal{V}_{\mathrm{tot}}}{\partial v^{*2}_\beta}=\fr{\partial^2 \mathcal{V}_{\mathrm{tot}}}{\partial v^2_\beta}\equiv \delta^2_{\beta}\, (\beta =H, \phi, \varphi, \chi,\rho,\eta)$, and the
scalar potential minimum condition becomes
\bea
&&\mu_H^2=-2 \left(\lambda^{H} v^2+\lambda^{H\chi} v_\chi^2+\lambda^{H\eta} v_\eta^2+3 \lambda^{H\phi} v_\phi^2+\lambda^{H\rho} v_\rho^2+\lambda^{H\varphi} v_\varphi^2\right), \\
&&\mu_\phi^2=-2 \left(\lambda^{H\phi} v^2+7 \lambda^{\phi} v_\phi^2+ \lambda^{\phi\chi} v_\chi^2+\lambda^{\phi\eta} v_\eta^2+\lambda^{\phi\rho} v_\rho^2+\frac{5}{3} \lambda^{\phi\varphi} v_\varphi^2\right)-\frac{\lambda^{\phi\varphi\chi\eta} v_\chi v_\eta v_\varphi}{3 v_\phi}, \\
&&\mu_\varphi^2=-2 \left(\lambda^{H\varphi} v^2+5 \lambda^{\phi\varphi} v_\phi^2+3 \lambda^{\varphi} v_\varphi^2+\lambda^{\varphi\chi} v_\chi^2+\lambda^{\varphi\eta} v_\eta^2+\lambda^{\varphi\rho} v_\rho^2\right)-\frac{\lambda^{\phi\varphi\chi\eta} v_\chi v_\eta v_\phi}{v_\varphi}, \\
&&\mu^2_\chi=-2 \left(\lambda^{\chi} v_\chi^2+\lambda^{\chi\eta} v_\eta^2+\lambda^{\chi\rho} v_\rho^2+\lambda^{H\chi} v^2+3 \lambda^{\phi\chi} v_\phi^2+ \lambda^{\varphi\chi} v_\varphi^2\right)-\frac{\lambda^{\phi\varphi\chi\eta} v_\eta v_\phi v_\varphi}{v_\chi}, \\
&&\mu^2_\rho=-2 \left(\lambda^{\chi\rho} v_\chi^2+\lambda^{H\rho} v^2+3 \lambda^{\phi\rho} v_\phi^2+\lambda^{\rho} v_\rho^2+\lambda^{\rho\eta} v_\eta^2+\lambda^{\varphi\rho} v_\varphi^2\right), \\
&&\mu^2_\eta=-2 \left(\lambda^{\chi\eta} v_\chi^2+\lambda^{\eta} v_\eta^2+\lambda^{H\eta} v^2+3 \lambda^{\phi\eta} v_\phi^2+\lambda^{\rho\eta} v_\rho^2+ \lambda^{\varphi\eta} v_\varphi^2\right)-\frac{\lambda^{\phi\varphi\chi\eta} v_\chi v_\phi v_\varphi}{v_\eta},
\eea
and 
\bea
&&\delta^2_H = 4\lambda^{H} v^2>0, \hs \delta^2_\rho =4 \lambda^{\rho} v_\rho^2>0, \label{ineq1}\\
&&\delta^2_\phi =84 \lambda^{\phi} v_\phi^2-\frac{\lambda^{\phi\varphi\chi\eta} v_\chi v_\eta v_\varphi}{v_\phi}>0, \hs \delta^2_\varphi =12 \lambda^{\varphi} v_\varphi^2-\frac{\lambda^{\phi\varphi\chi\eta} v_\chi v_\eta v_\phi}{v_\varphi}>0  \label{ineq2} \\
&&\delta^2_\chi =4 \lambda^{\chi} v_\chi^2-\frac{\lambda^{\phi\varphi\chi\eta} v_\eta v_\phi v_\varphi}{v_\chi}>0, \hs\hs
\delta^2_\eta =4 \lambda^{\eta} v_\eta^2-\frac{\lambda^{\phi\varphi\chi\eta} v_\chi v_\phi v_\varphi}{v_\eta}>0. \label{ineq3}
\eea
Using the VEVs of the scalar fields in Eq. (\ref{vevscales}), Eqs. (\ref{ineq1})-(\ref{ineq3}) yield the following conditions for the scalar potential stability conditions:
\bea
  &&\lambda^{H}>0, \hs \lambda^{\rho}>0, \hs \lambda^{\chi\rho\va} < 1.6 \times 10^5 \lambda^{\chi}, \hs \lambda^{\rho\eta\phi} < 10^3 \lambda^{\rho}, \label{condit1}\\
  &&\lambda^{\eta}>0.625 \lambda^{\rho\eta\phi},  \hs \lambda^{\va}> 0.625\times 10^{-14}\lambda^{\chi\rho\va}. \label{condit3}
  \eea
Expressions (\ref{ineq1})-(\ref{ineq3}) together with the VEV in Eq. (\ref{vevscales}) imply that \bea
&&\delta^2_H=60516\lambda^{H}, \hs \delta^2_\rho = 10^{22}\lambda^{\rho}, \hs
10^{-22}\delta^2_\phi =84 \lambda^{\phi} - \lambda^{\phi\varphi\chi\eta}, \label{deltasq1}\\
&& 10^{-22}\delta^2_\varphi =12 \lambda^{\varphi}-\lambda^{\phi\varphi\chi\eta},\,10^{-22}\delta^2_\chi = 4 \lambda^{\chi} - \lambda^{\phi\varphi\chi\eta}, \, 10^{-22}\delta^2_\eta =4 \lambda^{\eta} - \lambda^{\phi\varphi\chi\eta}. \label{deltasq2}
  \label{deltasq3}
\eea
i.e., $\delta^2_H$ \, ($\delta^2_\rho$) depends on only one parameter $\lambda^{H}\, (\lambda^{\rho})$ while $\delta^2_{\phi, \varphi,\chi,\eta}$ depends on $\lambda^{\phi, \varphi,\chi,\eta}$ and $\lambda^{\phi\varphi\chi\eta}$. We find possible ranges of $\lambda^{\beta}$ and $\lambda^{\phi\varphi\chi\eta}$ such that $\delta^2_\beta$ in Eqs.(\ref{deltasq1})--(\ref{deltasq3}) are  always positive as plotted in Figs. \ref{deltaPhi} and \ref{deltaphivarchieta}, i.e., the
minimization condition of the scalar potential is always satisfied, which demonstrate the VEV in Eq. (\ref{scalarvev}) is the natural solution of the minimum condition of $\mathcal{V}_{\mathrm{tot}}$ as expected.
\section{\label{lepton}Lepton masses and mixings}

\subsection{\label{lepanaly}Analytical result for lepton sector}
Using the tensor product of $A_4$ group \cite{A420Ishimori10}, after symmetry breaking, we get the charged lepton masses and the corresponding left-and right-mixing matrices as follows:
\bea
&&m_l = \lambda_l \frac{\sqrt{3} v v_{\phi}}{\Lambda},\,\, \mathbf{V}^\+_L=\fr{1}{\sqrt{3}}\left(%
\begin{array}{ccc}
  1 &\,\,\, 1 &\,\,\, 1 \\
  1 &\,\,\, \om^2 &\,\,\, \om \\
  1 &\,\,\, \om &\,\,\, \om^2 \\
\end{array}%
\right),\, \mathbf{V}_R=\mathbf{I}_{3\times 3} \,\left(l=e,\mu,\tau;\, \om=e^{\frac{i2\pi}{3}}\right). \label{Vclep}\eea

Now, using the tensor product of $A_4$ group \cite{A420Ishimori10}, after symmetry breaking, we get the following Majorana neutrino mass matrices:
 \bea
&&m^R_{\nu N} = x_{1\nu} v_\rho \textbf{I}_{3\times 3}\equiv \mathbf{x_1} \textbf{I}_{3\times 3},
\hs\hs M^R_{\nu S} = x_{2\nu} v_\rho \textbf{I}_{3\times 3}\equiv \mathbf{x_2} \textbf{I}_{3\times 3}, \label{submatrix1}\\
&&m^{'R}_{\nu N}=  x_{3\nu}  v_\rho \textbf{I}_{3\times 3}\equiv \mathbf{x_3} \textbf{I}_{3\times 3}, \hs\hs  M^{'R}_{\nu S} = x_{4\nu}  v_\rho \textbf{I}_{3\times 3}\equiv \mathbf{x_4} \textbf{I}_{3\times 3}, \label{submatrix2} \\
&&M^R_{\mathrm{NN}}=\left(%
\begin{array}{ccc}
y_{1\nu} v_\chi & 0  & y_{2\nu} v_\varphi  \\
0 & y_{1\nu} v_\chi &0 \\
y_{2\nu} v_\varphi  & 0& y_{1\nu} v_\chi \\
\end{array}%
\right)\equiv  \left(%
\begin{array}{ccc}
\mathbf{y_1} & 0 & \mathbf{y_2} \\
0 & \mathbf{y_1} & 0 \\
\mathbf{y_2} & 0  & \mathbf{y_1} \\
\end{array}%
\right), \label{submatrix3}\\
&&M^R_{\mathrm{NS}}= \left(%
\begin{array}{ccc}
z_{1\nu} v_\chi & 0  & (z_{2\nu}-z_{3\nu}) v_\varphi  \\
  0 & z_{1\nu} v_\chi &0 \\
(z_{2\nu}+z_{3\nu}) v_\varphi  & 0& z_{1\nu} v_\chi \\
\end{array}%
\right)\equiv  \left(%
\begin{array}{ccc}
 \mathbf{z_1} & 0 & \mathbf{z_2} - \mathbf{z_3} \\
  0 & \mathbf{z_1} & 0 \\
\mathbf{z_2} + \mathbf{z_3} & 0  & \mathbf{z_1} \\
\end{array}%
\right), \label{submatrix4}\\
&&M^{'R}_{\mathrm{NS}} =\left(%
\begin{array}{ccc}
t_{1\nu} v_\chi & 0  & (t_{2\nu}-t_{3\nu}) v_\varphi  \\
0 & t_{1\nu} v_\chi &0 \\
(t_{2\nu}+t_{3\nu}) v_\varphi  & 0& t_{1\nu} v_\chi \\
\end{array}%
\right)\,\, \equiv \,\, \left(%
\begin{array}{ccc}
\mathbf{t_1} & 0 & \mathbf{t_2} - \mathbf{t_3} \\
0 & \mathbf{t_1} & 0 \\
\mathbf{t_2} + \mathbf{t_3} & 0  & \mathbf{t_1} \\
\end{array}%
\right),\hs\hs \label{submatrix5}\crn
&&M^R_{\mathrm{SS}}=\left(%
\begin{array}{ccc}
w_{1\nu} v_\chi & 0  & w_{2\nu} v_\varphi  \\
0 & w_{1\nu} v_\chi &0 \\
w_{2\nu} v_\varphi  & 0& w_{1\nu} v_\chi \\
\end{array}%
\right)\equiv  \left(%
\begin{array}{ccc}
\mathbf{w_1} & 0 & \mathbf{w_2} \\
0 & \mathbf{w_1} & 0 \\
\mathbf{w_2} & 0  & \mathbf{w_1} \\
\end{array}%
\right)\hspace{-0.1 cm}.\hspace{0.7 cm}\label{submatrix6}\eea
The mass Lagrangian for neutrino in the linear  seesaw mechanism, in the basis $(\nu_L, \nu_R^c,  N_L, N_R^C, S_L, S_R^C)$, is given by:
\bea
 -\mathcal{L}^{\mathrm{mass}}_\nu = \frac{1}{2}\left(\overline{n}_L \hs\hs \overline{n^C_R}\right)\left(%
 \begin{array}{ccc}
 m_{L} & m_{D} \\
 m_{D}^T & m_R \\
\end{array}%
\right)\left(%
\begin{array}{c}
n^C_L \\
n_R\\
\end{array}%
\right), \label{masslagrangian}
\eea
where
\bea
&&n=\left(\begin{array}{c}
\nu \\
 N \\
 S \\
\end{array}%
\right), 
\hs m_L=\left(%
\begin{array}{ccc}
m^L_{\nu\nu}&\,\,\, m^L_{\nu N} &\,\,\, M^L_{\nu S} \\
 m^{'L}_{\nu N} & M^L_{NN} &\,\,\, M^L_{NS} \\
 M^{'L}_{\nu S} & \,\,\, M^{'L}_{NS}  &M^L_{SS}\\
\end{array}%
\right), \crn
&&m_D=\left(%
\begin{array}{ccc}
m^D_{\nu\nu}&\,\,\, m^D_{\nu N} &\,\,\, M^D_{\nu S} \\
 m^{'D}_{\nu N} & M^{D}_{NN} &\,\,\, M^{D}_{NS} \\
 M^{'D}_{\nu S} & \,\,\, M^{'D}_{NS}  &M^{D}_{SS}\\
\end{array}%
\right), \hs m_R=\left(%
\begin{array}{ccc}
m^R_{\nu\nu}&\,\,\, m^R_{\nu N} &\,\,\, M^R_{\nu S} \\
 m^{'R}_{\nu N} & M^R_{NN} &\,\,\, M^R_{NS} \\
 M^{'R}_{\nu S} & \,\,\, M^{'R}_{NS}  &M^R_{SS}\\
\end{array}%
\right). \label{massmatrices}
\eea
The additional symmetries $A_4$, $U(1)_L$, $Z_4, Z_3$ and $Z_2$ make $m_L=0,\, m_D=0$ and $m^R_{\nu\nu}=0$; thus, in the considered model, the mass Lagrangian for neutrino in Eqs. (\ref{masslagrangian}) and (\ref{massmatrices}) becomes:
\bea
 -\mathcal{L}^{\mathrm{mass}}_\nu&=&\frac{1}{2}\left(\overline{\nu^c_R} \hs\hs \overline{N^c_R} \hs\hs \overline{S^c_R}\right)\left(%
\begin{array}{ccc}
0 &\,\,\, m^R_{\nu N} &\,\,\, M^R_{\nu S} \\
m^{'R}_{\nu N} & M^R_{NN} &\,\,\, M^R_{NS} \\
M^{'R}_{\nu S} & \,\,\, M^{'R}_{NS}  &M^R_{SS}\\
\end{array}%
\right)\left(%
\begin{array}{c}
  \nu_{R} \\
  N_{R} \\
  S_{R} \\
\end{array}%
\right)\equiv \frac{1}{2} \mathrm{M}_{\mathrm{eff}}\bar{n}^c_R n_R,
\eea
where
  \bea
&&M_{\mathrm{eff}}\equiv  m_R=\left(%
\begin{array}{ccc}
0&\,\,\, m^R_{\nu N} &\,\,\, M^R_{\nu S} \\
 m^{'R}_{\nu N} & M^R_{NN} &\,\,\, M^R_{NS} \\
 M^{'R}_{\nu S} & \,\,\, M^{'R}_{NS}  &M^R_{SS}\\
\end{array}%
\right) =
\left(%
 \begin{array}{ccc}
  0&M_D \\
 M^T_D & M_R \\
\end{array}%
\right),  \label{Meff0}\\ 
&&M_D =(m^R_{\nu N} \hs \,\,\, M^R_{\nu S}),\,\, M^T_D= \left(%
 \begin{array}{c}
  m^{'R}_{\nu N}\\
 M^{'R}_{\nu S} \\
\end{array}%
\right), \,\, M_R= \left(%
 \begin{array}{cc}
M^R_{NN} & M^R_{NS} \\
M^{'R}_{NS}  & M^R_{SS} \\
\end{array}%
\right),
   \eea
and the submatrix of $M_{\mathrm{eff}}$ are identified in Eqs. (\ref{submatrix1}) - (\ref{submatrix6}).
The active Majorana neutrino mass matrix, $m_\nu=- M_D M^{-1}_R M^T_D$, then gets the following form:
\bea
   M_\nu &=& \left(%
\begin{array}{ccc}
 X &0 &T\\
 0&Y&0  \\
 V &0 &Z \\
\end{array}%
\right),  \label{mnu1}
\eea
where
\bea
P&=& P_{11} x_1+P_{12} x_2 \,\, (P=X, Y,Z),\crn
Q&=& Q_{13} x_1 x_3 + Q_{14} x_1 x_4 + Q_{23} x_2 x_3 + Q_{24} x_2 x_4\,\, (Q=T, V),  \label{PQ}
\eea
with $P_{ij}$ and $Q_{mn}\, (ij=11, 12; mn=13, 14, 23, 24)$ are explicitly defined in Appendix \ref{AppenPQ}.

We now define a Hermitian matrix, $\mathbf{M}^2_\nu=M_{\nu} M^\+_{\nu}$, which gets the following form:
\bea
\mathbf{M}^2_\nu&=& \left(
\begin{array}{ccc}
 A_0 & 0 & D_0 e^{-i \al} \\
 0 & B_0 & 0 \\
 D_0 e^{i \al} & 0 & C_0 \\
\end{array}
\right),\label{Mnusq}
\eea
where
\bea
&&A_0 =T_0^2 + X_0^2, \hs
B_0 = Y_0^2, \hs
C_0 = V_0^2 + Z_0^2, \crn
&&D_0 = \sqrt{V_0^2 X_0^2 + T_0^2 Z_0^2 + 2 T_0 V_0 X_0 Z_0 \cos(\alpha_V-\alpha_X +  \alpha_T- \alpha_Z)}, \\
&&\al=\arctan \left(\frac{V_0 X_0 \sin(\alpha_V - \alpha_X) - T_0 Z_0 \sin(\alpha_T - \alpha_Z)}
{V_0 X_0 \cos(\alpha_V - \alpha_X) + T_0 Z_0 \cos(\alpha_T - \alpha_Z)}\right),  \label{alpha}\\
&&\alpha_\Omega=\mathrm{arg}(\Omega) \hs (\Omega=X, Z, T, V).\eea
and $X_{0}=|X|,\, Y_{0}=|Y|, \, Z_{0}=|Z|, \, T_{0}=|T|,\, V_{0}=|V|$ are positive and real parameters.

The matrix $\mathbf{M}^2_\nu$ owns three real positive eigenvalues
\bea
&&m^2_1=X_\nu-Y_\nu,\hs m^2_2=B_0,\hs m^2_3=X_\nu+Y_\nu, \label{m1m2m3sq}
\eea
and three respective eigenvectors are:
\bea
&&\Phi_1=\left(c_\nu \hs\hs  0 \hs\hs  s_\nu\, e^{i \al}\right)^T,\hs\hs \Phi_2=\left(0 \hs\,  1 \hs\,  0\right)^T, \hs\hs
\Phi_3=\left(-s_\nu\, e^{-i \al} \hs\hs  0 \hs\hs  c_\nu\right)^T,
\eea
where
\bea
&&X_{\nu} =\frac{A_0 + C_0}{2}, \hs Y_\nu=\frac{\sqrt{\left(A_0 - C_0\right)^2 + 4 D_0^2}}{2}, \\
&&c_\nu=\cos\varphi_\nu, \hs s_\nu=\sin\varphi_\nu, \label{scnu}
\eea
with
\bea
&&\varphi_\nu=\arctan\left(\frac{D_0}{m^2_{1}-C_0}\right)=\arctan\left(\frac{D_0}{A_0-m^2_{3}}\right).  \label{vaphinu}
\eea
Currently, the neutrino mass hierarchy is undetermined which can be inverted ordering $(m_{3}< m_1< m_{2})$ or normal ordering $(m_1< m_{2}< m_{3})$. 
Since the eigenvalue $m_2^2=B_{0}$ matches with the eigenvector $\Phi_2$, the neutrino mass ordering could be either $\left(m^2_1,\, m^2_2,\, m^2_3\right)$ or $\left(m^2_3,\, m^2_2,\, m^2_1\right)$. Therefore, eigenvalues and eigenvectors of $\mathbf{M}^2_\nu$ 
corresponding to two neutrino mass orderings are:
\bea
&\mathbf{U}_{\nu}^\+ \mathbf{M}^2_\nu \mathbf{U}_{\nu} =\left\{
\begin{array}{l}
\mathrm{diag}\left(m^2_1 \hs\hs m^2_{2} \hs\hs m^2_{3}\right),\hspace{0.25cm} \mathbf{U}_{\nu}=\left(\Phi_1\hs\hs \Phi_2 \hs\hs \Phi_3\right)  \hspace{0.3cm}\mbox{for  NO},    \\
\mathrm{diag}\left(m^2_3 \hs\hs m^2_{2} \hs\hs m^2_{1}\right),\hspace{0.25cm} \mathbf{U}_{\nu}=\left(\Phi_3\hs\hs \Phi_1\hs\hs \Phi_2\right)  \hspace{0.25cm}\,\mbox{for  IO}.
\end{array}%
\right. \label{Unuv}
\eea
The leptonic mixing matrix of the model, $\mathbf{U}_{\mathrm{Lep}}=\mathbf{V}_{L}^{\dag} \mathbf{U}_{\nu }$,
then gets the following form:
\bea
\mathbf{U}_{\mathrm{Lep}}=
\left\{
\begin{array}{l}
\hspace{-0.15 cm}\fr{1}{\sqrt{3}}\left( \begin{array}{ccc}
\hspace{-0.1 cm} c_\nu + s_\nu\, e^{i \al}      &1 & c_\nu - s_\nu\, e^{-i \al}  \\
\hspace{-0.1 cm}c_\nu + \om s_\nu \, e^{i \al}\hs &\om^2 &\hs \om c_\nu - s_\nu \, e^{-i \al}  \\
\hspace{-0.1 cm} c_\nu + \om^2 s_\nu \, e^{i \psi}&\om  & \om^2 c_\nu - s_\nu \, e^{-i \al} \\
\end{array}\hspace{-0.1 cm}\right) \hspace{0.1cm}\mbox{for NO},  \label{Ulep}  \\
\hspace{-0.15 cm}\fr{1}{\sqrt{3}}\left( \begin{array}{ccc}
\hspace{-0.1 cm} c_\nu -s_\nu \, e^{-i \al}      &1 & c_\nu + s_\nu \, e^{i \al} \\
\hspace{-0.1 cm}\om c_\nu -s_\nu \, e^{-i \al}\hs &\om^2 &\hs  c_\nu + \om s_\nu \, e^{i \al}  \\
\hspace{-0.1 cm} \om^2 c_\nu -s_\nu \, e^{-i \al} &\om  & c_\nu + \om^2 s_\nu \, e^{i \al} \\
\end{array}\hspace{-0.1 cm}\right) \hspace{0.1cm}\mbox{for IO}.
\end{array}%
\right.
\eea
The lepton mixing
 matrix in the standard parameterization takes the form
\bea
&& \mathbf{U}_{\mathrm{PMNS}} =\left(
\begin{array}{ccc}
c_{12} c_{13} & c_{13} s_{12} & e^{-i \delta} s_{13} \\
 -c_{12} s_{13} s_{23} e^{i \delta}-c_{23} s_{12} & c_{12} c_{23}-e^{i \delta} s_{12} s_{13} s_{23} & c_{13} s_{23} \\
 s_{12} s_{23}-c_{12} c_{23} e^{i \delta} s_{13} & -c_{23} s_{12} s_{13} e^{i \delta}-c_{12} s_{23} & c_{13} c_{23} \\
\end{array}
\right)\left(
\begin{array}{ccc}
e^{i\eta_{1}} & 0 & 0 \\
0 & 1 & 0 \\
0 & 0 & e^{i\eta_{2}} \\
\end{array}
\right), \label{Ulepg}
\eea where $s_{ij}=\sin \theta_{ij}$, $c_{ij}=\cos \theta_{ij}$ with
$\theta_{12}$, $\theta_{23}$ and $\theta_{13}$, 
$\delta$ is the Dirac CP phase while $\eta_{1,2}$ are two Majorana CP phases.
From the obtained lepton mixing matrix and that of parameterization in Eqs. (\ref{Ulep}) and (\ref{Ulepg}), the lepton mixing angles are defined as \cite{PDG2020}:
\bea
&&s_{12}^2 =\frac{\left| \mathbf{U}_{1 2}\right|^2}{1-\left| \mathbf{U}_{1 3}\right|^2}=
\frac{1}{3(1-s^2_{13})}\hspace{0.25cm}\,\mbox{for \, both \,NO\, and \, IO}, \label{s12sq}\\
&&s_{13}^2=\left|\mathbf{U}_{1 3}\right|^2=\left\{
\begin{array}{l}
\fr{1}{3}\left(1 - 2 s_\nu c_\nu c_\al\right)\hspace{0.25cm}\mbox{for \, NO},    \\
\fr{1}{3}\left(1 + 2 s_\nu c_\nu c_\al\right)\hspace{0.25cm}\,\mbox{for \, IO},
\end{array}%
\right.  \label{s13sq}\\
&& s_{23}^2=\frac{\left| \mathbf{U}_{2 3}\right|^2}{1-\left| \mathbf{U}_{1 3}\right|^2}=
\left\{
\begin{array}{l}
\fr{1}{2}\left(1+\frac{\sqrt{3} s_\nu c_\nu  s_\al}{c_\nu s_\nu s_\al+1}\right)\hspace{0.3cm}\mbox{for \, NO},    \\
\fr{1}{2}\left(1+\frac{\sqrt{3} s_\nu c_\nu  s_\al}{c_\nu s_\nu s_\al-1}\right)\hspace{0.25cm}\,\mbox{for \, IO},
\end{array}%
\right. \label{s23sq}\eea
where $s_\alpha=\sin\alpha, \, c_\alpha=\cos\alpha$ 
and $s_\nu, c_\nu$ are defined in Eq. (\ref{scnu}).\\
Comparing the Jarlskog invariant\cite{Jarlskog, PDG2022} in the standard parametrization\footnote{Hereafter, we use the notation $s^'_{ij}=\sin 2\theta_{ij}$, $c^'_{ij}=\cos 2\theta_{ij}$, $s^'_{\al}=\sin 2 \al$, $c^'_{\al}=\cos 2 \al$, $s^'_{\nu}=\sin 2 \varphi_\nu$, $c^'_{\nu}=\cos 2 \varphi_\nu$ and $s_\delta=\sin\de$.} $J^{S}_{CP}=
\frac{1}{8} s^'_{12} s^'_{23} s^'_{13} c_{13} \sin\delta$ and
those of the model obtained from Eq. (\ref{Ulep}), $J^{M}_{CP}=\frac{1-2 c^2_\nu}{6 \sqrt{3}}$ for NO and $J^{M}_{CP}=\frac{2 c^2_\nu-1}{6 \sqrt{3}}$ for IO, we get
\bea
s_\delta=\left\{
\begin{array}{l}
\fr{1-2c^2_\nu}{6\sqrt{3} c_{12} c_{13}^2 c_{23} s_{12} s_{13} s_{23}} \hspace{0.3cm}\mbox{for \, NO},    \\
\fr{2c^2_\nu-1}{6\sqrt{3} c_{12} c_{13}^2 c_{23} s_{12} s_{13} s_{23}}  \hspace{0.25cm}\,\mbox{for \, IO}.
\end{array}%
\right. \label{sd}
\eea \\
Equations (\ref{s13sq})-(\ref{sd}) yield the following relations:
\bea
&&c_\nu =\left\{
\begin{array}{l}
\frac{1}{\sqrt{2}}\sqrt{1+\sqrt{3 \left(c^4_{13} s^{'2}_{23}-c^{'2}_{13}\right)}} \hspace{0.25cm}\mbox{for  NO},    \\
\frac{1}{\sqrt{2}}\sqrt{1-\sqrt{3 \left(c^4_{13} s^{'2}_{23}-c^{'2}_{13}\right)}} \hspace{0.25cm}\,\mbox{for  IO},
\end{array}%
\right.  \label{costhetas12s23}\\
&&c_\al=\left\{
\begin{array}{l}
\frac{1 - 3 s^2_{13}}{2\sqrt{1 - 3 s^2_{13} c^2_{13} - 3 c^4_{13} s^2_{23} c^2_{23}}} \hspace{0.25cm}\mbox{for  NO},    \\
 \frac{3 s^2_{13}-1}{2\sqrt{1 - 3 s^2_{13} c^2_{13} - 3 c^4_{13} s^2_{23} c^2_{23}}} \hspace{0.25cm}\,\mbox{for  IO},
\end{array}%
\right. \label{cosalphas12s23}\\
&&s_\delta  =-\frac{\sqrt{4 c^4_{13} s^2_{23} c^2_{23}-c^{'2}_{13}}}{6 s_{12}c_{12} c^2_{13} s_{13}  s_{23} c_{23}} \hspace{0.25cm}\mbox{for both NO and IO}.\label{sds12s23}\eea
Comparing two corresponding entries, $"11"$ and $"13"$, of $\mathbf{U}_{\mathrm{Lep}}$ and $\mathbf{U}_{\mathrm{PMNS}}$ in Eqs. (\ref{Ulep}) and (\ref{Ulepg}) yields the following solution:
\bea
&&\eta_{1} = \left\{
\begin{array}{l}
\sec ^{-1}\left(\frac{\sqrt{3} c_{13} c_{12} }{c_\nu+c_\al s_\nu}\right) \hspace{0.75 cm}\mbox{for \, NO},    \\
\sec ^{-1}\left(\frac{\sqrt{3} c_{13} c_{12}}{c_\nu-c_\alpha s_\nu}\right) \hspace{0.74cm}\,\mbox{for \, IO},
\end{array}%
\right.  \\
&&\eta_{2}=
 \left\{
\begin{array}{l}
 \delta + \cos ^{-1}\left(\frac{c_\nu-c_\alpha s_\nu}{\sqrt{3} s_{13}}\right) \hspace{0.2 cm}\mbox{for  NO},    \\
 \delta+\cos ^{-1}\left(\frac{c_\nu+c_\alpha s_\nu}{\sqrt{3} s_{13}}\right) \hspace{0.2 cm} \mbox{for  IO}.
\end{array}%
\right.  \label{Majoranaphases}
\eea
Next, by comparing neutrino masses in Eqs. (\ref{m1m2m3sq}) and the two observed parameters $\De  m^2_{21}$ and $\De  m^2_{31}$, we get:
\bea
&&X_\nu =B_{0}-\Delta m^2_{21}+\frac{\Delta m^2_{31}}{2} \hspace{0.2 cm}\mbox{(both  NO and IO)}, \hs Y_\nu = \left\{
\begin{array}{l}
\hspace{0.1cm}\frac{\Delta m^2_{31}}{2}  \hspace{0.3cm}\mbox{for  NO},    \\
\hspace{-0.15cm}-\frac{\Delta m^2_{31}}{2}\hspace{0.2cm}\mbox{for  IO},
\end{array}%
\right. \label{XYnu}\\
&&m_1=\sqrt{B_{0}-\Delta m^2_{21}}, \,\, m_2=\sqrt{B_0}, \,\, m_3=\sqrt{B_{0}-\Delta m^2_{21}+\Delta m^2_{31}} \hs \mbox{(both  NO and IO)}, \label{m1m2m3ana} \\
&&\sum_\nu m_\nu=\sqrt{B_{0}} + \sqrt{B_{0}-\Delta m^2_{21}} + \sqrt{B_{0}-\Delta m^2_{21} +\Delta m^2_{31}} \hs \mbox{(both  NO and IO)}. 
\label{sum}
\eea
The effective Majorana neutrino mass related to neutrinoless double beta ($0\nu \beta \beta$) decay for Majorana neutrino is given by 
\bea
&&\hspace{-0.75 cm}\langle m_{ee}\rangle = \left| \sum^3_{i=1} \left(\mathbf{U}_{\mathrm{Lep}}\right)^2_{1i} m_i \right| 
=\left\{
\begin{array}{l}
\frac{1}{3}\Big\{\big[m_2 + (c^2_\nu +s^2_\nu c^'_{\al})(m_1 + m_3) +
   2 c_\nu (m_1 - m_3) s_\nu c_\al\big]^2 \\
   \hspace{1.75 cm} + \, 4 s^2_\nu \big[c_\nu (m_1 + m_3) + (m_1 - m_3) s_\nu c_\alpha\big]^2 s^2_\alpha\Big\}^{\frac{1}{2}} \hspace{0.15cm} \mbox{(NO),} \\
\frac{1}{3}\Big\{\big[m_2 + (c^2_\nu +s^2_\nu c^'_{\al})(m_1 + m_3) +
   2 c_\nu (m_3 - m_1) s_\nu c_\al\big]^2 \\
   \hspace{1.75 cm} + \, 4 s^2_\nu \big[c_\nu (m_1 + m_3) + (m_3 - m_1) s_\nu c_\alpha\big]^2 s^2_\alpha\Big\}^{\frac{1}{2}} \hspace{0.15cm} \mbox{(IO),}
\end{array}%
\right. \label{meeef}
\eea
where $m_{1,2,3}$ are three active 
neutrinos masses defined in Eq. (\ref{m1m2m3sq}), $\left(\mathbf{U}_{\mathrm{Lep}}\right)^2_{1i} \, (i=1,2,3)$ are the leptonic mixing matrix elements given in Eq. (\ref{Ulep}), and $c_\nu$ and $c_\alpha$ are given in Eqs.(\ref{costhetas12s23}) and (\ref{cosalphas12s23}), respectively.

Expressions (\ref{m1m2m3sq}), (\ref{costhetas12s23}), (\ref{cosalphas12s23}), (\ref{XYnu})-(\ref{meeef}) show that $m_1$ depends on two parameters $B_{0}$ and $\Delta m^2_{21}$; $m_3$ and the sum of neutrino mass $\sum_\nu m_\nu$ depend on three parameters $B_0$, $\Delta m^2_{21}$ and $\Delta m^2_{31}$ whereas $\langle m_{ee}\rangle$ depends on five parameters $B_0$, $\Delta m^2_{21}$, $\Delta m^2_{31}$, $\theta_{13}$ and $\theta_{23}$ in which $B_0$ is the model parameter while $\Delta m^2_{21}$, $\Delta m^2_{31}$, $\theta_{13}$ and $\theta_{23}$ are experimental parameters which have been measured with high accuracy \cite{Salas2021}.

\subsection{\label{lepnumer} Numerical analysis for lepton sector}
In the lepton sector, the model parameters are expressed in terms of the observable parameters including two neutrino squared mass differences $\Delta m^2_{21}, \Delta m^2_{31}$ and three neutrino mixing angles $\theta_{12}, \theta_{23}$ and $\theta_{13}$  as shown in Eqs. (\ref{costhetas12s23})-(\ref{Majoranaphases}) and (\ref{m1m2m3ana})-(\ref{meeef}).
 To calculate the values of the model parameters in the lepton sector, we use the observables $\Delta m^2_{21}, \Delta m^2_{31}$  and the atmospheric ($\theta_{23}$) and the reactor ($\theta_{13}$) angles whose experimental values are shown in Tab. \ref{Salas2021T}.
Equations (\ref{s12sq}), (\ref{costhetas12s23})-(\ref{Majoranaphases}) and (\ref{m1m2m3ana})-(\ref{meeef}) indicate that $\theta_{12}$ depends on $\theta_{13}$; $c_{\nu}, c_{\alpha}, s_{\delta}$ and two Majorana phases $\eta_{1,2}$ depend on $\theta_{13}$ and $\theta_{23}$  while three active neutrino masses $m_{1,2,3}$ and their sum depend on $\Delta m^2_{21}, \Delta m^2_{31}$ and $B_0$, and the effective Majorana neutrino mass $\langle m_{ee}\rangle$ depends on $\Delta m^2_{21}, \Delta m^2_{31}$, $B_0$, $\theta_{13}$ and $\theta_{23}$. At the best fit points of two neutrino squared mass differences and three neutrino mixing angles, three active neutrino masses $m_{1,2,3}$, the sum of neutrino mass $\sum_\nu m_\nu$ as well as the effective Majorana neutrino mass $\langle m_{ee}\rangle$ depends on only one parameter $B_0$. We find the possible ranges of $B_0$ with $B_{0} \in (75, 10^4) \,\mathrm{meV}^2$ for NO and $B_{0} \in (2.525\times 10^3, 10^4) \, \mathrm{meV}^2$ for IO such that $\sum_\nu m_\nu$ and $\langle m_{ee}\rangle$ get the predictive ranges being consistent with their experimental constraints as shown in Tabs. \ref{sumconstraint} and \ref{meeconstraint}, respectively.

\begin{figure}[ht]
\begin{center}
\hspace{-6.25 cm}
\includegraphics[width=0.55\textwidth]{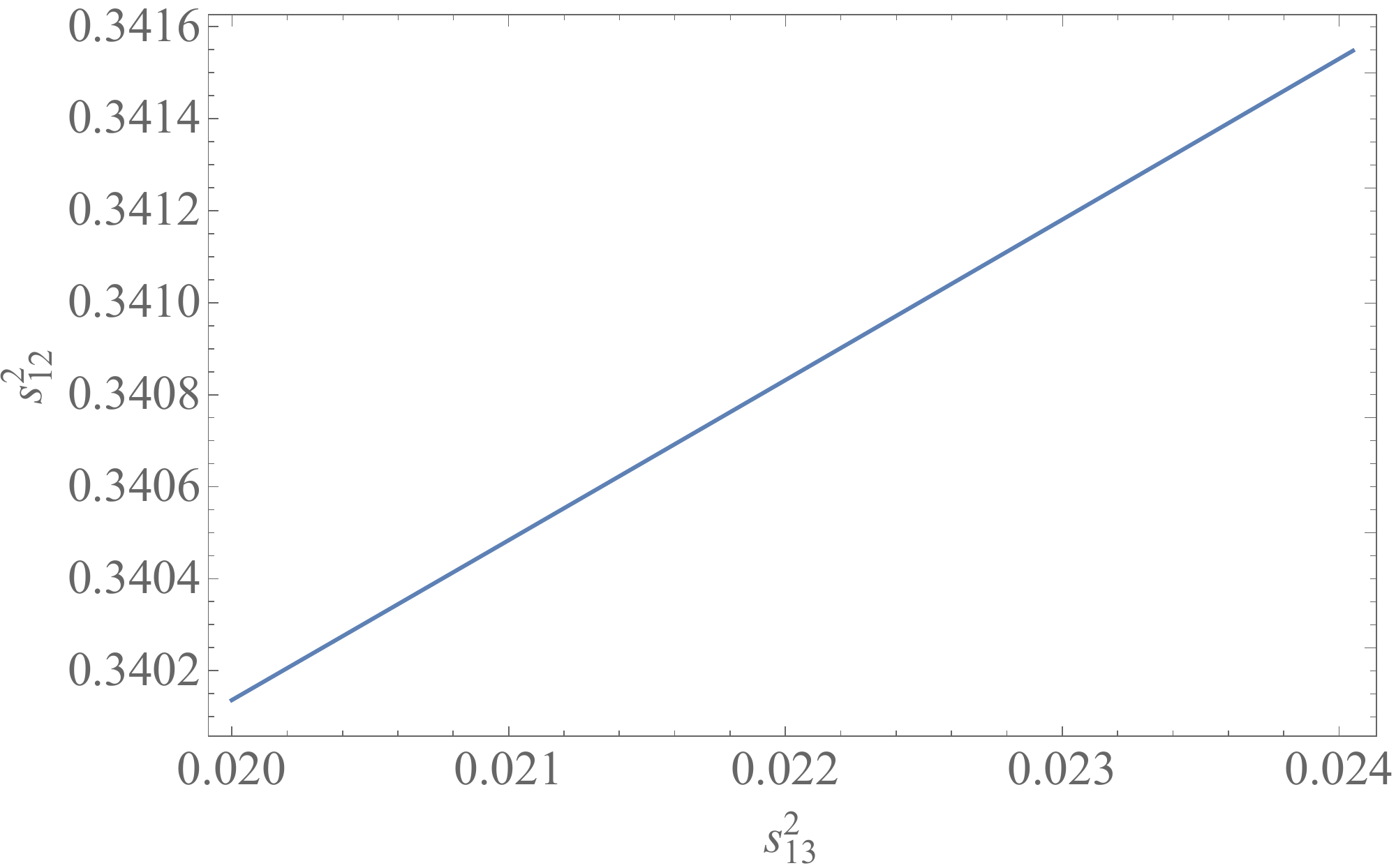}\hspace{0.0 cm}
\vspace{-0.95 cm}
\hspace*{-6.25 cm}
\end{center}
\vspace{0.25 cm}
\caption[$s^2_{12}$ versus $s^2_{13}$ with $s^2_{13}\in (2.000, 2.405) 10^{-2}$.
]{$s^2_{12}$ versus $s^2_{13}$ with $s^2_{13}\in (2.000, 2.405) 10^{-2}$.
}
\label{s12s13sqrelat}
\vspace{-0.25 cm}
\end{figure}

\begin{figure}[ht]
\begin{center}
\vspace{-0.5 cm}
\hspace{-6.0 cm}
\includegraphics[width=0.825\textwidth]{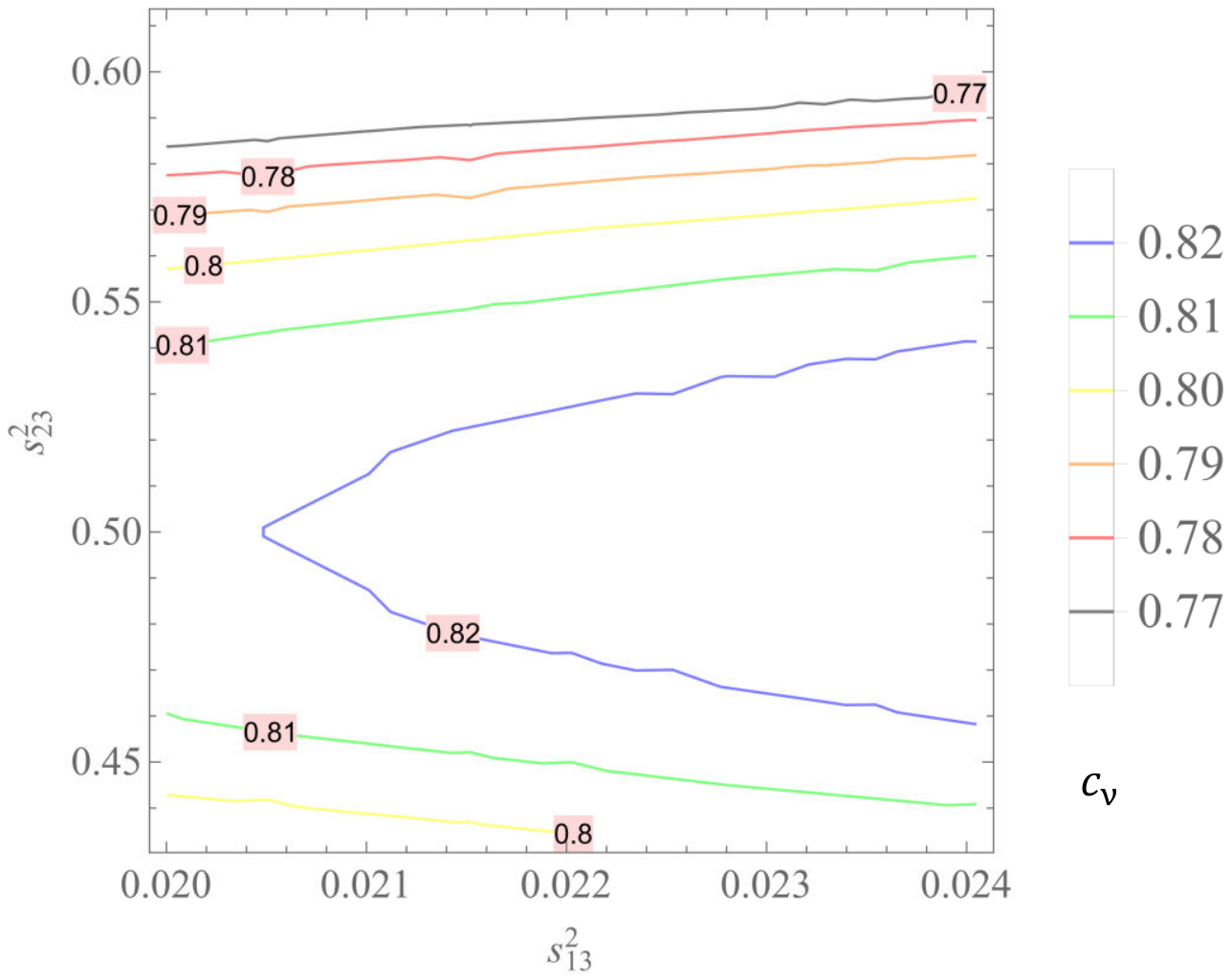}\hspace{-5.25 cm}
\vspace{-0.95 cm}
\includegraphics[width=0.825\textwidth]{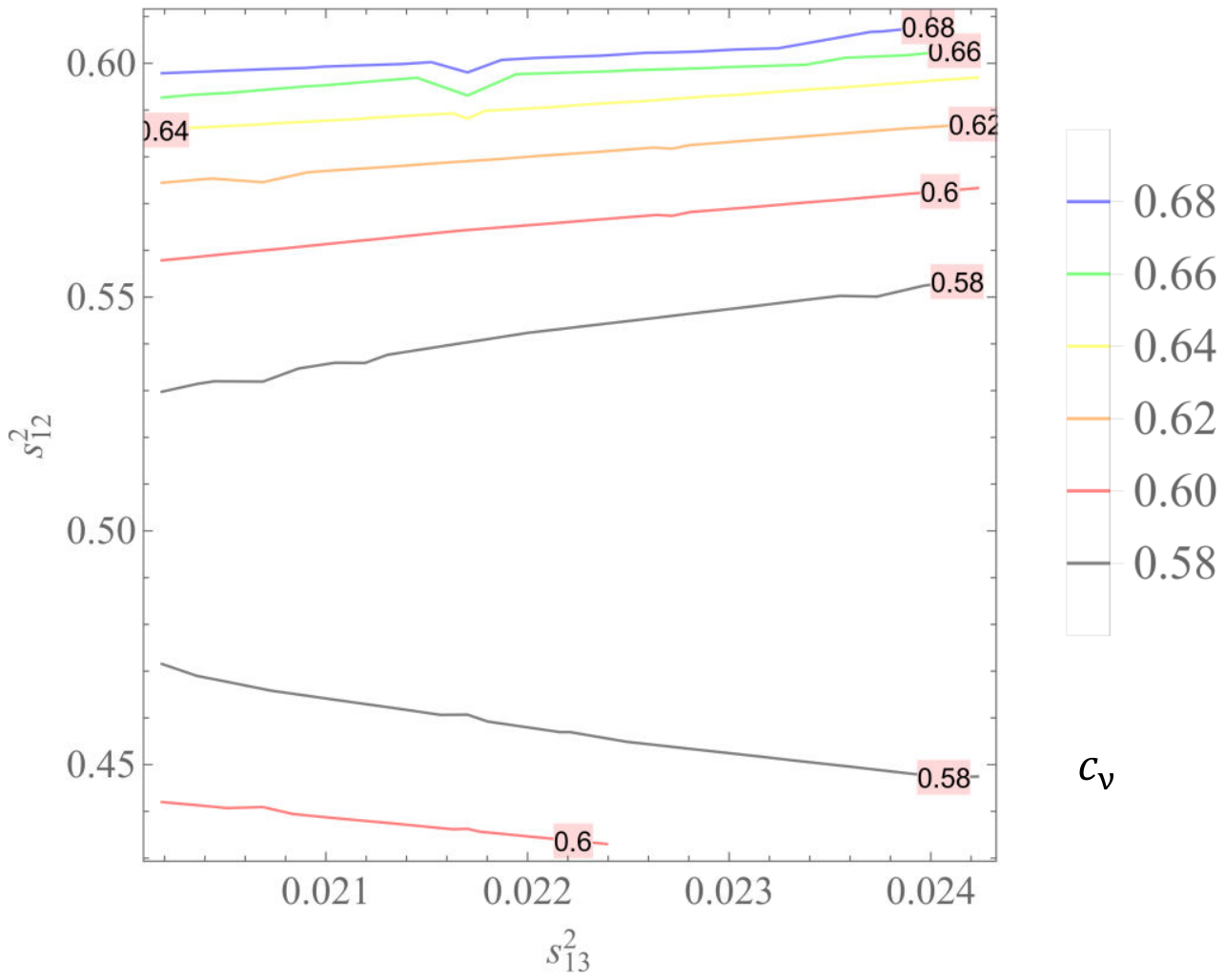}\hspace*{-5.25 cm}
\end{center}
\vspace{-9.5 cm}
\caption[(Colored lines) $c_\nu$ depends on $s^2_{23}$ and $s^2_{13}$ with $s^2_{23}\in (0.434,\, 0.610)$ and $s^2_{13}\in (2.000,\, 2.405) 10^{-2}$ for NO (left panel) while $s^2_{23}\in (0.433,\, 0.608)$  and $s^2_{13}\in (2.018,\, 2.424) 10^{-2}$ for IO (right panel).]{(Colored lines) $c_\nu$ depends on $s^2_{23}$ and $s^2_{13}$ with $s^2_{23}\in (0.434,\, 0.610)$ and $s^2_{13}\in (2.000,\, 2.405) 10^{-2}$ for NO (left panel) while $s^2_{23}\in (0.433,\, 0.608)$ and $s^2_{13}\in (2.018,\, 2.424) 10^{-2}$ for IO (right panel).}
\label{cosnuF}
\vspace{-0.25 cm}
\end{figure}

\begin{figure}[ht]
\begin{center}
\vspace{-0.5 cm}
\hspace{-6.0 cm}
\includegraphics[width=0.825\textwidth]{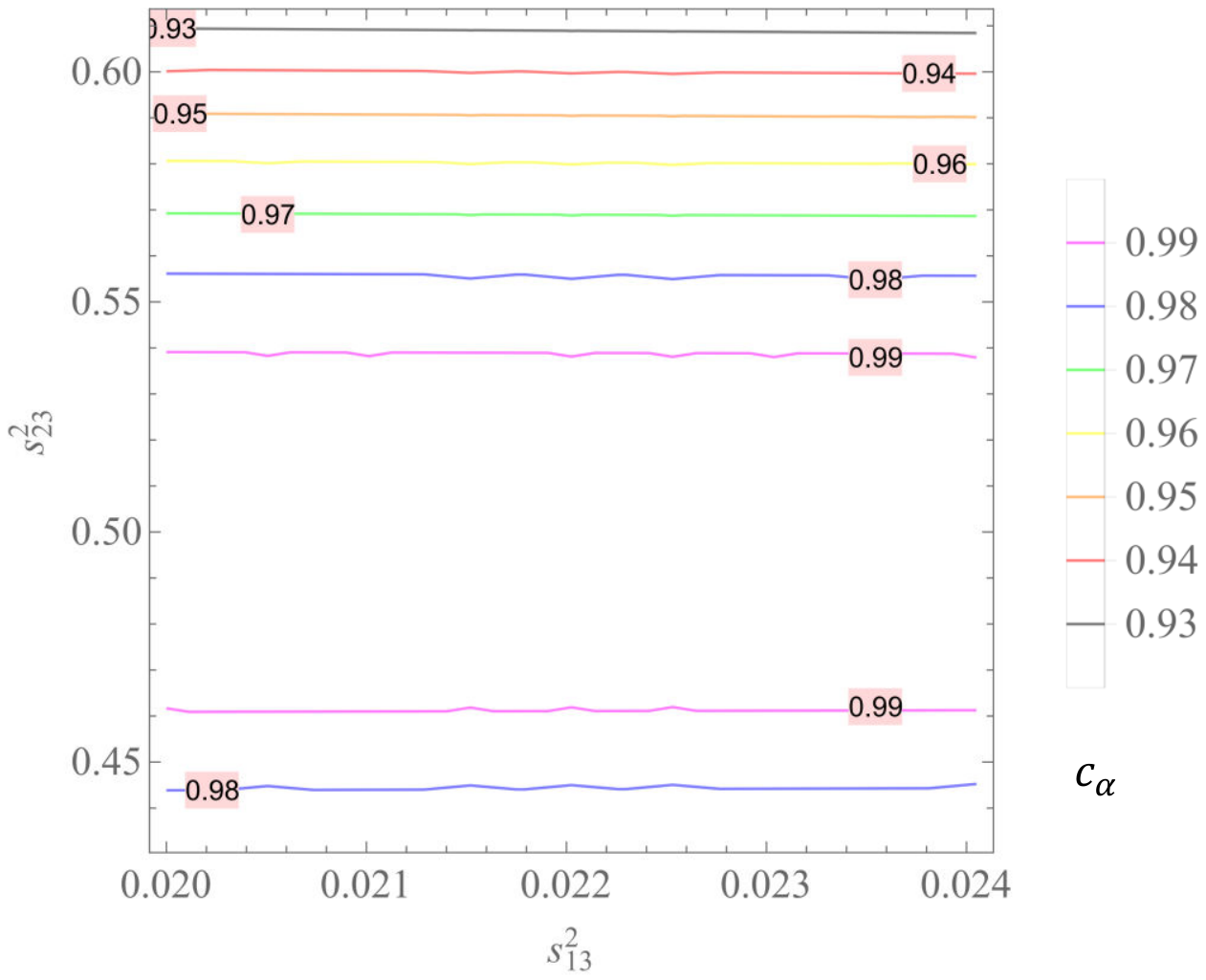}\hspace{-5.25 cm}
\vspace{-0.95 cm}
\includegraphics[width=0.825\textwidth]{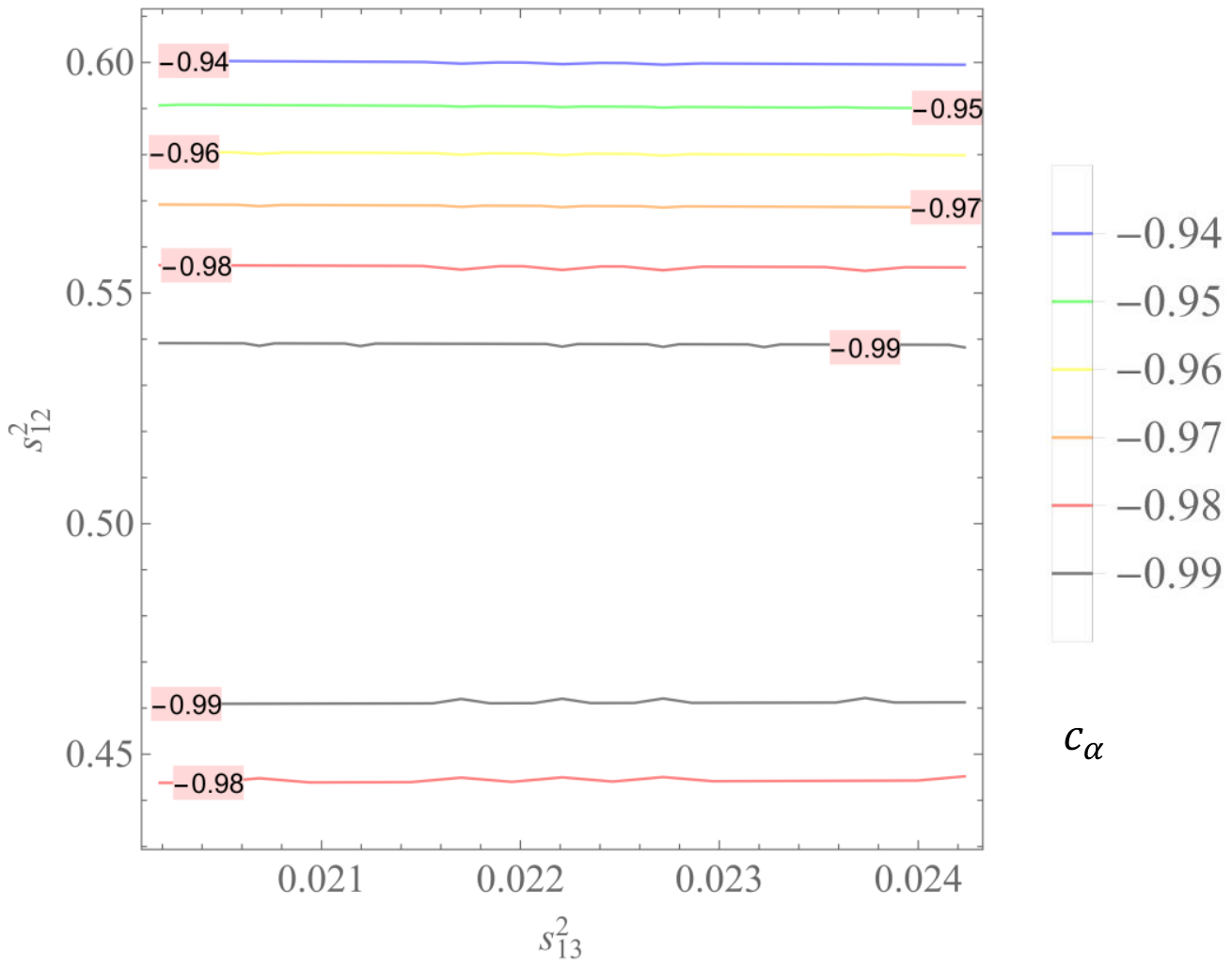}\hspace*{-5.25 cm}
\end{center}
\vspace{-9.5 cm}
\caption[(Colored lines) $c_\alpha$
depends on $s^2_{23}$ and $s^2_{13}$ with $s^2_{23}\in (0.434,\, 0.610)$ and $s^2_{13}\in (2.000,\, 2.405) 10^{-2}$ for NO (left panel) while $s^2_{23}\in (0.433,\, 0.608)$ $s^2_{13}\in (2.018,\, 2.424) 10^{-2}$ for IO (right panel).]
{(Colored lines) $c_\alpha$ depends on $s^2_{23}$ and $s^2_{13}$ with $s^2_{23}\in (0.434,\, 0.610)$ and $s^2_{13}\in (2.000,\, 2.405) 10^{-2}$ for NO (left panel) while $s^2_{23}\in (0.433,\, 0.608)$ $s^2_{13}\in (2.018,\, 2.424) 10^{-2}$ for IO (right panel).}
\label{cosalF}
\vspace{-0.25 cm}
\end{figure}

\begin{figure}[ht]
\begin{center}
\vspace{-1.0 cm}
\hspace{-3.0 cm} \includegraphics[width=0.825\textwidth]{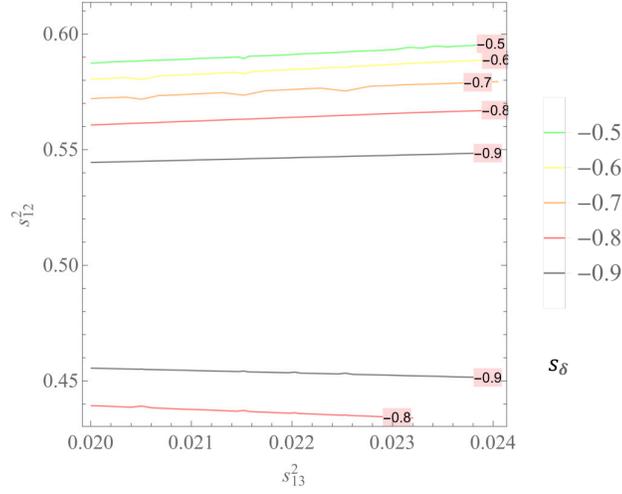}\hspace*{-3.0 cm}
\end{center}
\vspace{-10.5 cm}
\caption[(Colored lines) $\sin\delta$ depends on $s^2_{23}$ and $s^2_{13}$ with $s^2_{23}\in (0.434,\, 0.610)$ and $s^2_{13}\in (2.000,\, 2.405) 10^{-2}$. 
]
{(Colored lines) $\sin\delta$ depends on $s^2_{23}$ and $s^2_{13}$ with $s^2_{23}\in (0.434,\, 0.610)$ and $s^2_{13}\in (2.000,\, 2.405) 10^{-2}$.
}
\label{sdeltaF}
\end{figure}

\begin{figure}[ht]
\begin{center}
\vspace{-1.75 cm}
\hspace{-6.0 cm}
\includegraphics[width=0.825\textwidth]{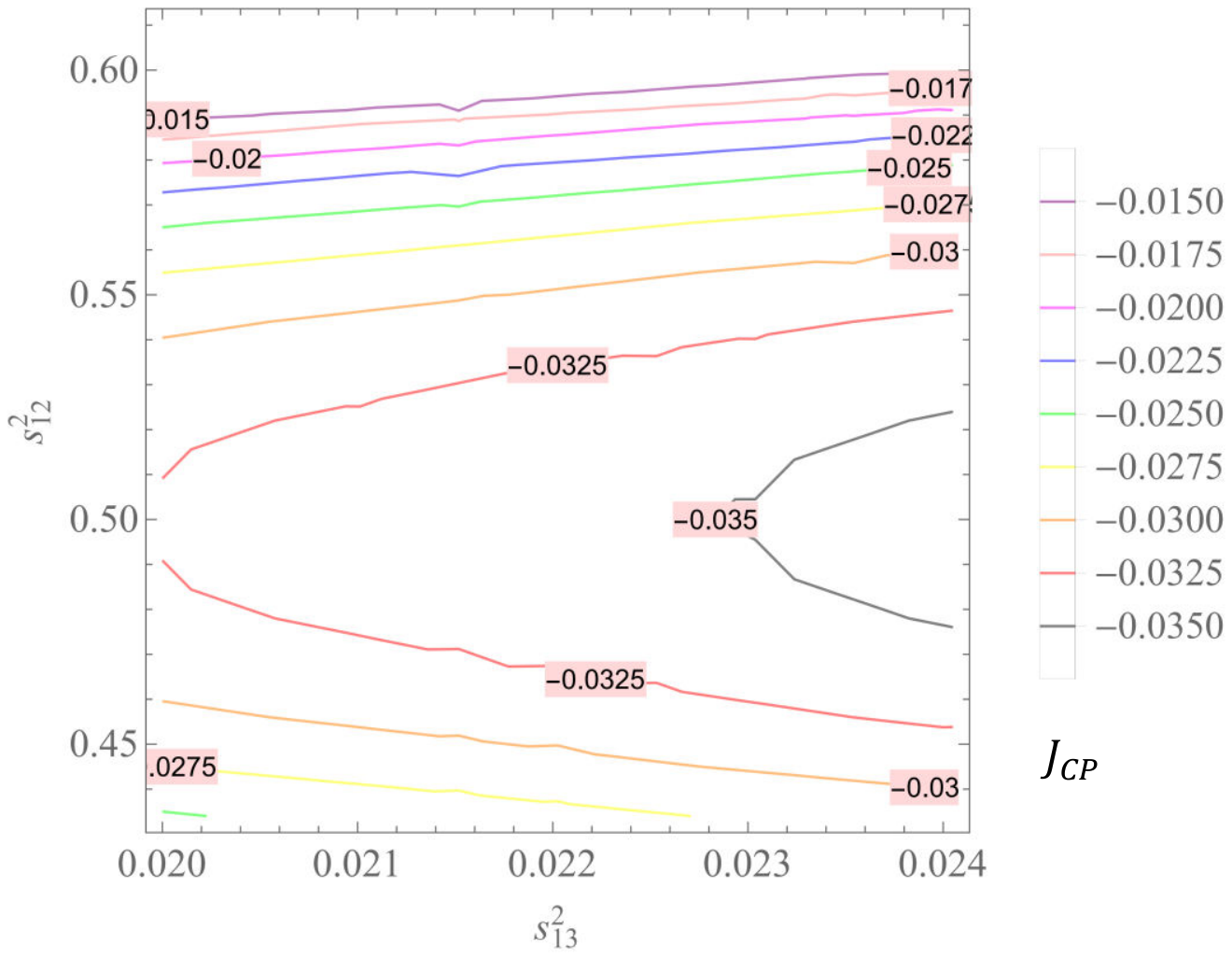}\hspace{-5.2 cm}
\vspace{-0.5 cm}
\includegraphics[width=0.825\textwidth]{countourplotjn.pdf}\hspace*{-5.25 cm}
\end{center}
\vspace{-10.0 cm}
\caption[(Colored lines) $J_{CP}$ depends on $s^2_{23}$ and $s^2_{13}$ with $s^2_{23}\in (0.434,\, 0.610)$ and $s^2_{13}\in (2.000,\, 2.405) 10^{-2}$ for NO (left panel) while $s^2_{23}\in (0.433,\, 0.608)$ $s^2_{13}\in (2.018,\, 2.424) 10^{-2}$ for IO (right panel).]
{(Colored lines) $J_{CP}$ depends on $s^2_{23}$ and $s^2_{13}$ with $s^2_{23}\in (0.434,\, 0.610)$ and $s^2_{13}\in (2.000,\, 2.405) 10^{-2}$ for NO (left panel) while $s^2_{23}\in (0.433,\, 0.608)$ $s^2_{13}\in (2.018,\, 2.424) 10^{-2}$ for IO (right panel).}
\label{JcpF}
\vspace{-0.5 cm}
\end{figure}

Expression (\ref{s12sq}) shows the relation between $s^2_{12}$ and $s^2_{13}$. Since the experimental result for $s^2_{13}$ have a higher accuracy than that of $s^2_{12}$ \cite{Salas2021}, we will determine the possible region\footnote{In fact, there is very slight difference between the experimental limits of $s_{13}$ for NO and IO, Eq. (\ref{s12sq}) indicates that the predictive ranges of $s_{12}$ are the same for NO and IO, we thus plot in Fig. \ref{s12s13sqrelat} only for the NO.} of $s^2_{12}$ based on the experimental region of $s^2_{13}$ at $3\, \si$ range, i.e., $s^2_{13}\in (2.000, 2.405) 10^{-2}$, 
as plotted in Fig. \ref{s12s13sqrelat}, which implies
\bea
s^2_{12} \in (0.3402, 0.3416) \hs \mathrm{,i.e.,}\hs \theta_{12}\in (35.68, 35.77)^\circ. \label{t12sqrange}
\eea
Expressions (\ref{costhetas12s23})-(\ref{sds12s23}) show that $c_\nu, c_\alpha$ and $s_\delta$ depend on two parameters $s^2_{23}$ and $s^2_{13}$ which are plotted in Figs. \ref{cosnuF}, \ref{cosalF} and \ref{sdeltaF}, respectively. 
\\
Figures \ref{cosnuF} and  \ref{cosalF} imply that the predictive ranges $c_\nu$ and $c_\alpha$ are
\bea
&&c_\nu  \in \left\{
\begin{array}{l}
(0.770, 0.820) \hspace{0.3cm}\mbox{for  NO},    \\
(0.580, 0.680) \hspace{0.25cm}\,\mbox{for  IO},
\end{array}%
\right. \,\,\,\,\,\mathrm{i.e.}, \,\, \theta_\nu (^\circ) \in \left\{
\begin{array}{l}
(34.920,\, 39.650) \hspace{0.3cm}\mbox{for  NO},    \\
(47.160,\, 54.550) \hspace{0.25cm}\,\mbox{for  IO}.
\end{array}%
\right. \hs\hs \label{thetanu1}\\
&&c_\alpha \hspace{0.1 cm}\in \left\{
\begin{array}{l}
(0.930, 0.990) \hspace{0.3cm}\mbox{for  NO},    \\
(-0.990, -0.940) \hspace{0.05cm}\,\mbox{for  IO},
\end{array}%
\right. \hspace{0.075 cm}\mathrm{i.e.}, \psi (^\circ) \in \left\{
\begin{array}{l}
(8.110,\, 21.570) \hspace{0.75cm}\mbox{for NO},    \\
(160.100,\, 171.900) \hspace{0.1 cm}\,\mbox{for  IO}.
\end{array}%
\right. \label{psi1}
\eea
Besides, Fig.\ref{sdeltaF} implies that the Dirac CP phase is estimated to be\footnote{In fact, there are very slight differences between the experimental ranges of $s_{13}$ and $s_{23}$ for NO and IO, Eq. (\ref{sds12s23}) tells us that the predictive ranges of $\sin\delta$ are the same for NO and IO, we thus plot in Fig. \ref{sdeltaF} only for the NO.}
\bea
\sin\delta  \in (-0.90, -0.50),\hs \mathrm{i.e.}, \hs \delta(^\circ)  \in (259.80, 330.00) \hs \mbox{for both NO and IO,}\label{delta1}
\eea
which are in consistent with the $3\, \si $ range 
taken from Ref. \cite{Salas2021}.
The Jarlskog invariant $J_{CP}$ depends on $s^2_{23}$ and $s^2_{13}$, plotted in Fig. \ref{JcpF}, which implies 
\bea
J_{CP}&=& (-3.50, -1.50) 10^{-2} \hspace{0.3cm}\mbox{for both NO and IO}.
\eea

\begin{figure}[ht]
\begin{center}
\vspace{-0.5 cm}
\hspace{-6.0 cm}
\includegraphics[width=0.825\textwidth]{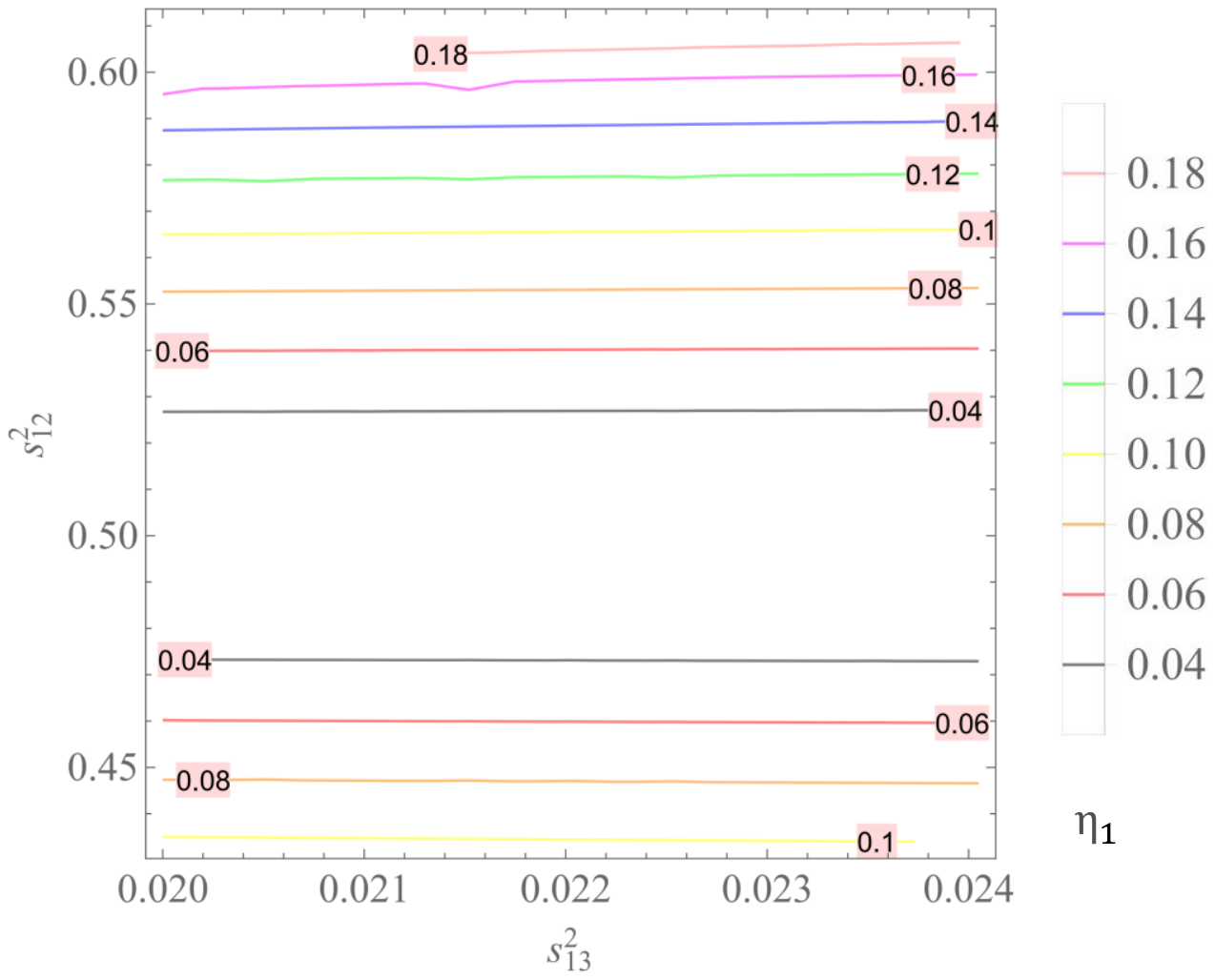}\hspace{-5.25 cm}
\vspace{-0.95 cm}
\includegraphics[width=0.825\textwidth]{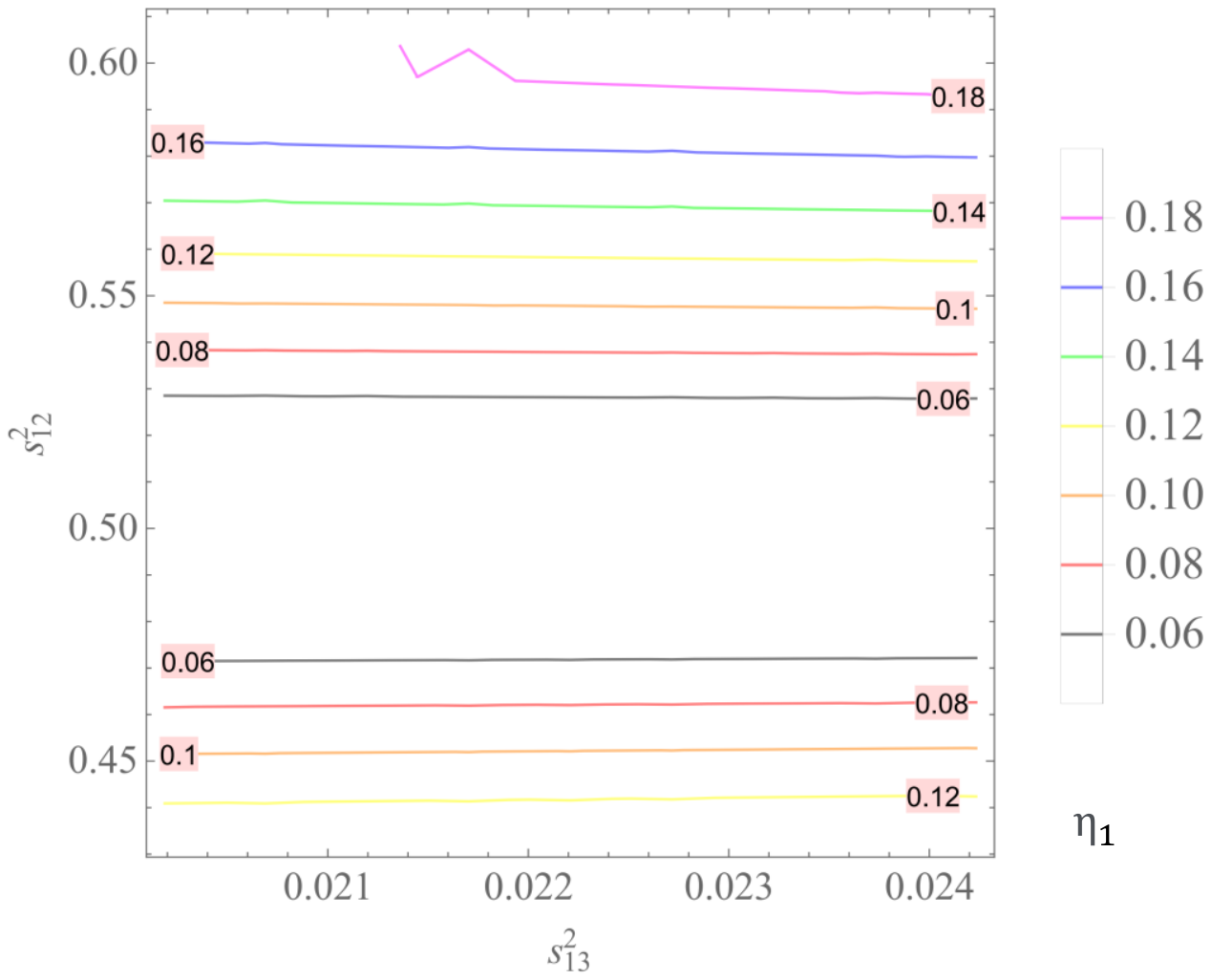}\hspace*{-5.25 cm}
\end{center}
\vspace{-9.5 cm}
\caption[(Colored lines) $\eta_1$ depends on $s^2_{23}$ and $s^2_{13}$ with $s^2_{23}\in (0.434,\, 0.610)$ and $s^2_{13}\in (2.000,\, 2.405) 10^{-2}$ for NO (left panel) while $s^2_{23}\in (0.433,\, 0.608)$  and $s^2_{13}\in (2.018,\, 2.424) 10^{-2}$ for IO (right panel).]{(Colored lines) $\eta_1$ depends on $s^2_{23}$ and $s^2_{13}$ with $s^2_{23}\in (0.434,\, 0.610)$ and $s^2_{13}\in (2.000,\, 2.405) 10^{-2}$ for NO (left panel) while $s^2_{23}\in (0.433,\, 0.608)$ and $s^2_{13}\in (2.018,\, 2.424) 10^{-2}$ for IO (right panel).}
\label{eta1F}
\vspace{-0.25 cm}
\end{figure}

\begin{figure}[ht]
\begin{center}
\vspace{-0.5 cm}
\hspace{-6.0 cm}
\includegraphics[width=0.825\textwidth]{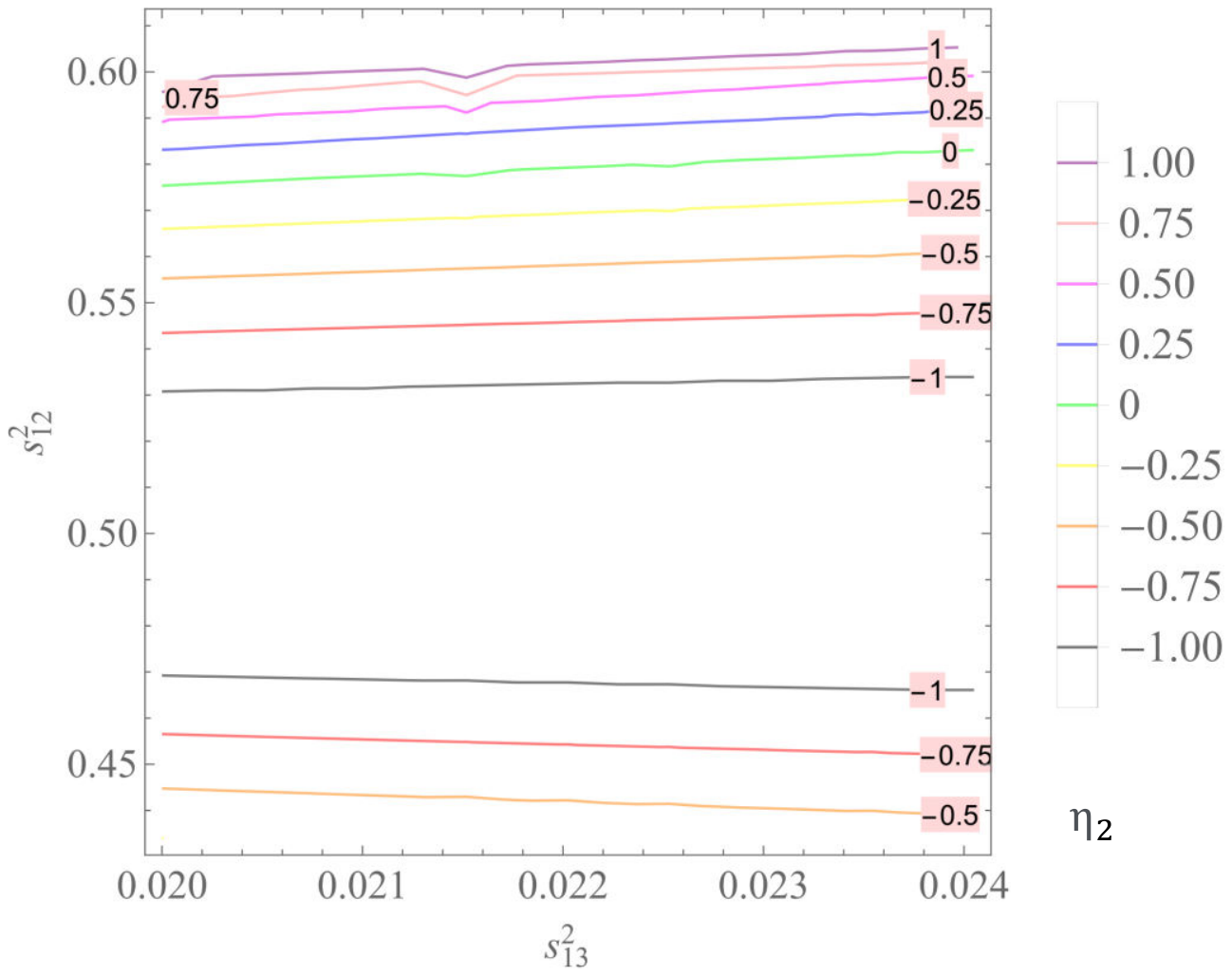}\hspace{-5.25 cm}
\vspace{-0.95 cm}
\includegraphics[width=0.825\textwidth]{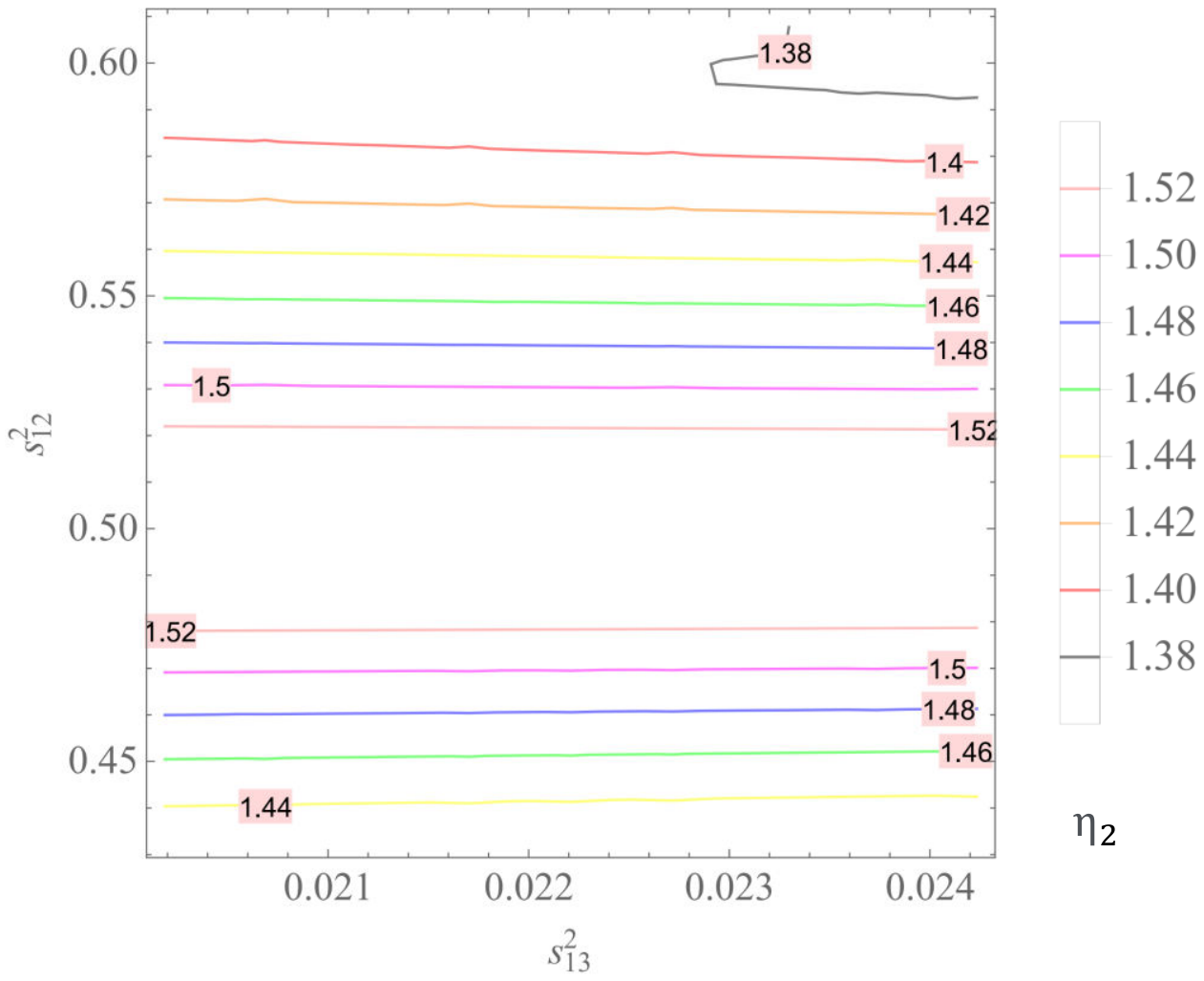}\hspace*{-5.25 cm}
\end{center}
\vspace{-9.5 cm}
\caption[(Colored lines) $\eta_2$ depends on $s^2_{23}$ and $s^2_{13}$ with $s^2_{23}\in (0.434,\, 0.610)$ and $s^2_{13}\in (2.000,\, 2.405) 10^{-2}$ for NO (left panel) while $s^2_{23}\in (0.433,\, 0.608)$  and $s^2_{13}\in (2.018,\, 2.424) 10^{-2}$ for IO (right panel).]{(Colored lines) $\eta_2$ depends on $s^2_{23}$ and $s^2_{13}$ with $s^2_{23}\in (0.434,\, 0.610)$ and $s^2_{13}\in (2.000,\, 2.405) 10^{-2}$ for NO (left panel) while $s^2_{23}\in (0.433,\, 0.608)$ and $s^2_{13}\in (2.018,\, 2.424) 10^{-2}$ for IO (right panel).}
\label{eta2F}
\vspace{-0.25 cm}
\end{figure}

The entries of $\mathbf{U}_{\mathrm{Lep}}$ 
in Eq. (\ref{Ulep}) get the following ranges:
\bea
|\mathbf{U}_{\mathrm{Lep}}|\in \left\{
\begin{array}{l}
\left(%
\begin{array}{ccc}
0.802\rightarrow 0.806    &\hs \frac{1}{\sqrt{3}} \hs&0.145\rightarrow0.165  \\
0.280\rightarrow0.380  &\hs \frac{1}{\sqrt{3}} \hs&0.720\rightarrow 0.790 \\
0.450\rightarrow 0.530&\hs \frac{1}{\sqrt{3}} \hs& 0.580\rightarrow 0.680 \\
\end{array}%
\hspace{0.05cm}\right) \hspace{0.35cm}\mbox{for  NO},    \\
\left(%
\begin{array}{ccc}
0.799\rightarrow 0.804    &\hs \frac{1}{\sqrt{3}} \hs&0.130\rightarrow0.155  \\
0.460\rightarrow0.0.560  &\hs \frac{1}{\sqrt{3}} \hs &0.620\rightarrow 0.680 \\
0.240\rightarrow 0.380&\hs \frac{1}{\sqrt{3}} \hs& 0.720\rightarrow 0.770 \\
\end{array}%
\right)\hspace{0.1cm}\mbox{for  IO.}
\end{array}%
\right. \label{Ulepranges}
\vspace{-0.5 cm}
\eea
Expressions (\ref{costhetas12s23})-(\ref{Majoranaphases}) show that $\eta_{1,2}$ depend on two parameters $s^2_{23}$ and $s^2_{13}$ which are plotted in Figs. \ref{eta1F} and \ref{eta2F}, respectively.
Figures \ref{eta1F} and \ref{eta2F} tell us that the possible regions of $\eta_{1,2}$ are
\bea
&&\eta_{1} \in \left\{
\begin{array}{l}
(0.04, 0.18)\, \mathrm{rad}\sim (2.29, 10.31)^\circ  \hspace{0.2cm}\mbox{for  NO},    \\
(0.06, 0.18)\, \mathrm{rad}\sim (3.44, 10.31)^\circ \hspace{0.2cm}\mbox{for  IO},
\end{array}%
\right. \label{eta1range}\\
&&\eta_{2} \in \left\{
\begin{array}{l}
(-1.00, 1.00)\, \mathrm{rad}\sim (57.30, 302.70)^\circ \hspace{0.1cm}\mbox{for  NO},    \\
(1.38, 1.52)\,\, \mathrm{rad}\,\,\sim\, (79.07, 87.09)^\circ \hspace{0.3cm}\mbox{for  IO}.
\end{array}%
\right. \label{eta2range}
\eea
The predicted Majorana phases in Eqs. (\ref{eta1range}) and (\ref{eta2range}) are acceptable since they are assumed to be in $[0,\, 2\pi]$ \cite{PDG2022}.

Here are some comments related to lepton mixing angles:
\begin{itemize}
  \item [(1)] The obtained solar neutrino mixing angle $\theta_{12}$ and the obtained Dirac CP violating phase are in $2\sigma$ range for both NO and IO.
  \item [(2)] The obtained Majorana violating phases, $\eta_{1} \in (2.29, 10.31)^\circ$ and $\eta_{2} \in (57.30, 302.70)^\circ $ for NO while $\eta_1\in (3.44, 10.31)^\circ$ and $\eta_{2} \in(79.07, 87.09)^\circ$ for  IO which are feasible since they are assumed to be in $[0,\, 2\pi]$.
\end{itemize}

\begin{figure}[h]
\begin{center}
\vspace*{0.25 cm}
\hspace*{-0.5cm}
\includegraphics[width=0.5\textwidth]{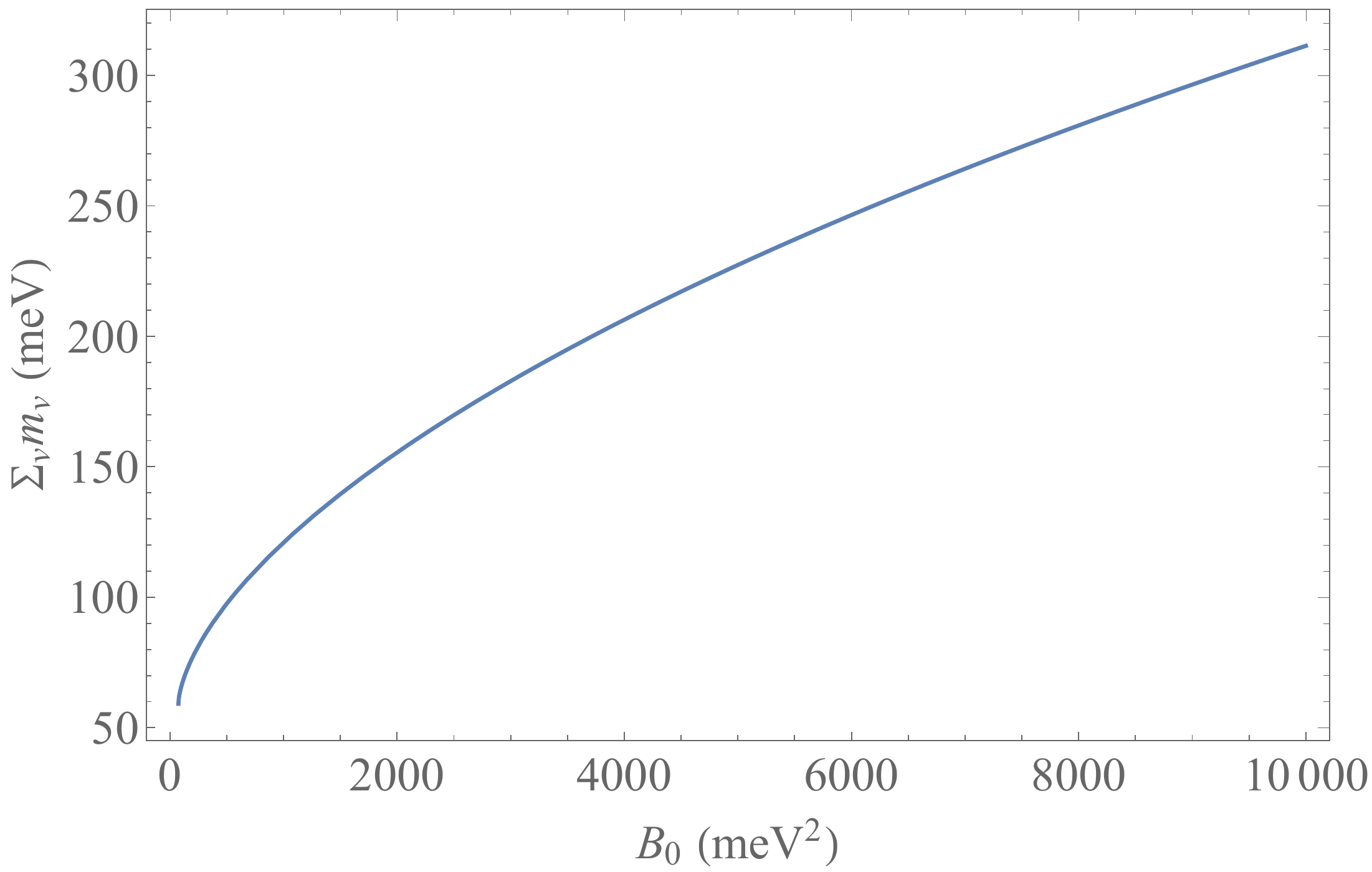}\hspace*{0.1cm}
\includegraphics[width=0.5\textwidth]{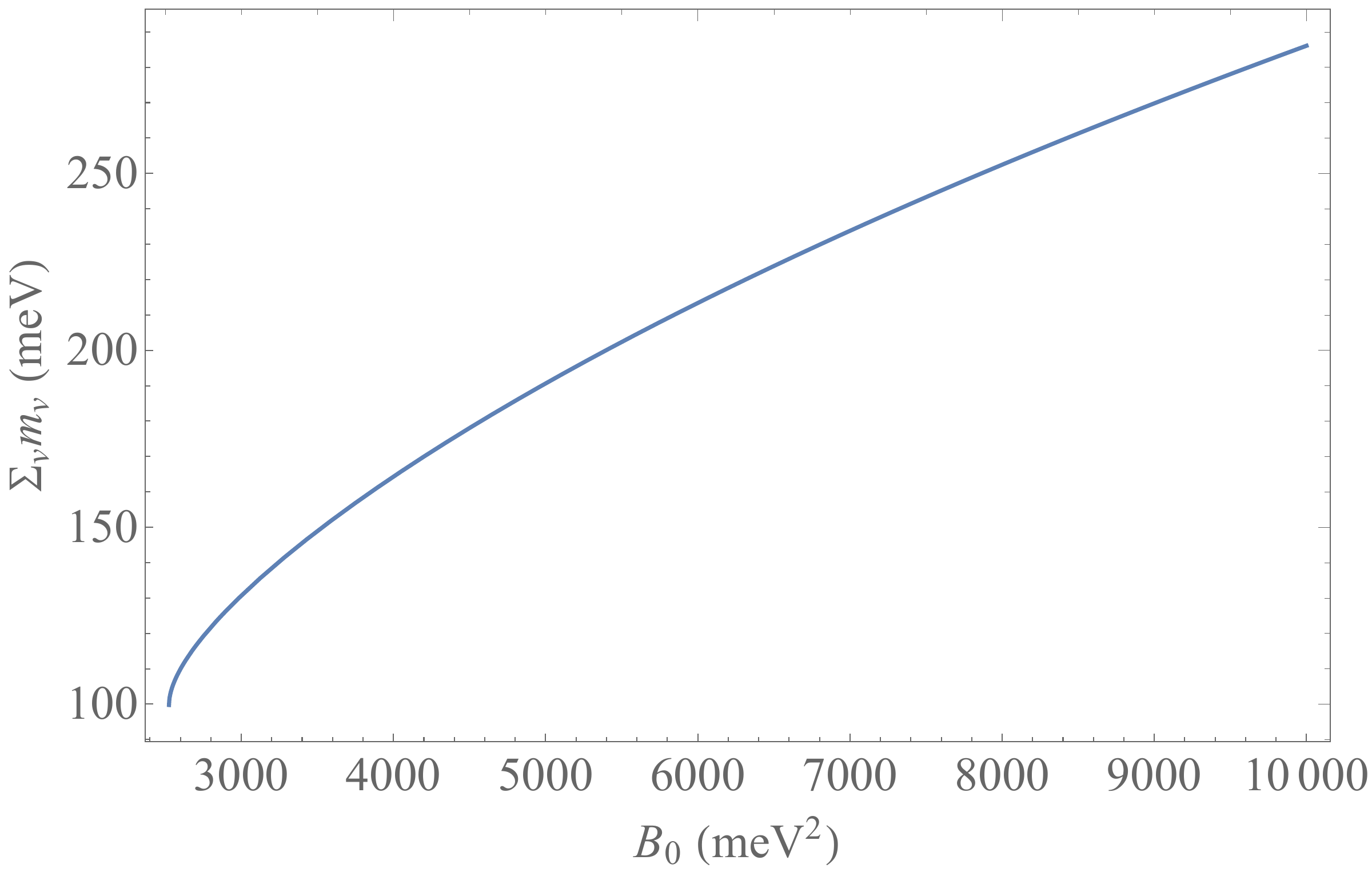}
\vspace*{-0.2cm}
\caption[$\sum_\nu m_\nu$\, (in meV) depends on $B_{0}$ with $B_{0} \in (75, 10^4) \,\mathrm{meV}^2$ for NO (left panel), and $B_{0} \in (2.525\times 10^3, 10^4) \, \mathrm{meV}^2$ for IO (right panel).]{$\sum_\nu m_\nu$\, (in meV) depends on $B_{0}$ with $B_{0} \in (75, 10^4) \,\mathrm{meV}^2$ for NO (left panel), and $B_{0} \in (2.525\times 10^3, 10^4) \, \mathrm{meV}^2$ for IO (right panel).}\label{sumF}
\vspace*{-0.5cm}
\end{center}
\end{figure}

\begin{figure}[h]
\begin{center}
\vspace*{0.25 cm}
\hspace*{-0.5cm}
\includegraphics[width=0.5\textwidth]{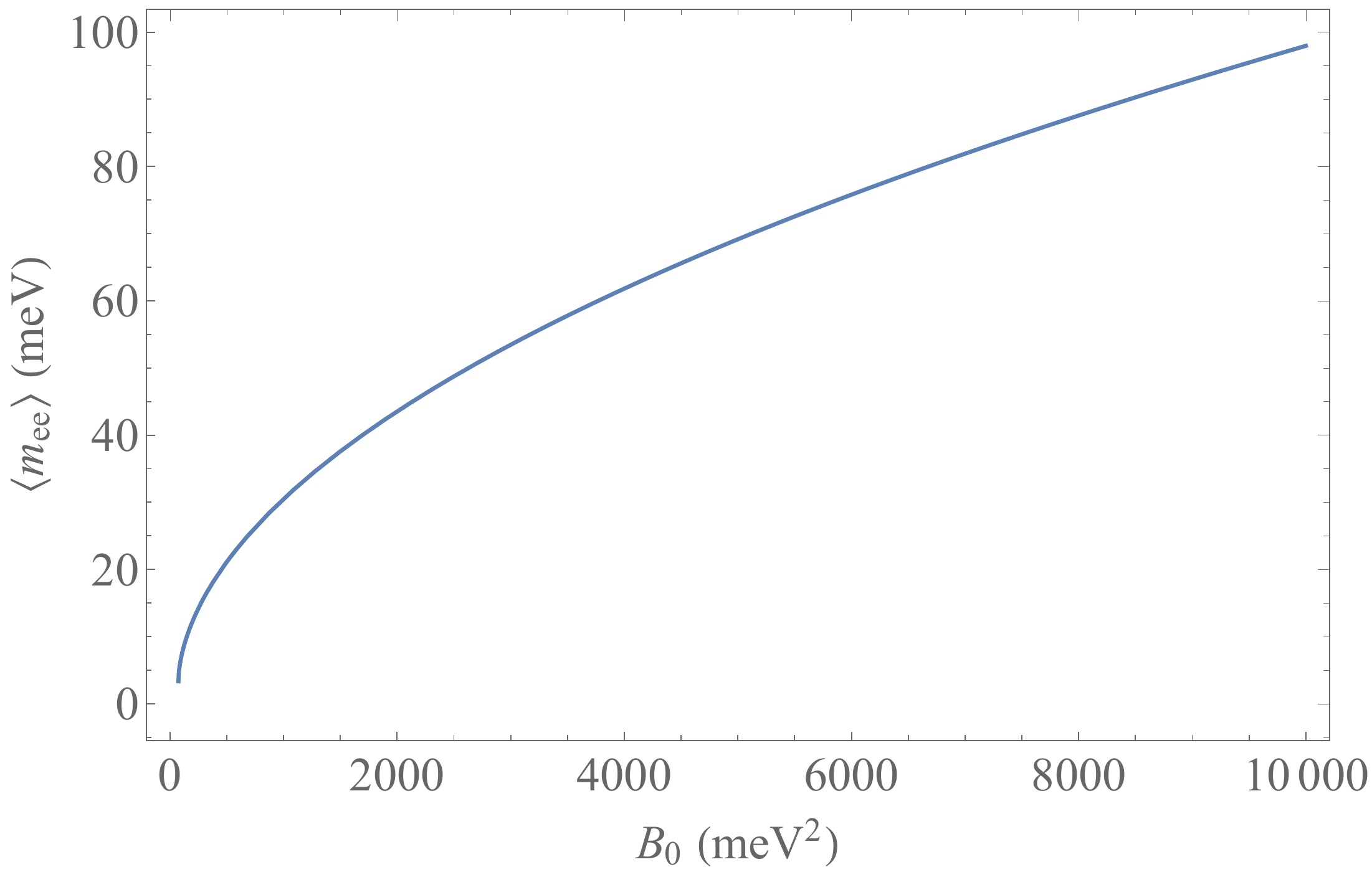}\hspace*{0.1cm}
\includegraphics[width=0.5\textwidth]{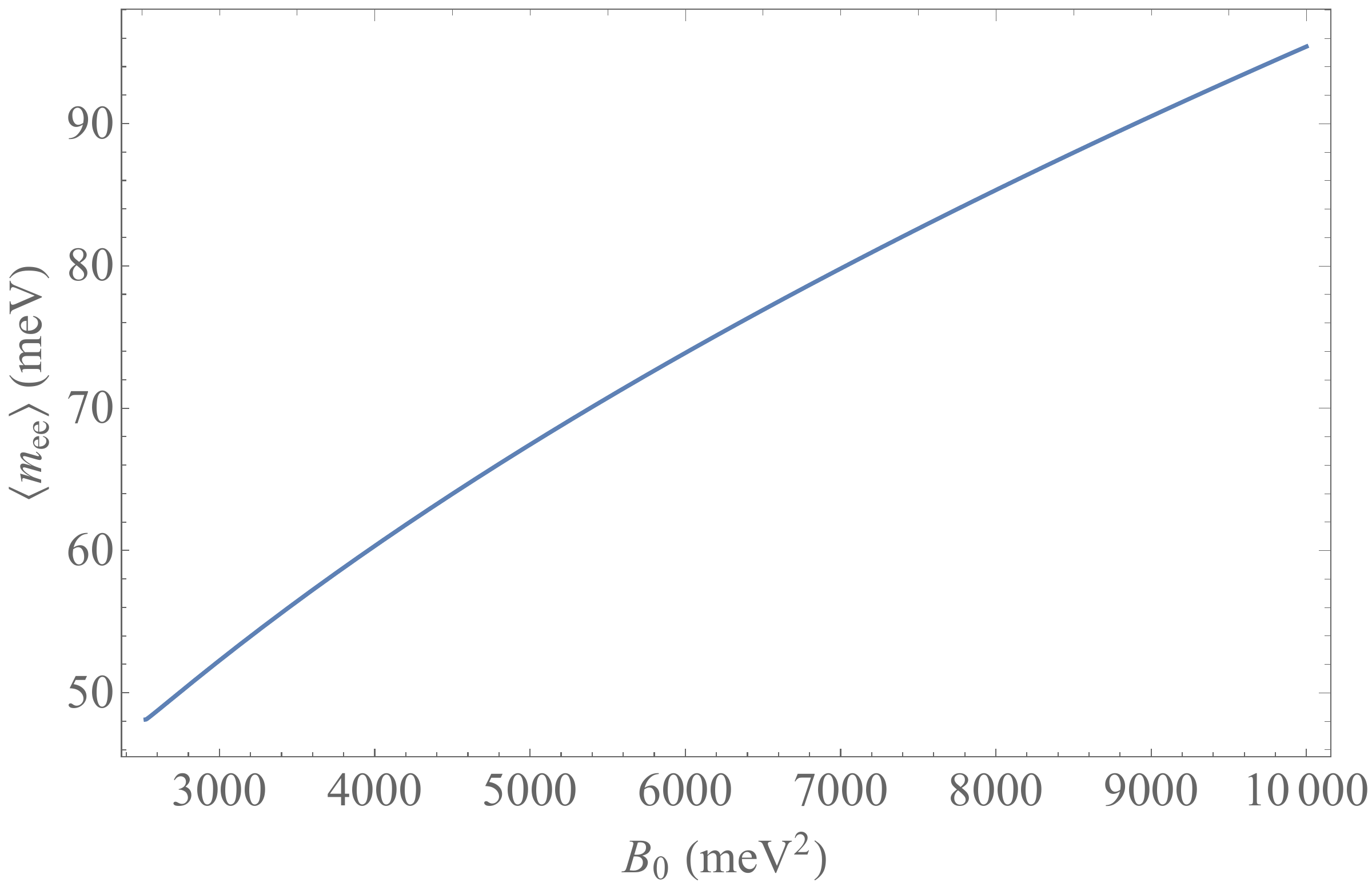}
\vspace*{-0.2cm}
\caption[$\langle m_{ee} \rangle$\, (in meV) depends on $B_{0}$ with $B_{0} \in (75, 10^4) \,\mathrm{meV}^2$ for NO (left panel), and $B_{0} \in (2.525\times 10^3, 10^4) \, \mathrm{meV}^2$ for IO (right panel).]{$\langle m_{ee} \rangle$\, (in meV) depends on $B_{0}$ with $B_{0} \in (75, 10^4) \,\mathrm{meV}^2$ for NO (left panel), and $B_{0} \in (2.525\times 10^3, 10^4) \, \mathrm{meV}^2$ for IO (right panel).}\label{meeF}
\vspace*{-0.5cm}
\end{center}
\end{figure}

We now come back to the neutrino mass issue. Currently, the value of the neutrino mass is still unknown, however, the constraints on $\sum_\nu m_\nu$ and constraints on effective neutrino mass for $0\nu \beta \beta$ decay $\langle m_{ee} \rangle$ have been indicated by recent works as shown in Tabs. \ref{sumconstraint} and \ref{meeconstraint} in which the upper limit on $\sum_\nu m_\nu$ is from 120 meV to 690 meV while the upper limit on $\langle m_{ee} \rangle$ is from about  1 meV to 490 meV.
In order to obtain the possible ranges of $\sum_\nu m_\nu$ and $\langle m_{ee} \rangle$, we consider $B_0\in (75,\, 10^4)\,$ meV while $\Delta m^2_{21}$, $\Delta m^2_{31}$, $\theta_{13}$ and $\theta_{23}$ get their best fit values taken from Ref. \cite{Salas2021} as given in Tab. \ref{Salas2021T}, i.e., $\Delta m^2_{21}=75 \,\mathrm{meV}^2$ and $\Delta m^2_{31} =2.55\times 10^3 \,\mathrm{meV}^2,\, s^2_{13} = 2.20\times 10^{-2},\, s^2_{23}=0.574$ for NO whereas $\Delta m^2_{31}=-2.45\times 10^3 \,\mathrm{meV}^2, \,s^2_{13} = 2.225\times 10^{-2},\, s^2_{23}=0.578$ for IO. At this benchmark point, $\sum_\nu m_\nu$ and $\langle m_{ee}\rangle$ depends on $B_0$, with\footnote{For NO ($m_1< m_2< m_3$), we choose $B_{0} \in (75, 10^4) \, \mathrm{meV}^2$ because Eq. (\ref{m1m2m3ana}) implies that $B_0\geq \Delta m^2_{21}\simeq 75\, \mathrm{meV}^2$ and $m_2\geq \sqrt{\Delta m^2_{21}}\simeq 8.66\, \mathrm{meV}$ \cite{Salas2021, PDG2022}. 
For IO ($m_3< m_1< m_2$), we choose $B_{0} \in (2.525\times 10^{3}, 10^4) \, \mathrm{meV}^2$ because Eq. (\ref{m1m2m3ana}) implies that $B_0\geq 2.525\times 10^{3}\mathrm{meV}^2=\Delta m^2_{21}-\Delta m^2_{31}$ and $m_2\geq \sqrt{\Delta m^2_{21}-\Delta m^2_{31}}=50.30\, \mathrm{meV}$ \cite{Salas2021, PDG2022}. With $B_{0} >10^4 \, \mathrm{meV}^2$ the neutrino mass spectra (NO and IO) are almost degenerate.} $B_{0} \in (75, 10^4) \, \mathrm{meV}^2$ for NO and $B_{0} \in (2.525\times 10^3, 10^4) \, \mathrm{meV}^2$ for IO, are plotted in Figs. \ref{sumF} and \ref{meeF}, respectively.
  \\
Figures (\ref{sumF}) and (\ref{meeF}) indicate the predicted values for $\sum_\nu m_\nu$ and $\langle m_{ee} \rangle$, respectively, are
\bea
&&\sum_\nu m_\nu \, (\mathrm{meV}) \in\left\{
\begin{array}{l}
(60.00, 300.0)\, \hspace{0.25cm} \mbox{for  NO,} \\
(100.00, 286.00)\, \hspace{0.25cm} \mbox{for  IO},
\end{array}%
\right. \label{sumrange}\\
&&\langle m_{ee} \rangle \, (\mathrm{meV}) \in\left\{
\begin{array}{l}
(3.00, 100.00)\,  \hspace{0.25cm} \mbox{for  NO,} \\
(48.00, 95.00)\,  \hspace{0.25cm} \mbox{for  IO},
\end{array}%
\right. \label{meerange}
\eea
which are in good agreement with the recent upper bounds for the sum of neutrino masses and the effective neutrino mass for $0\nu \beta \beta$ decay which are summarized in Tabs. \ref{sumconstraint} and \ref{meeconstraint}, respectively.
\begin{table}[ht]
\begin{center}
\caption{\label{sumconstraint} Constraints on the sum of neutrino masses $\sum_\nu m_\nu$.}
\vspace{0.25cm}
\begin{tabular}{|c|c|c|c|c|c|c|c|c|c|c|}\hline
References & Constraints   \\ \hline
\multirow{2}{2.3cm}{\hfill \cite{Salas2021}  \hfill }& $\sum_\nu m_\nu< 130\, \mathrm{meV}\, \mathrm{(NO)}$ \\
\cline{2-2}   & $\sum_\nu m_\nu< 150\, \mathrm{meV}\, \mathrm{(NO)}$  \\
  \hline
\cite{Capozzi20}& $\sum_\nu m_\nu< 120\div 690 \, \mathrm{meV}$ \\
 \hline
 \cite{nuboundAghanim20, nuboundShadab21}&  $\sum_\nu m_\nu< 120 \, \mathrm{meV}$  \\ 
 \hline
 \multirow{2}{2.3cm}{\hfill \cite{MikhailPRD2020}  \hfill }& $\sum_\nu m_\nu< 180\, \mathrm{meV}\, \mathrm{(NO)}$ \\
\cline{2-2}   & $\sum_\nu m_\nu< 210 \, \mathrm{meV}\, \mathrm{(NO)}$  \\ 
  \hline
 \cite{LorenzPRD2021}&  $\sum_\nu m_\nu< 210\, \mathrm{meV}$ \\ 
 \hline
 \cite{GuillermoJHEP21}&  $\sum_\nu m_\nu< 420\, \mathrm{meV}$ \\ 
 \hline
\end{tabular}
\end{center}
\vspace{-0.25cm}
\end{table}
\begin{table}[ht]
\begin{center}
\caption{\label{meeconstraint} Constraints on the effective neutrino mass for $0\nu \beta \beta$ decay $\langle m_{ee}\rangle$ .}
\vspace{0.25cm}
\begin{tabular}{|c|c|c|c|c|c|c|c|c|c|c|}\hline
References & Constraints    \\ \hline
\cite{AbeKamLAND2022}& $\langle m_{ee}\rangle < 36\div 156 \, \mathrm{meV}$ \\
 \hline
\cite{Arnquist2023}&$\langle m_{ee} \rangle < 113\div 269\, \mathrm{meV}$\\ \hline
\cite{KamLAND16}& $\langle m_{ee} \rangle <61 \div 165\, \mathrm{meV}$\\ \hline
\cite{GERDA19} & $\langle m_{ee} \rangle < 104\div 228\, \mathrm{meV}$ \\ \hline
\cite{CUORE20} &$\langle m_{ee} \rangle < 75 \div 350 \,\mathrm{meV}$\\ \hline
 \cite{AgostiniPRL20}&  $\langle m_{ee}\rangle < 79 \div 180 \, \mathrm{meV}$ \\  \hline
 \cite{AzzoliniPRL20}&  $\langle m_{ee}\rangle < 263\div 545\, \mathrm{meV}$ \\
 \hline
 \cite{Augierepjc22}&  $\langle m_{ee}\rangle < 280\div 490 \, \mathrm{meV}$ \\ 
 \hline
 \cite{Adamsnatrue22}& $\langle m_{ee}\rangle < 90\div 305\, \mathrm{meV}$ \\
 \hline
 \cite{Penedoplb2018,Caocpc2020, Geprd2017, HuangZhoujhep21}&  $\langle m_{ee}\rangle \sim 1.0 \div 10 \, \mathrm{meV}$ \\
 \hline
\end{tabular}
\end{center}
\vspace{-0.25cm}
\end{table}

\begin{figure}[h]
\begin{center}
\vspace*{0.25 cm}
\hspace*{-0.5cm}
\includegraphics[width=0.5\textwidth]{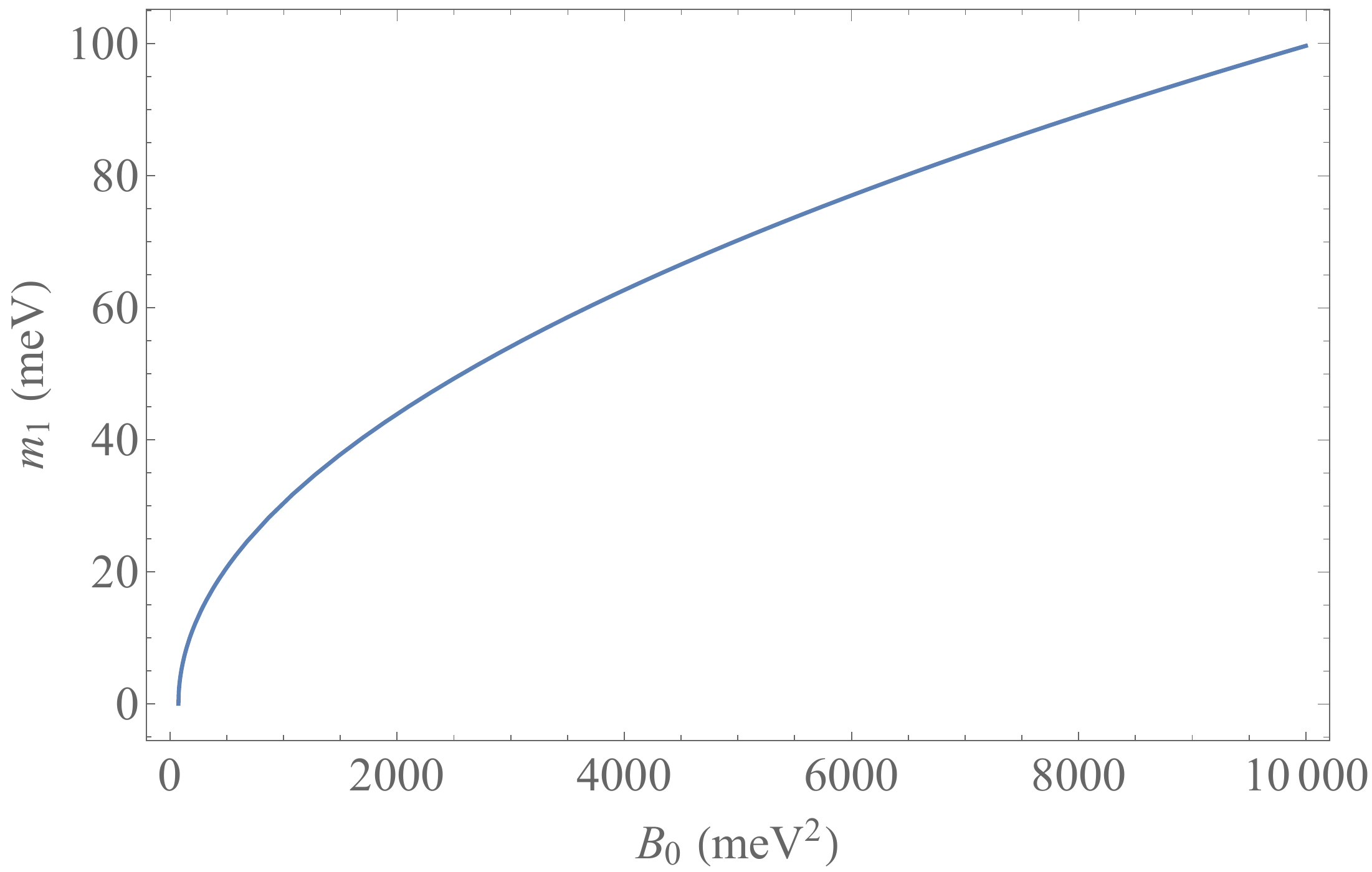}\hspace*{0.1cm}
\includegraphics[width=0.5\textwidth]{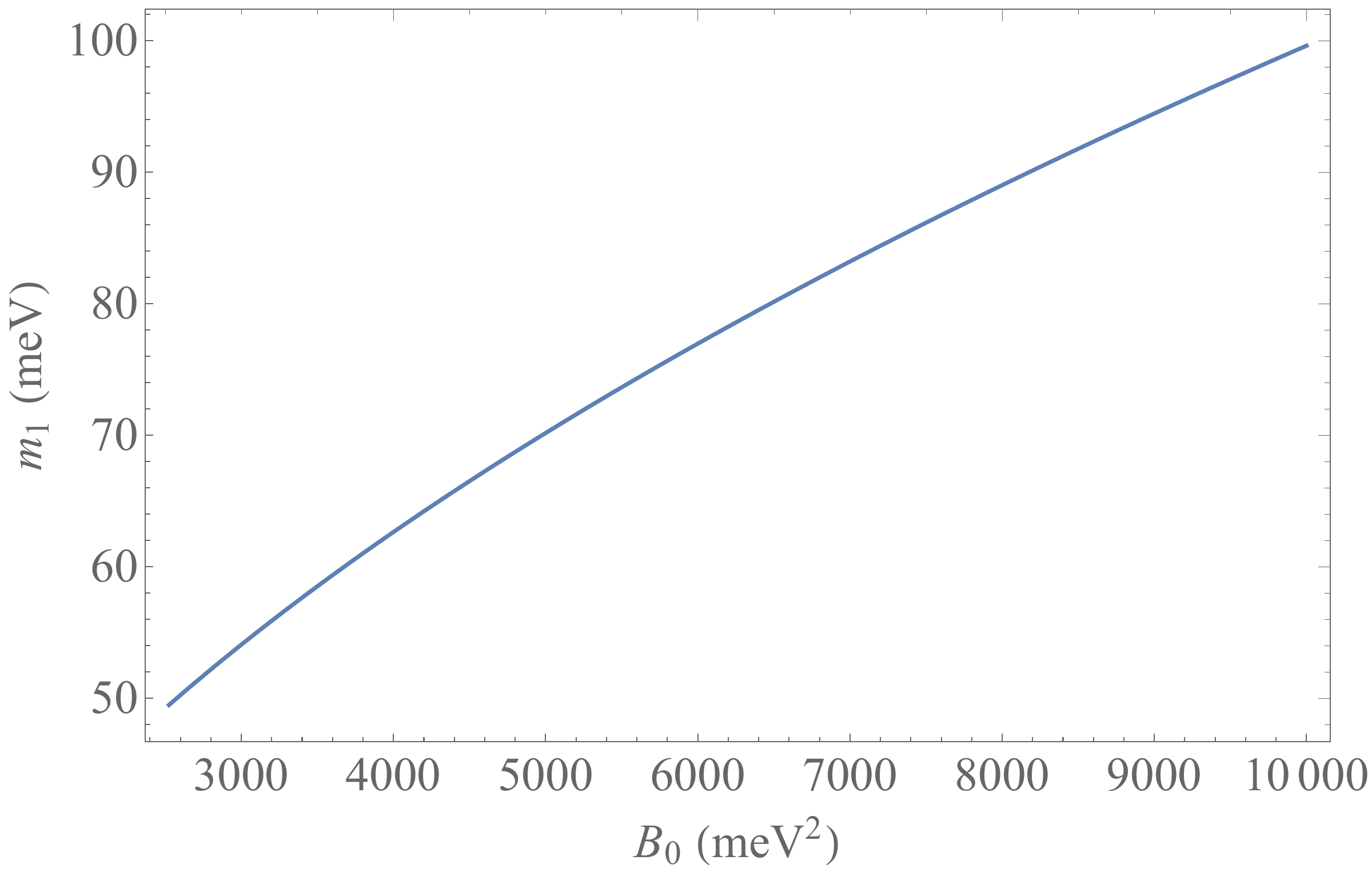}
\vspace*{-0.2cm}
\caption[$m_1$\, (in meV) depends on $B_{0}$ with $B_{0} \in (75, 10^4) \,\mathrm{meV}^2$ for NO (left panel), and $B_{0} \in (2.525\times 10^3, 10^4) \, \mathrm{meV}^2$ for IO (right panel).]{$m_1$\, (in meV) depends on $B_{0}$ with $B_{0} \in (75, 10^4) \,\mathrm{meV}^2$ for NO (left panel), and $B_{0} \in (2.525\times 10^3, 10^4) \, \mathrm{meV}^2$ for IO (right panel).}\label{m1F}
\vspace*{-0.5cm}
\end{center}
\end{figure}

\begin{figure}[h]
\begin{center}
\vspace*{0.25 cm}
\hspace*{-0.5cm}
\includegraphics[width=0.5\textwidth]{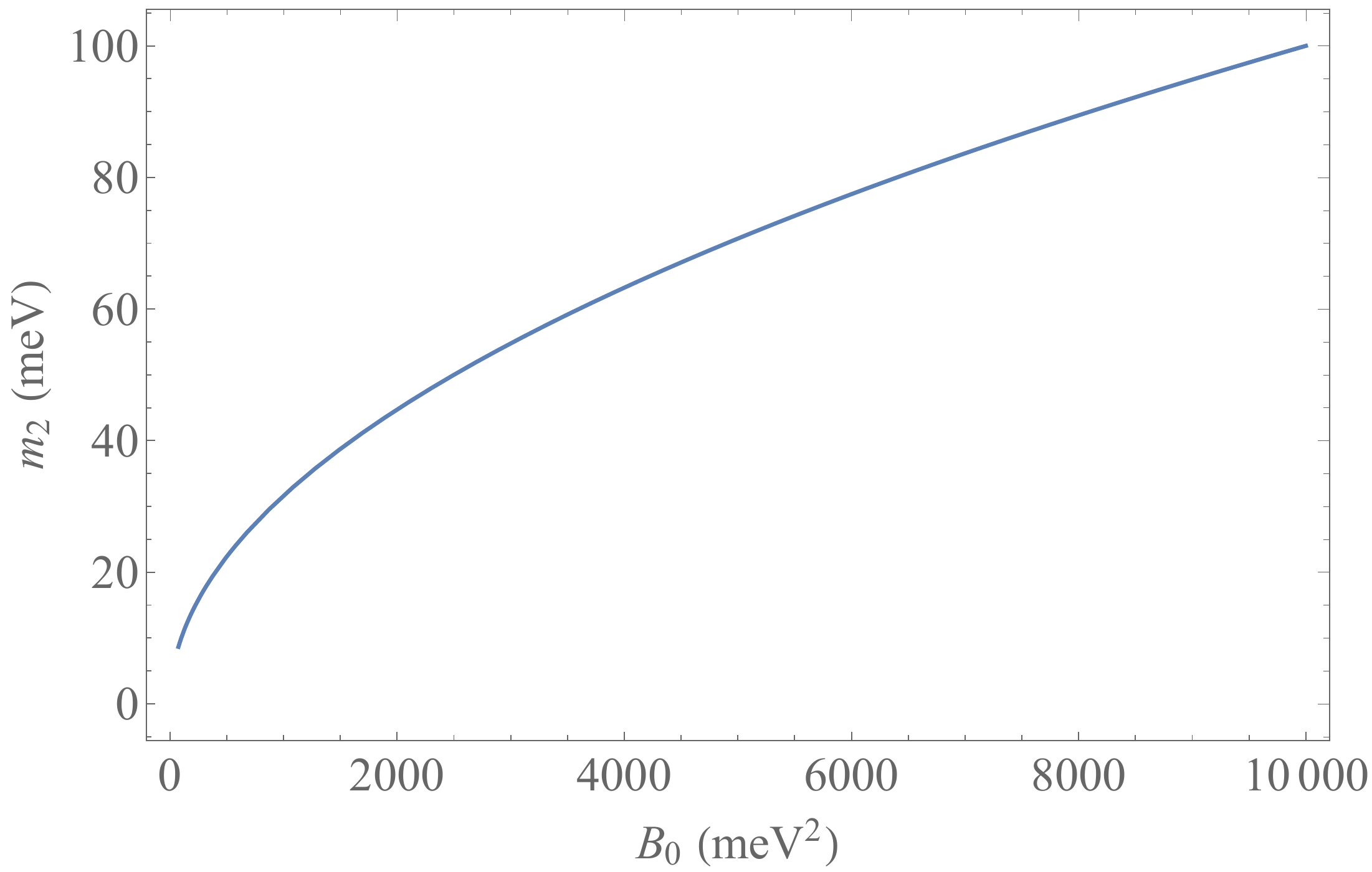}\hspace*{0.1cm}
\includegraphics[width=0.5\textwidth]{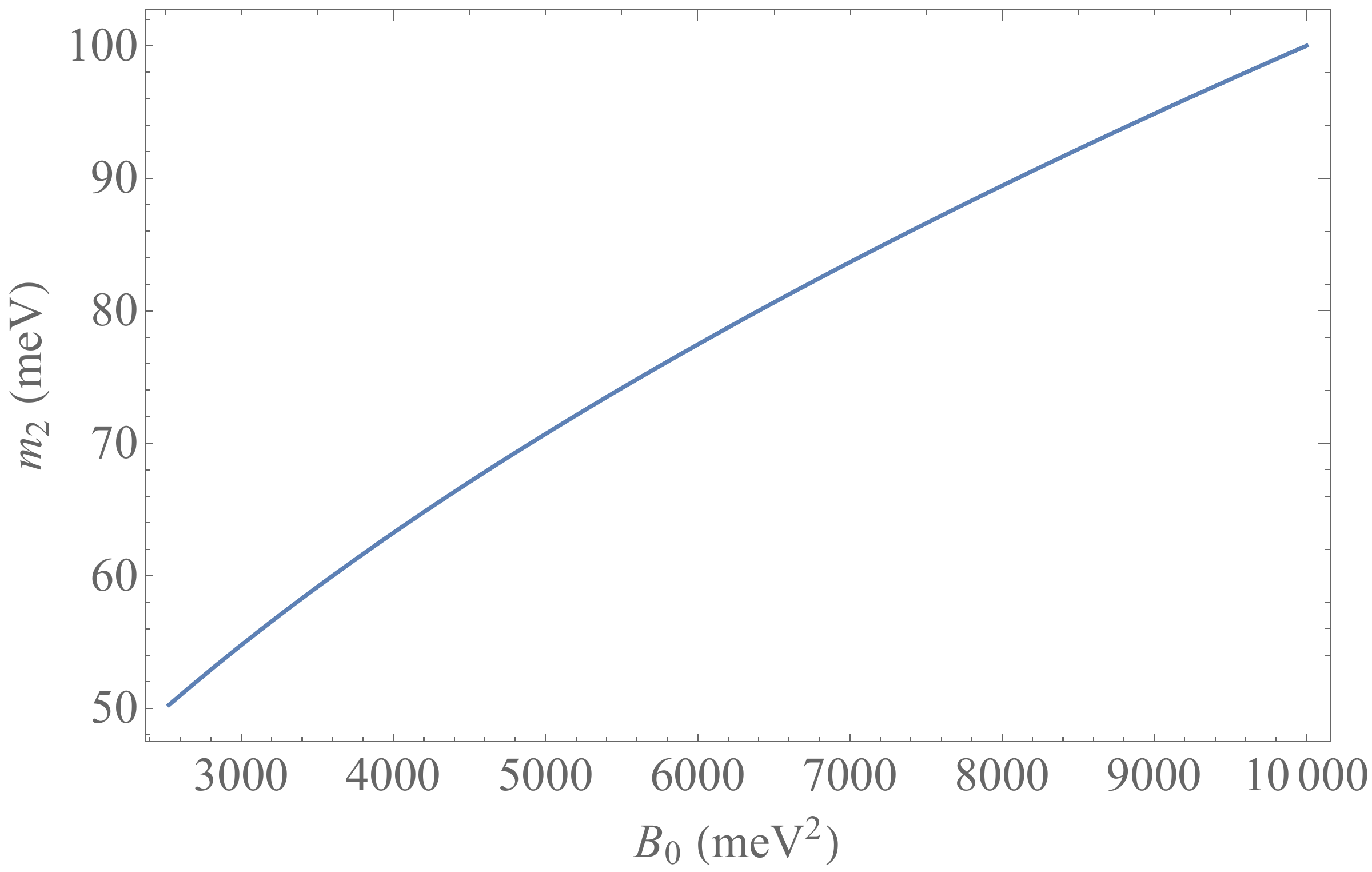}
\vspace*{-0.2cm}
\caption[$m_2$\, (in meV) depends on $B_{0}$ with $B_{0} \in (75, 10^4) \,\mathrm{meV}^2$ for NO (left panel), and $B_{0} \in (2.525\times 10^3, 10^4) \, \mathrm{meV}^2$ for IO (right panel).]{$m_2$\, (in meV) depends on $B_{0}$ with $B_{0} \in (75, 10^4) \,\mathrm{meV}^2$ for NO (left panel), and $B_{0} \in (2.525\times 10^3, 10^4) \, \mathrm{meV}^2$ for IO (right panel).}\label{m2F}
\vspace*{-0.5cm}
\end{center}
\end{figure}

\begin{figure}[h]
\begin{center}
\vspace*{0.25 cm}
\hspace*{-0.5cm}
\includegraphics[width=0.5\textwidth]{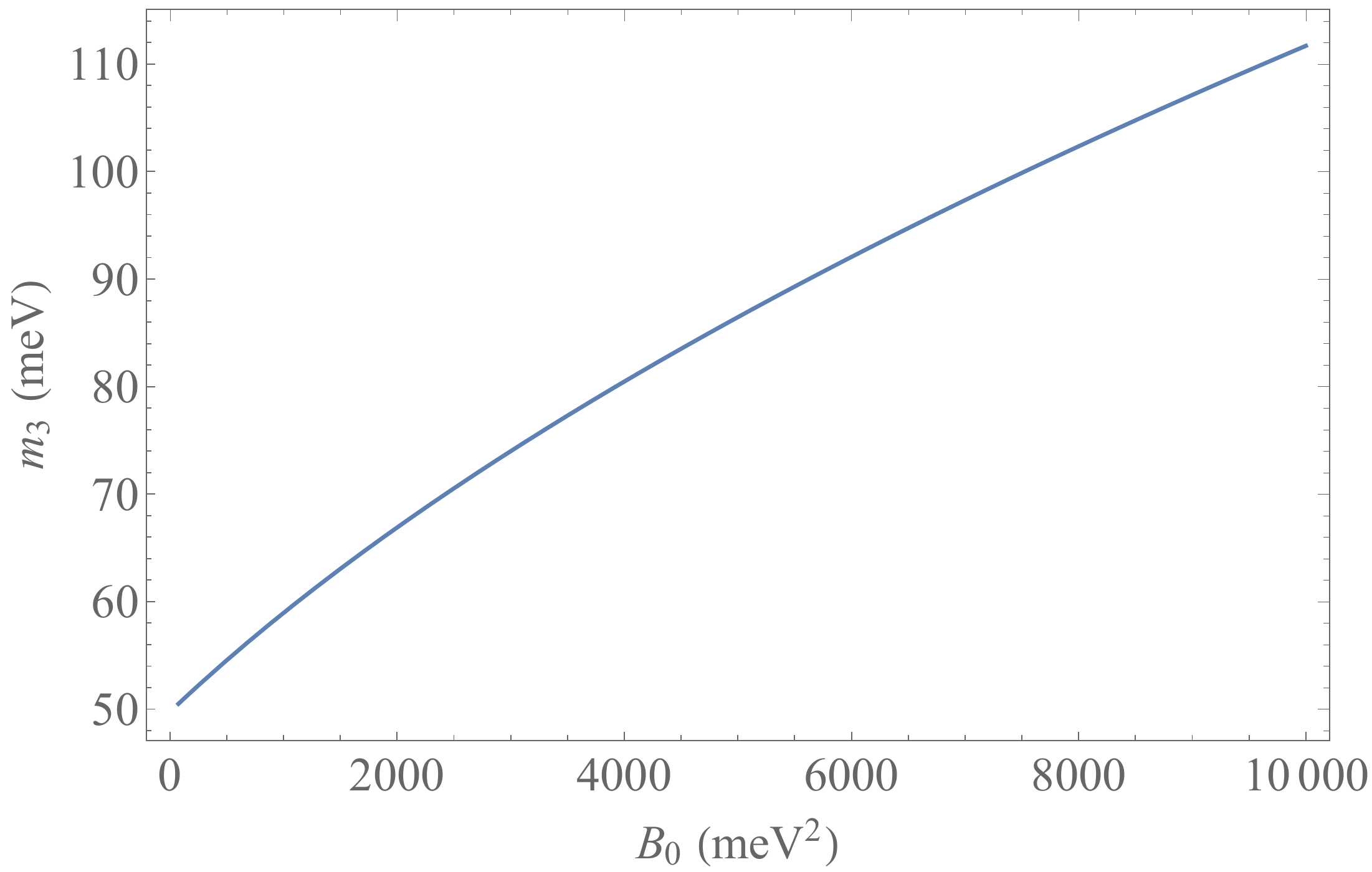}\hspace*{0.1cm}
\includegraphics[width=0.5\textwidth]{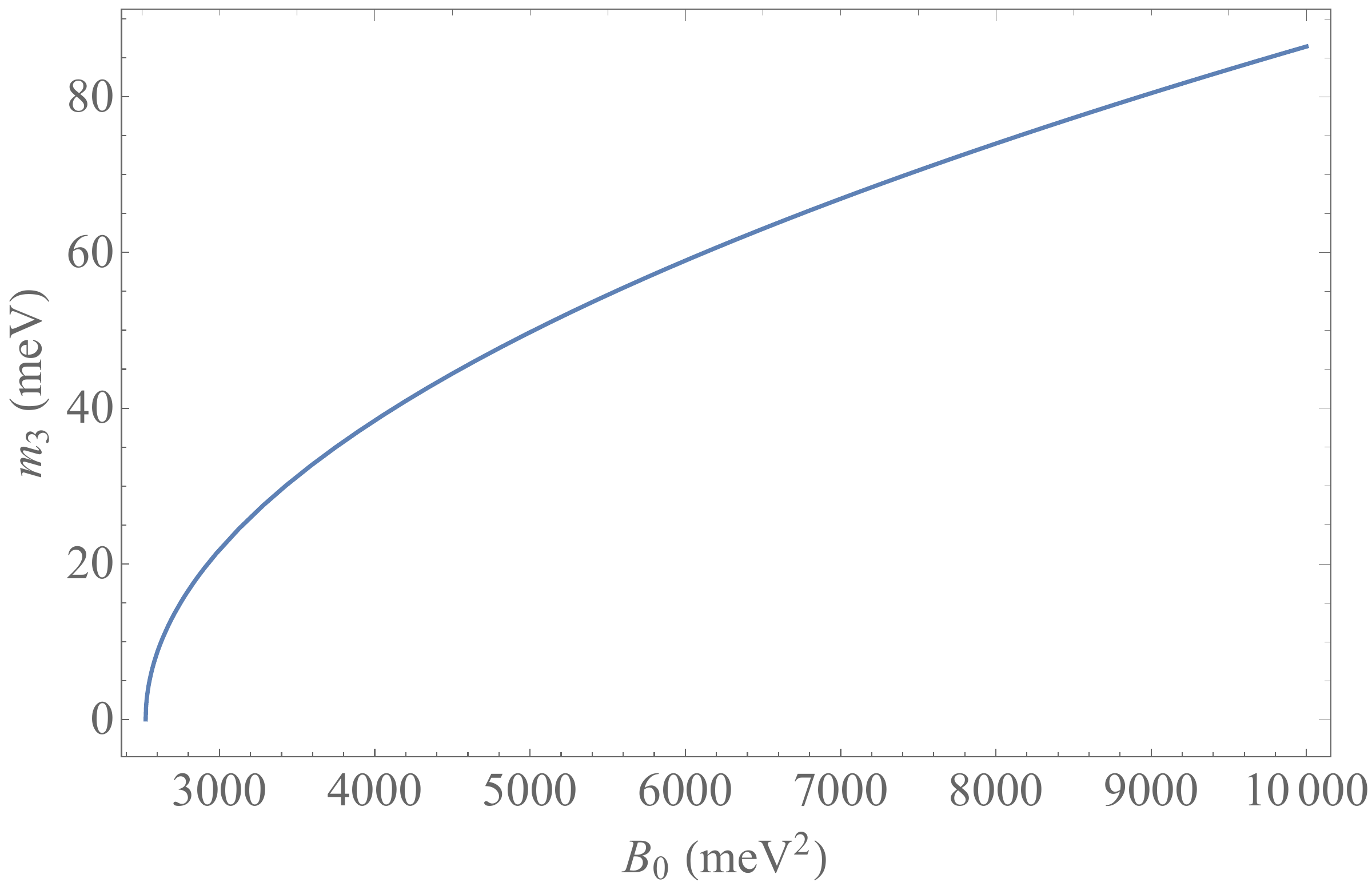}
\vspace*{-0.2cm}
\caption[$m_3$\, (in meV) depends on $B_{0}$ with $B_{0} \in (75, 10^4) \,\mathrm{meV}^2$ for NO (left panel), and $B_{0} \in (2.525\times 10^3, 10^4) \, \mathrm{meV}^2$ for IO (right panel).]{$m_3$\, (in meV) depends on $B_{0}$ with $B_{0} \in (75, 10^4) \,\mathrm{meV}^2$ for NO (left panel), and $B_{0} \in (2.525\times 10^3, 10^4) \, \mathrm{meV}^2$ for IO (right panel).}\label{m3F}
\vspace*{-0.5cm}
\end{center}
\end{figure}

Three light neutrino masses $m_1, m_2, m_3$ (in meV) are, respectively, predicted to be
\bea
&&\hspace{-0.2cm}\left\{
\begin{array}{l}
m_1 \, (\mathrm{meV})\in (0, 100.0), \hspace{0.1cm} m_{2}  \, (\mathrm{meV}) \in  (8.66, 100.0),  \hspace{0.1cm} m_{3}  \, (\mathrm{meV}) \in (50.50, 110.0) \hspace{0.2cm} \mbox{for  NO,} \\
m_1 \, (\mathrm{meV})\in (49.50, 99.50), \hspace{0.1cm} m_{2}  \, (\mathrm{meV}) \in  (50.0, 100.0),  \hspace{0.1cm} m_{3}  \, (\mathrm{meV}) \in (0.0, 86.50) \hspace{0.25cm} \mbox{for  IO},
\end{array}%
\right. \label{m1m2m3ranges}\eea
which are depicted in Figs. \ref{m1F}, \ref{m2F}  and \ref{m3F}, respectively.

Here are some comments related to the neutrino mass issue:
\begin{itemize}
  \item [(1)] The obtained neutrino masses and the sum of neutrino mass in Eqs. (\ref{m1m2m3ranges}) and \ref{sumrange} are in consistent with the recent upper limits on the sum of neutrino mass as given in Tab. \ref{sumconstraint}.
  \item [(2)] The obtained effective Majorana neutrino mass in Eq. (\ref{meerange}) is in consistent with the recent upper limits on $\langle m_{ee}\rangle$ as given in Tab. \ref{meeconstraint}.
\end{itemize}

\section{\label{quark} Quark sector}
\subsection{\label{quarkanal} Analytical result for quark sector}
The quark Yukawa terms in Eq.(\ref{Yquark}) and the
tensor products of the $A_4$ group \cite{A420Ishimori10}, together with the VEV alignment in Eq. (\ref{scalarvev}), yields:
\bea M_{q} &=& M^{(0)}_{q}+\Delta M_{q} \hs (q=u,d),
\label{Mq}\eea
where
\bea &&M_{q} =\left(%
\begin{array}{ccc}
A^q_{11} & A^q_{12} & B^q_{13} \\
A^q_{12} & A^q_{22} & B^q_{23}\\
B^q_{31} & B^q_{32}   & A^q_{33} \\
\end{array}%
\right), \hs M^{(0)}_{q} =\left(%
\begin{array}{ccc}
A^q_{11} & A^q_{12} & 0 \\
A^q_{12} & A^q_{22} & 0\\
   0     &      0   & A^q_{33} \\
\end{array}%
\right), \hs  \Delta M_{q}=\left(%
\begin{array}{ccc}
0        & 0          & B^q_{13} \\
0        & 0          & B^q_{23}\\
B^q_{31} & B^q_{32}   & 0 \\
\end{array}%
\right), \label{M0dMq}\eea
with
\bea
&&A^u_{11}=x^u_{11} v, \hs A^u_{22}=x^u_{22} v, \hs A^u_{33}=x^u_{33} v, \hs A^u_{12}=x^u_{12} v,\crn
&&B^u_{13}=\fr{x^u_{13}}{\La } v v_\eta, \, B^u_{31}=\fr{x^u_{31}}{\La } v v_{\eta^*}, \, B^u_{23}=\fr{x^u_{23}}{\La } v v_\eta, \, B^u_{32}=\fr{x^u_{32}}{\La} v v_{\eta^*}, \crn
&&A^d_{11}=x^d_{11} v, \hs A^d_{22}=x^d_{22} v, \hs A^d_{33}=x^d_{33} v, \hs A^d_{12}=x^d_{12} v,\crn
&&B^d_{13}=\fr{x^d_{13}}{\La } v v_\eta, \, B^d_{31}=\fr{x^d_{31}}{\La} v v_{\eta^*}, \, B^d_{23}=\fr{x^d_{23}}{\La} v v_\eta, \, B^d_{32}=\fr{x^d_{32}}{\La} v v_{\eta^*}.\label{AuAd}\eea
We found that  in Eqs.(\ref{Mq}), (\ref{M0dMq}) and (\ref{AuAd}), $M^{(0)}_{q} \, (q=u,d)$ due to the unique contribution of the $SU(2)_L$ doublet $H$ whereas $\Delta M_{q}\, (q=u,d)$ depend on contributions of $H$ and $\eta$. Hence, without the contribution of $\eta$, $\Delta M_{q}$ will vanish and $M_{q}\rightarrow M^{(0)}_{q}$ which are diagonalized as
\bea
U^{0\+}_{u L} M^{(0)}_{u} U^0_{u R}&=&\mathrm{diag} \left(m^{0}_u, \,\, m^{0}_c, \,\, m^{0}_t \right), \hs
U^{0\+}_{d L} M^{(0)}_{d} U^0_{d R}=\mathrm{diag} \left(m^{0}_d, \,\, m^{0}_s, \,\, m^{0}_b\right), \label{mudcstb0dia}\eea
where
\bea
m^{0}_{u,c} &=&X_{u} \mp Y_{u}, \hs m^{0}_{t}= A^u_{33}, \hs
 m^{0}_{d,s} = X_{d} \mp Y_{d}, \hs m^{0}_{b}= A^d_{33}, \label{mudcstb0} \\
U^0_{qL} &=&U^0_{qR}=\left(%
\begin{array}{ccc}
  \mathrm{cos} \psi_q  &\hs -\mathrm{sin} \psi_q & 0 \\
  \hs \mathrm{sin} \psi_q & \mathrm{cos} \psi_q &0 \\
  0 & 0 &  1\\
\end{array}%
\right),  \label{Uq0}
\eea
with
\bea
&&X_q=\frac{1}{2}\left(A^u_{11}+A^u_{22}\right), \hs  Y_q=\frac{1}{2} \sqrt{\left(A^u_{11}-A^u_{22}\right)^2+4 (A^u_{12})^2}, \label{XYq}\\
&&\psi_q = 
\arctan\left(K^{-1}_q\right),  \label{psiq}\\
&& K_{q}=\frac{A^q_{11}-A^q_{22}-\sqrt{\left(A^q_{11}-A^q_{22}\right)^2+4 \left(A^q_{12}\right)^2}}{2 A^q_{12}}, \label{Kq}
\eea
and $A^q_{11}, A^q_{22}, A^q_{12}$ are given in Eq. (\ref{AuAd}).
The quark mixing matrix at the tree level is defined as
 \bea
 U^0_\mathrm{CKM}=U^0_{uL} U^{0 \dagger}_{dL}  =\left(%
\begin{array}{ccc}
   \mathrm{sin} \psi_{d} \mathrm{sin} \psi_{u}+\mathrm{cos} \psi_{d} \mathrm{cos} \psi_{u} & \mathrm{sin} \psi_{d} \mathrm{cos} \psi_{u}-\mathrm{cos} \psi_{d} \mathrm{sin} \psi_{u}& 0 \\
 - \mathrm{sin} \psi_{d} \mathrm{cos} \psi_{u}+ \mathrm{cos} \psi_{d} \mathrm{sin} \psi_{u} & \mathrm{sin} \psi_{d} \mathrm{sin} \psi_{u}+\mathrm{cos} \psi_{d} \mathrm{cos} \psi_{u} &0 \\
  0 & 0 &  1\\
\end{array}%
\right), \label{Uckm} \eea
which can be comparable to the approximate quark mixing matrix in Eq. (\ref{PDG22qmixapp}).

In fact, the quark mixing matrix is given in Eq. (\ref{PDG22qmix}), which is vey close to the one in Eq. (\ref{PDG22qmixapp}). Thus, we can consider the contribution of $\eta$ as a
small perturbation for generating the observed quark mixing angles. In this work, the calculations are made at the first order of the perturbation theory. At the first order of perturbation theory, $\Delta M_{q}\, (q=u,d)$ have no contribution to their eigenvalues but they contribute and thus change the eigenstates. The perturbed quark masses and the (left-and right-handed) up-and down- quark mixing matrices are determined as:
\bea
&&m_{u}=m^{0}_{u}, \hs m_{c}=m^{0}_{c}, \hs m_{t}=m^{0}_{t},
\hs m_{d}=m^{0}_{d}, \hs m_{s}=m^{0}_{s}, \hs m_{b}=m^{0}_{b},\label{mudcstb} \\
&&U_{qL} =U_{qR}=
\left(
\begin{array}{ccc}
 c_{q} & -s_{q} & \frac{c_{q} (B^q_{13} c_{q}+B^q_{23} s_{q})}{A^q_{33}-X_q+Y_q}-\frac{s_{q} (B^q_{13} s_{q}-B^q_{23} s_{q})}{-A^q_{33}+X_q+Y_q} \\
 s_{q} & s_{q} & \frac{s_{q} (B^q_{13} s_{q}+B^q_{23} s_{q})}{A^q_{33}-X_q+Y_q}-\frac{s_{q} (B^q_{23} c_{q}-B^q_{13} s_{q})}{-A^q_{33}+X_q+Y_q} \\
 -\frac{B^q_{31} c_{q} +B^q_{32} s_{q}}{A^q_{33}-X_q+Y_q} & \frac{B^q_{32} c_{q}-B^q_{31} s_{q}}{-A^q_{33}+X_q+Y_q} & 1 \\
\end{array}
\right),  \label{Uq0}
\eea
 where
$s_{q}=\sin \psi_q, c_{q}=\cos \psi_q\,\, (q=u,d)$.

Equations (\ref{mudcstb0}), (\ref{XYq}) and (\ref{mudcstb}) require that $A^q_{11}, A^q_{22}, A^q_{33}$ and $A^q_{12}$ are real numbers, i.e., the corresponding Yukawa-like couplings $x^q_{11}, x^q_{22}, x^q_{33}$ and $x^q_{12}$ are real numbers. The perturbed quark mixing
matrix, $V_\mathrm{CKM}= U_{uL} U^{\dagger}_{dL}$, has the following entries:
\bea
&&(V_\mathrm{CKM})_{11}= \left(\frac{B^d_{-} s_d}{X^d_{+}}+\frac{B^d_{+} c_d}{X^d_{-}}\right) \left(\frac{B^u_{-} s_u}{X^u_{+}}+\frac{B^u_{+} c_u}{X^u_{-}}\right)+c_{d} c_{u}+s_{d} s_{u},\crn
&&(V_\mathrm{CKM})_{12}=\left(\frac{B^d_{+} s_d}{X^d_{-}}-\frac{B^d_{-} c_d}{X^d_{+}}\right) \left(\frac{B^u_{-} s_u}{X^u_{+}}+\frac{B^u_{+} c_u}{X^u_{-}}\right)-c_d s_u+c_u s_d,\crn
&&(V_\mathrm{CKM})_{13}=-\frac{B^{'d}_{-} s_u}{X^{d}_{+}}+\frac{B^{u}_{-} s_u}{X^{u}_{+}}-\frac{B^{'d}_{+} c_u}{X^{d}_{-}}+\frac{B^{u}_{+} c_u}{X^{u}_{-}},\crn
&&(V_\mathrm{CKM})_{21}=\left(\frac{B^{d}_{-} s_d}{X^{d}_{+}}+\frac{B^{d}_{+} c_d}{X^{d}_{-}}\right) \left(\frac{B^{u}_{+} s_u}{X^{u}_{-}}-\frac{B^{u}_{+} c_u}{X^{u}_{+}}\right)+c_d s_u -c_u s_d,\crn
&&(V_\mathrm{CKM})_{22}=\left(\frac{B^{d}_{+} s_d}{X^{d}_{-}}-\frac{B^{d}_{-} c_d}{X^{d}_{+}}\right) \left(\frac{B^{u}_{+} s_u}{X^{u}_{-}}-\frac{B^{u}_{-} c_u}{X^{u}_{+}}\right)+c_d c_u+s_d s_u, \nonumber\eea
\bea 
&&(V_\mathrm{CKM})_{23}=\frac{B^{'d}_{-} c_u}{X^{d}_{+}}-\frac{B^{u}_{-} c_u}{X^{u}_{+}}-\frac{B^{'d}_{+} s_u}{X^{d}_{-}}+\frac{B^{u}_{+} s_u}{X^{u}_{-}}, \crn
&&(V_\mathrm{CKM})_{31}=\frac{B^{d}_{-} s_d}{X^{d}_{+}}-\frac{B^{'u}_{-} s_d}{X^{u}_{+}}+\frac{B^{d}_{+}c_d}{X^{d}_{-}}-\frac{B^{'u}_{+} c_d}{X^{u}_{-}},\crn
&&(V_\mathrm{CKM})_{32}=-\frac{B^{d}_{-} c_d}{X^{d}_{+}}+\frac{B^{'u}_{-} c_d}{X^{u}_{+}}
+\frac{B^{d}_{+} s_d}{X^{d}_{-}}-\frac{B^{'u}_{+} s_d}{X^{u}_{-}},\crn
&&(V_\mathrm{CKM})_{33}=\frac{B^{'d}_{-} B^{'u}_{-}}{X^{d}_{+} X^{u}_{+}}+\frac{B^{'d}_{+} B^{'u}_{+}}{X^{d}_{-} X^{u}_{-}}+1,\label{Vckmv}\eea
where
\bea 
&&B^q_{+} = B^q_{13} c_q + B^q_{23} s_q, \hs
B^q_{-} = B^q_{13} s_q - B^q_{23} c_q, \crn
&&B^{'q}_{+} = B^q_{31} c_q + B^q_{32} s_q, \hs
B^{'q}_{-} = B^q_{31} s_q - B^q_{32} c_q, \crn
&&X^q_{+} = A^q_{33} - (X_q + Y_q), \hs
X^q_{-} = A^q_{33} - (X_q - Y_q) \hspace{0.35 cm} (q=u, d). \label{BqXq}
\eea
Equation (\ref{mudcstb0}) and (\ref{XYq}) 
provide the following relations:
\bea
&&A^u_{33} = m_t, \hs X_u = \frac{m_c + m_u}{2}, \hs Y_u = \frac{m_c - m_u}{2}, \crn
&&A^u_{11}=\frac{1}{2} \left(m_u+m_c+\sqrt{(m_c-m_u)^2-4 \big(A^u_{12}\big)^2}\right), \crn
&&A^u_{22}= \frac{1}{2} \left(m_u+m_c-\sqrt{(m_c-m_u)^2-4 \big(A^u_{12}\big)^2}\right), \label{relationsq1}\\ 
&&A^d_{33} = m_b, \hs X_d = \frac{m_d + m_s}{2}, \hs Y_d = \frac{m_s-m_d}{2}, \crn
&&A^d_{11}=\frac{1}{2} \left(m_d+m_s+\sqrt{(m_c-m_u)^2-4 (A^d_{12})^2}\right), \crn
&&A^d_{22} =\frac{1}{2} \left(m_d+m_s-\sqrt{(m_c-m_u)^2-4 (A^d_{12})^2}\right). \label{relationsq2}
\eea
Comparison of the model results on the elements of the quark mixing matrix in Eqs.(\ref{Vckmv}) and (\ref{BqXq}) with their corresponding experimental values 
$(V^{exp}_\mathrm{CKM})_{ij}\,\, (i,j=1,2,3)$, we get the relations:
\bea
&&B^d_{13}=
-B^d_0 \Big\{\big[(m_b - m_d) s_d c_u  + (m_d - m_s) \big(V^{exp}_\mathrm{CKM}\big)_{12}\big] s_d c_d +(m_b - m_d) c_d^3 c_u  \crn
&& \hspace{0.75 cm}+\big[(m_b - m_s) s_d s_u - (m_b - m_d) (V^{exp}_\mathrm{CKM})_{11}\big] c_d^2 + (m_b - m_s) \big[s_d s_u -(V^{exp}_\mathrm{CKM})_{11}\big]s_d^2\Big\}, \label{B13q}\\
&&B^d_{23}=B^d_0\Big\{(m_d-m_b) c_u s_d + (m_b - m_s) s_u c_d  + (m_s-m_d) c_d  s_d (V^{exp}_\mathrm{CKM})_{11} \crn
&&\hspace{0.75 cm}+ \left[(m_b - m_s)c_d^2 + (m_b - m_d) s_d^2\right] (V^{exp}_\mathrm{CKM})_{12}\Big\}, \\
&&B^d_0=\frac{(m_c - m_t) (m_t - m_u)}{B^u_{13} [(m_u-m_t) s^2_u+(m_c - m_t) c_u^2]  + B^u_{23} (m_c - m_u) s_u c_u}, \label{Bd0} \\
&&B^u_{23}=B^u_{13}\Bigg(\frac{\varepsilon^{(1)}_{u}}{\varepsilon^{(2)}_{u}}\Bigg), \, B^u_{13}=\frac{\epsilon^{(1)}_u \epsilon^{(2)}_u}{\epsilon_0}, \, B^d_{31}=\frac{B^d_{32}\epsilon^{(1)}_d -\epsilon^{(2)}_d}{\epsilon_0}, \, B^u_{31}= \frac{\tau_4 + \sqrt{\tau^2_3 - 4 \tau_1\tau_2}}{2 \tau_0}, \\
&&B^d_{32}=\frac{\kappa_1 - \kappa_2 + (m_b - m_d) (m_c - m_t) s_d \kappa_3 +  (c^2_u  \kappa_4 +  s_u \kappa_5+ c_u \kappa_6 ) c_d}{\kappa_0},  \label{B32q} \eea
where $\varepsilon^{1,2}_{u}, \epsilon_0, \epsilon^{1,2}_{u,d}, \kappa_{l}\, (l=0\div 6)$ are defined in Appendices \ref{Appenvarepsilon} and \ref{Appenepsilon}. The analytic expressions of $\tau_s\, (s=0\div 4)$ are rather lengthy, thus we do not present their explicit forms, but sketch their dependence on $s_u$ and $s_d$ as shown in Figs. \ref{tau0F}- \ref{tau4F} in Appendix \ref{Appentau01234}.

\subsection{\label{quarknumer} Numerical analysis for quark sector}
\begin{figure}[ht]
\begin{center}
\vspace{-0.85 cm}
\hspace{-2.0 cm}
\includegraphics[width=0.825\textwidth]{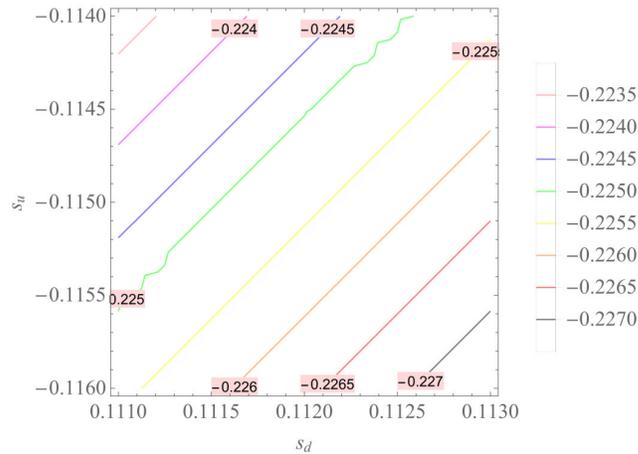}\hspace{-2.5 cm}
\end{center}
\vspace{-10.75 cm}
\caption[(Colored lines) $(V_{\mathrm{CKM}})_{21}$
versus $s_{u}$ and $s_{d}$ with $s_{u}\in (-0.116, -0.114) $ and $s_{d}\in (0.111, 0.113)$.]
{(Colored lines) $(V_{\mathrm{CKM}})_{21}$
versus $s_{u}$ and $s_{d}$ with $s_{u}\in (-0.116, -0.114) $ and $s_{d}\in (0.111, 0.113)$.}
\label{u21F}
\vspace{-0.25 cm}
\end{figure}

In the quark sector, the model parameters including $A^q_{11,22,33,12}$ and $B^q_{13, 31, 23, 32}$ and $\psi_q$ with $\, (q=u,d)$ are expressed in terms of the experimental parameters including six quark mass parameters $m_u, m_d,  m_c, m_s, m_t, m_b$, nine CKM elements $(V^{exp}_{\mathrm{CKM}})_{ij}\, (i,j=1,2,3)$ and two free parameters $s_u, s_d$ as given in Eqs. (\ref{XYq}), (\ref{psiq}), (\ref{relationsq1}), (\ref{B13q})-(\ref{B32q}). To calculate the values of the model parameters in the quark sector, we use the observables quark masses and the CKM elements whose experimental values are shown in Eqs.(\ref{PDG22qmass}) and (\ref{PDG22qmix}).
Equations (\ref{Vckmv})-(\ref{relationsq2}) indicate that $(V_\mathrm{CKM})_{21}$ depends on observables quark masses, the CKM elements $(V^{exp}_{\mathrm{CKM}})_{ij}\, (i,j=1,2,3)$ and $s_u, s_d$ while all other elements of $V_\mathrm{CKM}$ depend on observables quark masses, the CKM elements $(V^{exp}_{\mathrm{CKM}})_{ij}\, (i,j=1,2,3; ij\neq 21)$, $s_u, s_d$ and $B^u_{32}$. At the best fit points of the quark masses and the CKM elements as in Eqs.(\ref{PDG22qmass}) and (\ref{PDG22qmix}), $(V_\mathrm{CKM})_{21}$ depends on two parameters $s_u, s_d$ while all the other elements $(V^{exp}_{\mathrm{CKM}})_{ij}\, (i,j=1,2,3; ij\neq 21)$ depend on three parameters $s_u, s_d$ and $B^u_{32}$. We find the possible values of $s_u$ and $s_d$, with $s_{u}\in (-0.116, -0.114) $ and $s_{d}\in (0.111, 0.113)$, such that $(V_{\mathrm{CKM}})_{21}$ gets a value consistent with its experimental value as in Eq. (\ref{PDG22qmix}), $(V^{exp}_{\mathrm{CKM}})_{21}=-0.224876$, as plotted in Fig. \ref{u21F}.
In the case of $s_d = 0.111 \, (\psi_d=6.37^\circ)$ and $s_u = -0.115 \, (\psi_d=353.40^\circ)$, the element $(V_\mathrm{CKM})_{31}$ depends on only one parameter $B^u_{32}$. Next, comparing $\,(V_\mathrm{CKM})_{31}$ with its best-fit point \cite{PDG2022}, $(V^{exp}_\mathrm{CKM})_{31}=0.00792 - 0.00327 i$, we get:
\bea
&&B^u_{32}=3.16064-3.7931 i \, \mathrm{GeV}, \eea
and the model parameters in the quark sector are obtained as follows:
\bea
&&A^u_{33}=172.69 \, \mathrm{GeV}, \hs A^d_{33}=4.18 \, \mathrm{GeV}, \crn
&&B^u_{13}=-1.84812 + 1.69045 i \, \mathrm{GeV},\hs B^u_{23}=3.98863 - 3.64836i \, \mathrm{GeV},  \crn
&&B^u_{31}=(3.18377+ 6.09065i)10^{-1}  \, \mathrm{GeV}, \crn
&&B^d_{13}=(2.31286+2.11554 i) 10^{-2} \, \mathrm{GeV}, \hs
B^d_{23}=-(9.27631+8.48494 i)10^{-2} \, \mathrm{GeV}, \crn
&&B^d_{31}=(-3.24745 + 7.31785i) 10^{-2}\, \mathrm{GeV}, \,
B^d_{32}=-(8.52256+7.25109i)10^{-2} \, \mathrm{GeV}, \label{BuBdij}\\
&&X_u=0.63608 \, \mathrm{GeV}, \hs Y_u=0.63392 \, \mathrm{GeV}, \crn
&&X_d=4.9035\times 10^{-2} \, \mathrm{eV}, \hs Y_d=4.4365 \times10^{-2} \, \mathrm{eV}. \label{XYudvalues}\eea
Simultaneously, the predicted values for the quark mixing matrix elements are given in Table \ref{quarkmĩingpara}.
\begin{table}[tbh]
\caption{\label{quarkmĩingpara}The best-fit points for the elements of the quark mixing matrix taken from Ref.\cite{PDG2022} and the model prediction.}
\vspace{-0.25cm}
\begin{center}
\begin{tabular}{|c|c|c|c|c|}
\hline
Observable &The model prediction& Best-fit point &  Percent error $(\%)$ \\ \hline
$\left(V_{\mathrm{CKM}}\right)_{11}$ & \quad $0.97435$& \quad $0.97435$ &   $0$\\ \hline
$\left(V_{\mathrm{CKM}}\right)_{12}$ & \quad $0.22500$& \quad $0.22500$ &   $0$\\ \hline
$\left(V_{\mathrm{CKM}}\right)_{13}$ & \quad $0.0015275-0.003359 i$& \quad $0.0015275-0.003359 i$ & $0$ \\ \hline
$\left(V_{\mathrm{CKM}}\right)_{21}$ & \quad $-0.22487$& \quad $-0.22431$&  $0.24523$ \\ \hline
$\left(V_{\mathrm{CKM}}\right)_{22}$ & \quad $0.97349$& \quad $0.97349$&    $0$\\ \hline
$\left(V_{\mathrm{CKM}}\right)_{23}$ & \quad $0.04182$& \quad $0.04182$ &$0$\\ \hline
$\left(V_{\mathrm{CKM}}\right)_{31}$ & \quad $0.0079225 -0.00327 i$& \quad $0.0079225 -0.00327 i$& $0$\\ \hline
$\left(V_{\mathrm{CKM}}\right)_{32}$ & \quad $-0.041091$& \quad $-0.041091$&$0$\\ \hline
$\left(V_{\mathrm{CKM}}\right)_{33}$ & \quad $0.99912$& \quad $0.99912$ & $0$ \\ \hline
\end{tabular}%
\end{center}
\vspace{-0.5cm}
\end{table}

Equations (\ref{XYq})-(\ref{Kq}) and (\ref{XYudvalues}) provide the following solutions:
\bea
&&A^u_{11}=1.89272\times 10^{-2}\, \, \mathrm{GeV},\hs A^u_{12}= 0.14483\, \mathrm{GeV}, \hs A^u_{22}=1.25323\, \mathrm{GeV}\\
&&A^d_{11}=5.76324\times 10^{-3}\, \mathrm{GeV}, A^d_{12}= -9.78817\times 10^{-3}\, \mathrm{GeV}, A^d_{22}=9.23068\times 10^{-2}\, \mathrm{GeV}. \label{Audiivalues}
\eea
The analysis shows that the considered model can explain the observed pattern of quark masses and mixings in Eqs. (\ref{PDG22qmass}) and (\ref{PDG22qangles}) in which all the quark masses can get the best-fit values and all the elements of the quark mixing matrix are in good agreement with the experimental constraints except one element, $(V_{\mathrm{CKM}})_{21}$ with a deviation about $0.25\,\%$ deviation.
\section{\label{conclusion}Conclusions}
We have proposed a $U(1)_L$ model with $A_4$ symmetry in light of the linear seesaw for majorana neutrino that capable of generating the current lepton and quark mass and mixing patterns. The smallness of Majorana neutrino mass is reproduced through the linear seesaw mechanism. The model can accommodate the current observed patterns of lepton and quark mixing in which the solar neutrino mixing angle and the Dirac CP violating phase are in $2\sigma$ range for both NO and IO, the Majorana violating phases are predicted to be $\eta_{1} \in (2.29, 10.31)^\circ$ and $\eta_{2} \in (57.30, 302.70)^\circ $ for NO while $\eta_1\in (3.44, 10.31)^\circ$ and $\eta_{2} \in(79.07, 87.09)^\circ$ for  IO. The obtained sum of neutrino mass and the effective Majorana neutrino mass are in good consistent with the recent upper limits. For quark sector, all the quark masses can get the best-fit values and all the elements of the quark mixing matrix are in agreement with the experimental constraints except one element, $(V_{\mathrm{CKM}})_{21}$, with a deviation about $0.25\,\%$ deviation.

\newpage
\appendix
\section{\label{preventedterms} Yukawa terms forbidden by additional symmetries $U(1)_L, Z_2, Z_3$ and $Z_4$}
\vspace{-0.5 cm}
 \begin{table}[h]
 \caption{\label{preventedtermsT} Forbidden terms}
  \vspace{0.5 cm}
 \begin{tabular}{|c|c|c|c|c|} \hline
Forbidden terms &Forbidden by \\ \hline
$(\overline{\psi}_{L} N^c_{L})_{\underline{3}_s} (\widetilde{H} \varphi^*)_{\underline{3}}, (\overline{\psi}_{L} N^c_{L})_{\underline{3}_a} (\widetilde{H} \varphi^*)_{\underline{3}}, (\overline{\psi}_{L} N^c_{L})_{\underline{1}} (\widetilde{H} \chi^*)_{\underline{1}};
(\overline{\psi}_{L} S^c_{L})_{\underline{3}_s} (\widetilde{H} \varphi^*)_{\underline{3}}, $&\multirow{15}{2.2 cm}{\hspace{0.55 cm}$U(1)_L$}  \\
$(\overline{\psi}_{L} S^c_{L})_{\underline{3}_a} (\widetilde{H} \varphi^*)_{\underline{3}}, (\overline{\psi}_{L} S^c_{L})_{\underline{1}} (\widetilde{H} \varphi^*)_{\underline{1}};
(\overline{N}_L \psi^c_{L})_{\underline{3}_s} (\widetilde{H} \varphi^*)_{\underline{3}}, (\overline{N}_L \psi^c_{L})_{\underline{3}_a} (\widetilde{H} \varphi^*)_{\underline{3}},  $ &\\
$(\overline{N}_L \psi^c_{L})_{\underline{1}} (\widetilde{H} \chi^*)_{\underline{1}}; (\overline{S}_L \psi^c_{L})_{\underline{3}_s} (\widetilde{H} \varphi^*)_{\underline{3}}, (\overline{S}_L \psi^c_{L})_{\underline{3}_a} (\widetilde{H} \varphi^*)_{\underline{3}}, (\overline{S}_L \psi^c_{L})_{\underline{1}} (\widetilde{H} \chi^*)_{\underline{1}}; $&\\
$(\overline{N}_L \nu_{R})_{\underline{3}_s} \phi, (\overline{N}_L \nu_{R})_{\underline{3}_a} \phi, (\overline{S}_L \nu_{R})_{\underline{3}_s} \phi, (\overline{S}_L \nu_{R})_{\underline{3}_a} \phi; (\overline{N}_L N_{R})_{\underline{3}_s} (\phi\rho)_{\underline{3}},  $&\\
$(\overline{N}_L N_{R})_{\underline{3}_a} (\phi\rho)_{\underline{3}},  (\overline{N}_L S_{R})_{\underline{3}_s} (\phi\rho)_{\underline{3}}, (\overline{N}_L S_{R})_{\underline{3}_a} (\phi\rho)_{\underline{3}}, (\overline{S}_L N_{R})_{\underline{3}_s} (\phi\rho)_{\underline{3}},$&\\
$(\overline{S}_L N_{R})_{\underline{3}_a} (\phi\rho)_{\underline{3}}; (\overline{S}_L S_{R})_{\underline{3}_s} (\phi\rho)_{\underline{3}},(\overline{S}_L S_{R})_{\underline{3}_a} (\phi\rho)_{\underline{3}}; (\overline{\nu}^c_R \nu_{R})_{\underline{3}_s} (\varphi^*\chi)_{\underline{3}}, $&\\ 
$(\overline{\nu}^c_R \nu_{R})_{\underline{3}_a} (\varphi^*\chi)_{\underline{3}}, (\overline{\nu}^c_R \nu_{R})_{\underline{3}_s} (\varphi\chi^*)_{\underline{3}},  (\overline{\nu}^c_R \nu_{R})_{\underline{3}_a} (\varphi\chi^*)_{\underline{3}}, (\overline{\nu}^c_R \nu_{R})_{\underline{1}} (H\widetilde{H})_{\underline{1}}, $ & \\
$(\overline{\nu}^c_R \nu_{R})_{\underline{1}} (\chi\chi^*)_{\underline{1}}, (\overline{\nu}^c_R \nu_{R})_{\underline{1}} (\rho\rho^*)_{\underline{1}}, (\overline{\nu}^c_R \nu_{R})_{\underline{1}} (\eta\eta^*)_{\underline{1}}; (\overline{\nu}^c_R \nu_{R})_{\underline{1}} (\phi^2)_{\underline{1}}, (\overline{\nu}^c_R \nu_{R})_{\underline{1}} (\varphi\varphi^*)_{\underline{1}}, $&\\
$(\overline{\nu}^c_R \nu_{R})_{\underline{1}^'} (\phi^2)_{\underline{1}^{''}}, (\overline{\nu}^c_R \nu_{R})_{\underline{1}^'} (\varphi\varphi^*)_{\underline{1}^{''}}, (\overline{\nu}^c_R \nu_{R})_{\underline{1}^{''}} (\phi^2)_{\underline{1}^'}, (\overline{\nu}^c_R \nu_{R})_{\underline{1}^{''}} (\varphi\varphi^*)_{\underline{1}^'}, $&\\
$(\overline{\nu}^c_R \nu_{R})_{\underline{3}_s} (\phi^2)_{\underline{3}_s}, (\overline{\nu}^c_R \nu_{R})_{\underline{3}_s} (\varphi\varphi^*)_{\underline{3}_s},
(\overline{\nu}^c_R \nu_{R})_{\underline{3}_s} (\phi^2)_{\underline{3}_a}, (\overline{\nu}^c_R \nu_{R})_{\underline{3}_s} (\varphi\varphi^*)_{\underline{3}_a}, $&\\
$(\overline{\nu}^c_R \nu_{R})_{\underline{3}_a} (\phi^2)_{\underline{3}_s}, (\overline{\nu}^c_R \nu_{R})_{\underline{3}_a} (\varphi\varphi^*)_{\underline{3}_s},
(\overline{\nu}^c_R \nu_{R})_{\underline{3}_a} (\phi^2)_{\underline{3}_a}, (\overline{\nu}^c_R \nu_{R})_{\underline{3}_a} (\varphi\varphi^*)_{\underline{3}_a}, $&\\
$(\overline{\nu}^c_R N_{R})_{\underline{3}_s} (\varphi\rho^*)_{\underline{3}}, (\overline{\nu}^c_R N_{R})_{\underline{3}_a} (\varphi\rho^*)_{\underline{3}},
(\overline{\nu}^c_R S_{R})_{\underline{3}_s} (\varphi\rho^*)_{\underline{3}}, (\overline{\nu}^c_R S_{R})_{\underline{3}_a} (\varphi\rho^*)_{\underline{3}}, $&\\
$(\overline{N}^c_R \nu_{R})_{\underline{3}_s} (\varphi\rho^*)_{\underline{3}}, (\overline{N}^c_R \nu_{R})_{\underline{3}_a} (\varphi\rho^*)_{\underline{3}},
(\overline{S}^c_R \nu_{R})_{\underline{3}_s} (\varphi\rho^*)_{\underline{3}}, (\overline{S}^c_R \nu_{R})_{\underline{3}_a} (\varphi\rho^*)_{\underline{3}}, $&\\
$(\overline{\nu}^c_R N_{R})_{\underline{1}} (\chi\rho^*)_{\underline{1}}, (\overline{\nu}^c_R S_{R})_{\underline{1}} (\chi\rho^*)_{\underline{1}}, (\overline{N}^c_R \nu_{R})_{\underline{1}} (\chi\rho^*)_{\underline{1}},
(\overline{S}^c_R \nu_{R})_{\underline{1}} (\chi\rho^*)_{\underline{1}},$&\\
$(\overline{N}^c_R N_{R})_{\underline{1}} (\rho^2)_{\underline{1}},
(\overline{N}^c_R S_{R})_{\underline{1}} (\rho^2)_{\underline{1}},
(\overline{S}^c_R N_{R})_{\underline{1}} (\rho^2)_{\underline{1}},
(\overline{S}^c_R S_{R})_{\underline{1}} (\rho^2)_{\underline{1}}.$&\\\hline

$(\overline{\nu}^c_{R} N_{R})_{\underline{3}_{s,a}} (\phi\rho)_{\underline{3}},
(\overline{\nu}^c_{R} S_{R})_{\underline{3}_{s,a}} (\phi\rho)_{\underline{3}},
(\overline{N}^c_{R} \nu_{R})_{\underline{3}_{s,a}} (\phi\rho)_{\underline{3}},
(\overline{S}^c_{R} \nu_{R})_{\underline{3}_{s,a}} (\phi\rho)_{\underline{3}},$&\multirow{6}{1.0 cm}{\hspace{0.25 cm}$Z_2$}  \\
$(\overline{N}^c_{R} N_{R})_{\underline{1}} (\phi\varphi)_{\underline{1}},
(\overline{N}^c_{R} S_{R})_{\underline{1}} (\phi\varphi)_{\underline{1}},
(\overline{S}^c_{R} N_{R})_{\underline{1}} (\phi\varphi)_{\underline{1}},
(\overline{S}^c_{R} S_{R})_{\underline{1}} (\phi\varphi)_{\underline{1}},$ &\\
$(\overline{N}^c_{R} N_{R})_{\underline{1}^'} (\phi\varphi)_{\underline{1}^{''}},
(\overline{N}^c_{R} N_{R})_{\underline{1}^{''}} (\phi\varphi)_{\underline{1}^{'}},
(\overline{N}^c_{R} S_{R})_{\underline{1}^'} (\phi\varphi)_{\underline{1}^{''}},
(\overline{N}^c_{R} S_{R})_{\underline{1}^{''}} (\phi\varphi)_{\underline{1}^{'}},$&\\
$(\overline{S}^c_{R} N_{R})_{\underline{1}^'} (\phi\varphi)_{\underline{1}^{''}},
(\overline{S}^c_{R} N_{R})_{\underline{1}^{''}} (\phi\varphi)_{\underline{1}^{'}},
(\overline{S}^c_{R} S_{R})_{\underline{1}^'} (\phi\varphi)_{\underline{1}^{''}},
(\overline{S}^c_{R} S_{R})_{\underline{1}^{''}} (\phi\varphi)_{\underline{1}^{'}},$ &\\
$(\overline{N}^c_{R} N_{R})_{\underline{3}_{s,a}} (\phi\varphi)_{\underline{3}_{s,a}},
(\overline{N}^c_{R} N_{R})_{\underline{3}_{s,a}} (\phi\chi)_{\underline{3}},
(\overline{N}^c_{R} S_{R})_{\underline{3}_{s,a}} (\phi\varphi)_{\underline{3}_{s,a}},
(\overline{N}^c_{R} S_{R})_{\underline{3}_{s,a}} (\phi\chi)_{\underline{3}},$ &\\
$(\overline{S}^c_{R} N_{R})_{\underline{3}_{s,a}} (\phi\varphi)_{\underline{3}_{s,a}},
(\overline{S}^c_{R} N_{R})_{\underline{3}_{s,a}} (\phi\chi)_{\underline{3}},
(\overline{S}^c_{R} S_{R})_{\underline{3}_{s,a}} (\phi\varphi)_{\underline{3}_{s,a}},
(\overline{S}^c_{R} S_{R})_{\underline{3}_{s,a}} (\phi\chi)_{\underline{3}},$ &\\ \hline

$(\overline{N}_{L} N^c_{L})_{\underline{1}}\chi, (\overline{N}_{L} N^c_{L})_{\underline{1}}\chi^*,
(\overline{N}_{L} S^c_{L})_{\underline{1}}\chi, (\overline{N}_{L} S^c_{L})_{\underline{1}}\chi^*,
(\overline{S}_{L} N^c_{L})_{\underline{1}}\chi, (\overline{S}_{L} N^c_{L})_{\underline{1}}\chi^*,$&\multirow{4}{1.0 cm}{\hspace{0.25 cm}$Z_3$}  \\
$(\overline{S}_{L} S^c_{L})_{\underline{1}}\chi, (\overline{S}_{L} S^c_{L})_{\underline{1}}\chi^*; (\overline{N}^c_{R} N_{R})_{\underline{1}} \chi^*,
(\overline{N}^c_{R} S_{R})_{\underline{1}} \chi^*,
(\overline{S}^c_{R} N_{R})_{\underline{1}} \chi^*,
(\overline{S}^c_{R} S_{R})_{\underline{1}} \chi^*;$ &\\
$(\overline{Q}_{1L} u_{3R})_{\underline{1}} (\widetilde{H} \eta^*)_{\underline{1}},
(\overline{Q}_{2L} u_{3R})_{\underline{1}} (\widetilde{H} \eta^*)_{\underline{1}},
(\overline{Q}_{3L} u_{1R})_{\underline{1}} (\widetilde{H}\eta)_{\underline{1}},
(\overline{Q}_{3L} u_{2R})_{\underline{1}} (\widetilde{H}\eta)_{\underline{1}},$&\\
$(\overline{Q}_{1L} d_{3R})_{\underline{1}} (H \eta^*)_{\underline{1}},
(\overline{Q}_{2L} d_{3R})_{\underline{1}} (H \eta^*)_{\underline{1}},
(\overline{Q}_{3L} d_{1R})_{\underline{1}} (H\eta)_{\underline{1}},
(\overline{Q}_{3L} d_{2R})_{\underline{1}} (H \eta)_{\underline{1}}.$&\\\hline

\end{tabular}
\end{table}
\newpage
\begin{center}
Table \ref{preventedtermsT}. Forbidden terms (continued)
\end{center}
 \begin{table}[h]
 \vspace{-0.65 cm}
\vspace*{0.25 cm}
 \begin{tabular}{|c|c|c|c|c|} \hline
Forbidden terms &Forbidden by \\ \hline
$(\overline{N}_{L} N^c_{L})_{\underline{1}}(H\widetilde{H})_{\underline{1}},
(\overline{N}_{L} N^c_{L})_{\underline{1}}(\phi^2)_{\underline{1}},
(\overline{N}_{L} N^c_{L})_{\underline{1}}(\phi \phi^*)_{\underline{1}},
(\overline{N}_{L} N^c_{L})_{\underline{1}}(\varphi \varphi^*)_{\underline{1}},$&\multirow{31}{1.0 cm}{\hspace{0.25 cm}$Z_4$}  \\
$(\overline{N}_{L} N^c_{L})_{\underline{1}}(\chi \chi^*)_{\underline{1}},
(\overline{N}_{L} N^c_{L})_{\underline{1}}(\rho \rho^*)_{\underline{1}},
(\overline{N}_{L} S^c_{L})_{\underline{1}}(H\widetilde{H})_{\underline{1}},
(\overline{N}_{L} S^c_{L})_{\underline{1}}(\phi^2)_{\underline{1}},$&\\
$
(\overline{N}_{L} S^c_{L})_{\underline{1}}(\phi \phi^*)_{\underline{1}},
(\overline{N}_{L} S^c_{L})_{\underline{1}}(\varphi \varphi^*)_{\underline{1}},
(\overline{N}_{L} S^c_{L})_{\underline{1}}(\chi \chi^*)_{\underline{1}},
(\overline{N}_{L} S^c_{L})_{\underline{1}}(\rho \rho^*)_{\underline{1}},$ &\\
$
(\overline{S}_{L} N^c_{L})_{\underline{1}}(H\widetilde{H})_{\underline{1}},
(\overline{S}_{L} N^c_{L})_{\underline{1}}(\phi^2)_{\underline{1}},
(\overline{S}_{L} N^c_{L})_{\underline{1}}(\phi \phi^*)_{\underline{1}},
(\overline{S}_{L} N^c_{L})_{\underline{1}}(\varphi \varphi^*)_{\underline{1}},$ &\\
$
(\overline{S}_{L} N^c_{L})_{\underline{1}}(\chi \chi^*)_{\underline{1}},
(\overline{S}_{L} N^c_{L})_{\underline{1}}(\rho \rho^*)_{\underline{1}}, (\overline{S}_{L} S^c_{L})_{\underline{1}}(H\widetilde{H})_{\underline{1}},
(\overline{S}_{L} S^c_{L})_{\underline{1}}(\phi^2)_{\underline{1}},$ &\\
$(\overline{S}_{L} S^c_{L})_{\underline{1}}(\phi \phi^*)_{\underline{1}},
(\overline{S}_{L} S^c_{L})_{\underline{1}}(\varphi \varphi^*)_{\underline{1}},
(\overline{S}_{L} S^c_{L})_{\underline{1}}(\chi \chi^*)_{\underline{1}},
(\overline{S}_{L} S^c_{L})_{\underline{1}}(\rho \rho^*)_{\underline{1}},$ &\\
$(\overline{N}_{L} N^c_{L})_{\underline{1}^'}(\phi^2)_{\underline{1}^{''}},
(\overline{N}_{L} N^c_{L})_{\underline{1}^{'}}(\phi \phi^*)_{\underline{1}^{''}},
(\overline{N}_{L} N^c_{L})_{\underline{1}^{'}}(\varphi \varphi^*)_{\underline{1}^{''}},
(\overline{N}_{L} S^c_{L})_{\underline{1}^'}(\phi^2)_{\underline{1}^{''}},$ &\\
$
(\overline{N}_{L} S^c_{L})_{\underline{1}^{'}}(\phi \phi^*)_{\underline{1}^{''}},
(\overline{N}_{L} S^c_{L})_{\underline{1}^{'}}(\varphi \varphi^*)_{\underline{1}^{''}},
(\overline{S}_{L} N^c_{L})_{\underline{1}^'}(\phi^2)_{\underline{1}^{''}},
(\overline{S}_{L} N^c_{L})_{\underline{1}^{'}}(\phi \phi^*)_{\underline{1}^{''}},$&  \\
$(\overline{S}_{L} N^c_{L})_{\underline{1}^{'}}(\varphi \varphi^*)_{\underline{1}^{''}},
(\overline{S}_{L} S^c_{L})_{\underline{1}^'}(\phi^2)_{\underline{1}^{''}},
(\overline{S}_{L} S^c_{L})_{\underline{1}^{'}}(\phi \phi^*)_{\underline{1}^{''}},
(\overline{S}_{L} S^c_{L})_{\underline{1}^{'}}(\varphi \varphi^*)_{\underline{1}^{''}},$ &\\
$

(\overline{N}_{L} N^c_{L})_{\underline{1}^{''}}(\phi^2)_{\underline{1}^{'}},
(\overline{N}_{L} N^c_{L})_{\underline{1}^{''}}(\phi \phi^*)_{\underline{1}^{'}},
(\overline{N}_{L} N^c_{L})_{\underline{1}^{''}}(\varphi \varphi^*)_{\underline{1}^{'}},
(\overline{N}_{L} S^c_{L})_{\underline{1}^{''}}(\phi^2)_{\underline{1}^{'}},$ &\\

$
(\overline{N}_{L} S^c_{L})_{\underline{1}^{''}}(\phi \phi^*)_{\underline{1}^{'}},
(\overline{N}_{L} S^c_{L})_{\underline{1}^{''}}(\varphi \varphi^*)_{\underline{1}^{'}},
(\overline{S}_{L} N^c_{L})_{\underline{1}^{''}}(\phi^2)_{\underline{1}^{'}},
(\overline{S}_{L} N^c_{L})_{\underline{1}^{''}}(\phi \phi^*)_{\underline{1}^{'}},$&  \\
$

(\overline{S}_{L} N^c_{L})_{\underline{1}^{''}}(\varphi \varphi^*)_{\underline{1}^{'}},
(\overline{S}_{L} S^c_{L})_{\underline{1}^{''}}(\phi^2)_{\underline{1}^{'}},
(\overline{S}_{L} S^c_{L})_{\underline{1}^{''}}(\phi \phi^*)_{\underline{1}^{'}},
(\overline{S}_{L} S^c_{L})_{\underline{1}^{''}}(\varphi \varphi^*)_{\underline{1}^{'}},$ &\\
$

(\overline{N}_{L} N^c_{L})_{\underline{3}_{s,a}} (\varphi \chi^*)_{\underline{3}},
(\overline{N}_{L} N^c_{L})_{\underline{3}_{s,a}} (\varphi^* \chi)_{\underline{3}},
(\overline{N}_{L} N^c_{L})_{\underline{3}_{s,a}} (\phi^2)_{\underline{3}_{s,a}},
(\overline{N}_{L} N^c_{L})_{\underline{3}_{s,a}} (\phi\phi^*)_{\underline{3}_{s,a}},$ &\\ 

$
(\overline{N}_{L} N^c_{L})_{\underline{3}_{s,a}} (\varphi \varphi^*)_{\underline{3}_{s,a}},
(\overline{N}_{L} S^c_{L})_{\underline{3}_{s,a}} (\varphi \chi^*)_{\underline{3}},
(\overline{N}_{L} S^c_{L})_{\underline{3}_{s,a}} (\varphi^* \chi)_{\underline{3}},
(\overline{N}_{L} S^c_{L})_{\underline{3}_{s,a}} (\phi^2)_{\underline{3}_{s,a}},$& \\
$
(\overline{N}_{L} S^c_{L})_{\underline{3}_{s,a}} (\phi\phi^*)_{\underline{3}_{s,a}},
(\overline{N}_{L} S^c_{L})_{\underline{3}_{s,a}} (\varphi \varphi^*)_{\underline{3}_{s,a}},
(\overline{S}_{L} N^c_{L})_{\underline{3}_{s,a}} (\varphi \chi^*)_{\underline{3}},
(\overline{S}_{L} N^c_{L})_{\underline{3}_{s,a}} (\varphi^* \chi)_{\underline{3}},$ &\\
$

(\overline{S}_{L} N^c_{L})_{\underline{3}_{s,a}} (\phi^2)_{\underline{3}_{s,a}},
(\overline{S}_{L} N^c_{L})_{\underline{3}_{s,a}} (\phi\phi^*)_{\underline{3}_{s,a}},
(\overline{S}_{L} N^c_{L})_{\underline{3}_{s,a}} (\varphi \varphi^*)_{\underline{3}_{s,a}},
(\overline{S}_{L} S^c_{L})_{\underline{3}_{s,a}} (\varphi \chi^*)_{\underline{3}},$ &\\
$

(\overline{S}_{L} S^c_{L})_{\underline{3}_{s,a}} (\varphi^* \chi)_{\underline{3}},
(\overline{S}_{L} S^c_{L})_{\underline{3}_{s,a}} (\phi^2)_{\underline{3}_{s,a}},
(\overline{S}_{L} S^c_{L})_{\underline{3}_{s,a}} (\phi\phi^*)_{\underline{3}_{s,a}},
(\overline{S}_{L} S^c_{L})_{\underline{3}_{s,a}} (\varphi \varphi^*)_{\underline{3}_{s,a}},$ &\\
$
(\overline{N}_{L} \nu_{R})_{\underline{1}} (\chi\rho)_{\underline{1}},
(\overline{S}_{L} \nu_{R})_{\underline{1}} (\chi\rho)_{\underline{1}},
(\overline{N}_{L} \nu_{R})_{\underline{3}_{s,a}} (\varphi\rho)_{\underline{3}},
(\overline{S}_{L} \nu_{R})_{\underline{3}_{s,a}} (\varphi\rho)_{\underline{3}};
(\overline{N}_{L} N_{R})_{\underline{1}} \chi^*,$ &\\

$(\overline{N}_{L} S_{R})_{\underline{1}} \chi^*,
(\overline{S}_{L} N_{R})_{\underline{1}} \chi^*,
(\overline{S}_{L} S_{R})_{\underline{1}} \chi^*;
(\overline{N}_{L} N_{R})_{\underline{1}} (\varphi^2)_{\underline{1}},
(\overline{N}_{L} N_{R})_{\underline{1}} (\chi^2)_{\underline{1}},$ &  \\ 

$

(\overline{N}_{L} S_{R})_{\underline{1}} (\varphi^2)_{\underline{1}},
(\overline{N}_{L} S_{R})_{\underline{1}} (\chi^2)_{\underline{1}},
(\overline{S}_{L} N_{R})_{\underline{1}} (\varphi^2)_{\underline{1}},
(\overline{S}_{L} N_{R})_{\underline{1}} (\chi^2)_{\underline{1}},
(\overline{S}_{L} S_{R})_{\underline{1}} (\varphi^2)_{\underline{1}},$ & \\ 

$(\overline{S}_{L} S_{R})_{\underline{1}} (\chi^2)_{\underline{1}},
(\overline{N}_{L} N_{R})_{\underline{1}^{'}} (\varphi^2)_{\underline{1}^{''}},
(\overline{N}_{L} S_{R})_{\underline{1}^{'}} (\varphi^2)_{\underline{1}^{''}},
(\overline{S}_{L} N_{R})_{\underline{1}^{'}} (\varphi^2)_{\underline{1}^{''}},
(\overline{S}_{L} S_{R})_{\underline{1}^{'}} (\varphi^2)_{\underline{1}^{''}},$ &\\
$(\overline{N}_{L} N_{R})_{\underline{1}^{''}} (\varphi^2)_{\underline{1}^{'}},
(\overline{N}_{L} S_{R})_{\underline{1}^{''}} (\varphi^2)_{\underline{1}^{'}},
(\overline{S}_{L} N_{R})_{\underline{1}^{''}} (\varphi^2)_{\underline{1}^{'}},
(\overline{S}_{L} S_{R})_{\underline{1}^{''}} (\varphi^2)_{\underline{1}^{'}},
(\overline{N}_{L} N_{R})_{\underline{3}_{s,a}} \varphi^*,$& \\

$(\overline{N}_{L} S_{R})_{\underline{3}_{s,a}} \varphi^*,
(\overline{S}_{L} N_{R})_{\underline{3}_{s,a}} \varphi^*,
(\overline{S}_{L} S_{R})_{\underline{3}_{s,a}} \varphi^*,
(\overline{N}_{L} N_{R})_{\underline{3}_{s,a}} (\varphi\chi)_{\underline{3}},
(\overline{N}_{L} S_{R})_{\underline{3}_{s,a}} (\varphi\chi)_{\underline{3}},$ &   \\ 

$(\overline{S}_{L} N_{R})_{\underline{3}_{s,a}} (\varphi\chi)_{\underline{3}},
(\overline{S}_{L} S_{R})_{\underline{3}_{s,a}} (\varphi\chi)_{\underline{3}},
(\overline{N}_{L} N_{R})_{\underline{3}_{s,a}} (\varphi^2)_{\underline{3}_{s,a}},
(\overline{N}_{L} S_{R})_{\underline{3}_{s,a}} (\varphi^2)_{\underline{3}_{s,a}},$ &   \\ 

$(\overline{S}_{L} N_{R})_{\underline{3}_{s,a}} (\varphi^2)_{\underline{3}_{s,a}},
(\overline{S}_{L} S_{R})_{\underline{3}_{s,a}} (\varphi^2)_{\underline{3}_{s,a}},
(\overline{N}^c_{R} N_{R})_{\underline{1}} (\varphi^{*2})_{\underline{1}},
(\overline{N}^c_{R} N_{R})_{\underline{1}} (\chi^{*2})_{\underline{1}},$ &   \\  

$(\overline{N}^c_{R} S_{R})_{\underline{1}} (\varphi^{*2})_{\underline{1}},
(\overline{N}^c_{R} S_{R})_{\underline{1}} (\chi^{*2})_{\underline{1}},
(\overline{S}^c_{R} N_{R})_{\underline{1}} (\varphi^{*2})_{\underline{1}},
(\overline{S}^c_{R} N_{R})_{\underline{1}} (\chi^{*2})_{\underline{1}},
(\overline{S}^c_{R} S_{R})_{\underline{1}} (\varphi^{*2})_{\underline{1}},$&  \\
$(\overline{S}^c_{R} S_{R})_{\underline{1}} (\chi^{*2})_{\underline{1}};
(\overline{N}^c_{R} N_{R})_{\underline{1}^{'}} (\varphi^{*2})_{\underline{1}^{''}},
(\overline{N}^c_{R} N_{R})_{\underline{1}^{''}} (\varphi^{*2})_{\underline{1}^{'}},
(\overline{N}^c_{R} S_{R})_{\underline{1}^{'}} (\varphi^{*2})_{\underline{1}^{''}},$ &\\ 

$(\overline{N}^c_{R} S_{R})_{\underline{1}^{''}} (\varphi^{*2})_{\underline{1}^{'}},
(\overline{S}^c_{R} N_{R})_{\underline{1}^{'}} (\varphi^{*2})_{\underline{1}^{''}},
(\overline{S}^c_{R} N_{R})_{\underline{1}^{''}} (\varphi^{*2})_{\underline{1}^{'}},
(\overline{S}^c_{R} S_{R})_{\underline{1}^{'}} (\varphi^{*2})_{\underline{1}^{''}},$ &\\ 

$(\overline{S}^c_{R} S_{R})_{\underline{1}^{''}} (\varphi^{*2})_{\underline{1}^{'}},
(\overline{N}^c_{R} N_{R})_{\underline{3}_{s,a}} (\varphi^{*2})_{\underline{3}_{s,a}},
(\overline{N}^c_{R} N_{R})_{\underline{3}_{s,a}} (\varphi^{*}\chi^*)_{\underline{3}_{s,a}},
$& \\
$(\overline{N}^c_{R} S_{R})_{\underline{3}_{s,a}} (\varphi^{*2})_{\underline{3}_{s,a}},
(\overline{N}^c_{R} S_{R})_{\underline{3}_{s,a}} (\varphi^{*}\chi^*)_{\underline{3}_{s,a}},
(\overline{S}^c_{R} N_{R})_{\underline{3}_{s,a}} (\varphi^{*2})_{\underline{3}_{s,a}},$ &\\
$(\overline{S}^c_{R} N_{R})_{\underline{3}_{s,a}} (\varphi^{*}\chi^*)_{\underline{3}_{s,a}},
(\overline{S}^c_{R} S_{R})_{\underline{3}_{s,a}} (\varphi^{*2})_{\underline{3}_{s,a}},
(\overline{S}^c_{R} S_{R})_{\underline{3}_{s,a}} (\varphi^{*}\chi^*)_{\underline{3}_{s,a}}. $&\\ \hline
\end{tabular}
\vspace*{-1.5 cm}
\end{table}

\newpage

\section{\label{AppenPQ} Explicit expressions of $P_{11}, P_{12}, Q_{mn}\, (P=X,Y,Z; Q=T, V; mn=13, 14, 23, 24)$}
\vspace*{-0.5 cm}
\bea
X_{11}&=&-\frac{t_{1} x_{4}}{t_{1}^2-t_{2}^2+t_{3}^2}
+\frac{x_{4} \left[w_{1} y_{1} z_{1}-w_{1} y_{2} (z_{2}+z_{3})+w_{2} y_{1} (z_{3}-z_{2})+w_{2} y_{2} z_{1}\right]}{\left(w_{1}^2-w_{2}^2\right) \left(y_{1}^2-y_{2}^2\right)} \crn
&+& \frac{x_{3} \left[(t_1 w_1 -t_2w_2)\Omega_1  +(t_1 w_2 -t_2w_1)\Omega_2\right]}{(w^2_{1}-w^2_{2}) \left(y_{1}^2-y_{2}^2\right)^2} \crn
&+& \frac{x_{3} t_3 \left\{(y^2_1 - y^2_2) (w_2 z_1 - w_1 z_2) + \left[w_1 (y_1^2 + y_2^2)-2 w_2 y_1 y_2\right] z_3\right\}}{(w^2_{1}-w^2_{2}) \left(y_{1}^2-y_{2}^2\right)^2}, \\
X_{12}&=&x_3\left(\frac{t_1 (w_1 y_1 + w_2 y_2)-(t_2 + t_3) w_2 y_1- (t_2 - t_3) w_1 y_2}{(w_1^2 - w_2^2)(y_1^2 - y_2^2)}-\frac{z_1}{z_1^2 - z_2^2 + z_3^2}\right)\crn
&+& x_4\left(\frac{t_1 [y_1 z_1 - y_2 (z_2 - z_3)]-(t_2 + t_3) [y_2 z_1 - y_1 (z_2 - z_3)]}{(t_1^2 - t_2^2 + t_3^2) (z_1^2 - z_2^2 + z_3^2)}-\frac{w_1}{w_1^2 - w_2^2}\right),\\
Y_{11}&=&\frac{z_{1} (x_{4} y_{1}-t_{1} x_{3})}{w_{1} y_{1}^2}-\frac{x_{4}}{t_{1}},\hs\hs Y_{12}=\frac{(t_{1} z_{1}-w_{1} y_{1}) (t_{1} x_{3}-x_{4} y_{1})}{t_{1} w_{1} y_{1} z_{1}},\\
Z_{11}&=&X_{11}+\frac{2 t_3 x_3 (w_1 z_2-w_2 z_1)}{(w_1^2 - w_2^2) (y_1^2 - y_2^2)}+\frac{2 (t_1 w_2-t_2 w_1) x_3 z_3}{(w_1^2 - w_2^2) (y_1^2 - y_2^2)}+\frac{2 x_4 (w_1 y_2-w_2 y_1) z_3}{(w_1^2 - w_2^2) (y_1^2 - y_2^2)}, \\
Z_{12}&=&X_{12}+ \frac{2t_{3} x_{3}(w_2 y_1 - w_1 y_2)}{(w^2_{1}-w^2_{2}) (y^2_{1}-y^2_{2})}
+ \frac{2 x_{4} \left[y_{1}(t_{2} z_{3}-t_{3}z_{2})+ y_{2}(t_{3} z_{1}-t_{1} z_{3})\right]}{\left(t_{1}^2-t_{2}^2+t_{3}^2\right) \left(z_{1}^2-z_{2}^2+z_{3}^2\right)}, \\
T_{13}&=&
\frac{ \left(t_2w_1-t_1 w_2\right) \Omega_1+\left(t_2w_2-t_1 w_1\right) \Omega_2}{(w_1^2 - w_2^2)(y_1^2 - y_2^2)^2}\crn
   &+& \frac{t_3 \left\{(y_1^2 - y_2^2)(w_1 z_1 - w_2 z_2) + [w_2 (y_1^2 + y_2^2)-2 w_1 y_1 y_2] z_3\right\}}{(w_1^2 - w_2^2)(y_1^2 - y_2^2)^2}, \\
   T_{23}&=&\frac{w_1 (t_2 y_1 - t_3 y_1 - t_1 y_2) - w_2 (t_1 y_1 - t_2 y_2 - t_3 y_2)}{(w_1^2 - w_2^2) (y_1^2 - y_2^2)}+\frac{z_2-z_3}{z_1^2 - z_2^2 + z_3^2}, \\
   T_{14}&=&\frac{t_2 - t_3}{t_1^2 - t_2^2 + t_3^2}+\frac{w_1 (y_1 z_2 - y_1 z_3 - y_2 z_1) - w_2 (y_1 z_1 - y_2z_2 - y_2z_3)}{(w_1^2 - w_2^2) (y_1^2 - y_2^2)}, \\
   T_{24}&=&\frac{w_2}{w_1^2 - w_2^2}+\frac{t_1 (y_2 z_1 - y_1 z_2 + y_1 z_3) +(t_3-t_2) (y_1 z_1 - y_2 z_2 + y_2 z_3)}{(t_1^2 - t_2^2 + t_3^2) (z_1^2 - z_2^2 + z_3^2)}, \\
   V_{13}&=&T_{13}+\frac{2 [ (w_1 z_1 - w_2 z_2)t_3 + (t_1 w_1- t_2 w_2) z_3]}{(w_1^2 - w_2^2) (y_1^2 - y_2^2)}, \\
   V_{23}&=&T_{23}+\frac{2  (w_1 y_1 - w_2 y_2)t_3}{(w_1^2 - w_2^2) (y_1^2 - y_2^2)}+\frac{2z_3}{z_1^2 - z_2^2 + z_3^2}, \\
   V_{14}&=&T_{14}+\frac{2 (w_1 y_1 - w_2 y_2) z_3}{(w_1^2 - w_2^2) (y_1^2 - y_2^2)}+\frac{2 t_3}{t_1^2 - t_2^2 + t_3^2}, \\
   V_{24}&=&T_{24}-\frac{2 [t_3 (y_1 z_1 - y_2 z_2) + (t_1 y_1- t_2 y_2)z_3]}{(t_1^2 - t_2^2 + t_3^2) (z_1^2 - z_2^2 + z_3^2)},
   \eea
   with
   \bea
   \Omega_1&=&2y_1 y_2 z_2 -(y_1^2 + y_2^2) z_1, \Omega_2=y_1^2 (z_2 - z_3) + y_2^2 (z_2 + z_3)-2 y_1 y_2 z_1. \eea

\section{\label{Appenvarepsilon} Explicit expressions of $\varepsilon^{1,2}_{u}, \epsilon_0$ and $\epsilon^{1,2}_{u}$}
\bea
&&\varepsilon^{(1)}_{u}= (m_t-m_c)c_d c_u^3 +  (m_u-m_c) c_d^5 c_u s_u^2 + 2 (m_u-m_c) c_d^3 c_u  s_d^2 s_u^2 \crn
&&\hspace{0.7 cm} +\,   \big[m_t - m_u + (m_u-m_c) s_d^4\big] c_d c_us_u^2 + (m_c - m_u) \big[c_u s_d - (V^{exp}_\mathrm{CKM})_{12}\big] c_d^4 c_u s_u\crn
&&\hspace{0.7 cm} +\, 2 (m_c - m_u) \big[c_u s_d -(V^{exp}_\mathrm{CKM})_{12}\big] c_d^2 c_u s_d^2 s_u +
 (m_u-m_c)  (V^{exp}_\mathrm{CKM})_{12} c_u  s_d^4 s_u \crn
&&\hspace{0.7 cm} +\,  \big\{s_d [m_t-m_c + (m_c - m_u) s_d^4] s_u + (m_c - m_t) (V^{exp}_\mathrm{CKM})_{22}\big\}c_u^2\crn
&&\hspace{0.7 cm} +\, (m_t - m_u) \big[s_d s_u - (V^{exp}_\mathrm{CKM})_{22}\big]s_u^2, \\
 &&\varepsilon^{(2)}_{u}=(m_t - m_u) c_u^3  s_d^5 + \big[m_c - m_u + (m_u-m_t) s_d^4\big] c_d c_u^2  s_u + (m_c - m_t) c_d  s_d^4 s_u^3 \crn
&&\hspace{0.675 cm}+\, (m_c - m_u - m_c s_d^4 + m_t s_d^4) c_u s_d  s_u^2 +  \big[(m_u-m_t)c_u^2  + (m_c - m_t) s_u^2\big]c_d^5 s_u \crn
&&\hspace{0.675 cm}+\, 2 \big[(m_u-m_t)c_u^2  + (m_c - m_t) s_u^2\big] c_d^3 s_d^2 s_u+  (m_u-m_c) s_u c_u (V^{exp}_\mathrm{CKM})_{22} \crn
&&\hspace{0.675 cm}+\, \big[(m_t - m_u) c_u^2 + (m_t-m_c) s_u^2\big] \big[c_u s_d - (V^{exp}_\mathrm{CKM})_{12}\big] c_d^4 \crn
&&\hspace{0.675 cm}+\,  2 c_d^2 s_d^2 \big[(m_t - m_u)c_u^2  + (m_t-m_c) s_u^2\big] \big[c_u s_d - (V^{exp}_\mathrm{CKM})_{12}\big] \crn
&&\hspace{0.675 cm}+\, \big[c_u^2 (m_u-m_t) + (m_c - m_t) s_u^2\big] (V^{exp}_\mathrm{CKM})_{12} s_d^4, \\
&&\epsilon^{(1)}_u=(m_t - m_u) c_u^3 s_d^5 + \big[m_c - m_u + (m_u-m_t) s_d^4\big]c_d c_u^2  s_u \crn
&&\hspace{0.665 cm}+\, \big[m_c - m_u - (m_c - m_t) s_d^4\big] c_u s_d s_u^2 +  (m_c - m_t) c_d s_d^4 s_u^3 \crn
&&\hspace{0.665 cm}+\, \big[(m_u-m_t) c_u^2 + (m_c - m_t) s_u^2\big]c_d^5 s_u +
 2  \big[(m_u-m_t)c_u^2 + (m_c - m_t) s_u^2\big] c_d^3 s_d^2 s_u \crn
&&\hspace{0.665 cm}+\, c_d^4 \big[(m_t - m_u) c_u^2  + (m_t-m_c) s_u^2\big] \big[c_u s_d - (V^{exp}_\mathrm{CKM})_{12}\big] \crn
&&\hspace{0.665 cm}+\, 2 \big[(m_t - m_u) c_u^2 + (m_t-m_c) s_u^2\big] \big[c_u s_d - (V^{exp}_\mathrm{CKM})_{12}\big] c_d^2 s_d^2 \crn
&&\hspace{0.665 cm}+\,  \big[(m_u-m_t)c_u^2  + (m_c - m_t) s_u^2\big] (V^{exp}_\mathrm{CKM})_{12} s_d^4  + c_u (m_u-m_c) (V^{exp}_\mathrm{CKM})_{22} s_u, \\
&&\epsilon^{(2)}_u=B^d_{32} + \big[(m_b - m_d) c_u  s_d + (m_s-m_b) c_d s_u \big](V^{exp}_\mathrm{CKM})_{13} \crn
&&\hspace{0.665 cm}+\,  \big[(m_b - m_s)c_d c_u + (m_b - m_d) s_d s_u\big] (V^{exp}_\mathrm{CKM})_{23}, \\
&&\epsilon_0= \big[(m_b - m_d) c_u^2 s_d + 2 (m_b - m_s) s_d s_u^2 +
    (m_d-m_b ) c_u  (V^{exp}_\mathrm{CKM})_{12}\big] c_d^4 s_d  \crn
&&\hspace{0.665 cm}+\, \big[(m_d + m_s-2 m_b) c_u s_d + (m_b - m_s) (V^{exp}_\mathrm{CKM})_{12}\big] c_d^5 s_u  +(m_s-m_b) c_d c_u (V^{exp}_\mathrm{CKM})_{22} \crn
&&\hspace{0.665 cm}+\, 2 \big[ (m_d + m_s-2 m_b) c_u s_d + (m_b - m_s) (V^{exp}_\mathrm{CKM})_{12}\big] c_d^3 s_d^2 s_u \crn
&&\hspace{0.665 cm}+\,  \big\{\big[m_b - m_s + 2 (m_b - m_d) s_d^4\big]c_u^2 + (m_b - m_s) s_d^4 s_u^2 +
    2 (m_d-m_b) c_u s_d^3 (V^{exp}_\mathrm{CKM})_{12}\big\} c_d^2 \crn
&&\hspace{0.665 cm}+\, \big[c_u (2 m_b - m_d - m_s) s_d (1-s_d^4) + (m_b -
       m_s) s_d^4 (V^{exp}_\mathrm{CKM})_{12}\big] c_d s_u  \crn
&&\hspace{0.665 cm}+\, (m_b - m_d) \big\{c_u^2 s_d^5 - c_u s_d^4 (V^{exp}_\mathrm{CKM})_{12} +
    \big[s_d s_u - (V^{exp}_\mathrm{CKM})_{22}\big] s_u \big\} s_d+(m_b - m_s)c_d^6  s_u^2,
\eea
\newpage
\vspace*{-2.0 cm}
\section{\label{Appenepsilon}Explicit expressions of $\epsilon^{1,2}_{d}$ and $\kappa_{l}\, (l=1\div6)$}
\vspace{-0.75 cm}

\bea
&&\epsilon^{(1)}_d= (m_d-m_b) c_d^6 c_u s_u +
  \big[(m_b - m_d)c_u^2  s_d + (m_s-m_b) s_d s_u^2 + (m_d-m_b)c_u  (V^{exp}_\mathrm{CKM})_{12}\big]c_d^5 \crn
  &&\hspace{0.665 cm}+\, 2\big[(m_b - m_d) c_u^2  s_d + (m_s-m_b) s_d s_u^2 +
     (m_d-m_b) c_u (V^{exp}_\mathrm{CKM})_{12}\big]  c_d^3 s_d^2  \crn
    &&\hspace{0.665 cm}-\, \big[(m_b - 2 m_d + m_s) c_u  s_d + (m_b - m_s) (V^{exp}_\mathrm{CKM})_{12}\big] c_d^4 s_d s_u  \crn
 &&\hspace{0.665 cm}+\,  \big\{[m_b - m_d + (m_b + m_d - 2 m_s) s_d^4]c_u  +
    2 (m_s-m_b) s_d^3 (V^{exp}_\mathrm{CKM})_{12}\big\} c_d^2 s_u \crn
    &&\hspace{0.665 cm}+\, (m_b - m_s) \big\{\big[(s_d^4-1) s_d  s_u + (V^{exp}_\mathrm{CKM})_{22}\big]c_u-s_d^4 s_u (V^{exp}_\mathrm{CKM})_{12}\big\} s_d  \crn
    &&\hspace{0.665 cm}+\, \Big\{\big[m_s-m_b + (m_b - m_d) s_d^4\big] c_u^2 s_d  + (m_d-m_b) c_u s_d^4 (V^{exp}_\mathrm{CKM})_{12} \crn
    &&\hspace{0.665 cm}+\, \big[s_d (m_b - m_d - m_b s_d^4 + m_s s_d^4) s_u + (m_d-m_b) (V^{exp}_\mathrm{CKM})_{22}\big] s_u \Big\} c_d , \\
&&\epsilon^{(2)}_d=(m_b - m_d) (m_b - m_s) \big[(c_d c_u + s_d s_u)(V^{exp}_\mathrm{CKM})_{13}+(c_d s_u- c_u s_d)(V^{exp}_\mathrm{CKM})_{23}\crn
&&\hspace{0.675 cm}+\, (V^{exp}_\mathrm{CKM})_{12} (V^{exp}_\mathrm{CKM})_{23}- (V^{exp}_\mathrm{CKM})_{13} (V^{exp}_\mathrm{CKM})_{22}\big], \\
&&\kappa_0=(m_t-m_c) (m_t - m_u)(V^{exp}_\mathrm{CKM})_{32}+\big[(m_t-m_c) c_u s_d + (m_u-m_t) c_d s_u \big] B^u_{31}\crn
&&\hspace{0.675 cm}+\, \big[(m_t - m_u) c_d c_u +(m_t-m_c) s_d s_u\big] B^u_{32}, \\
&&\kappa_1=(m_c - m_t) (m_t - m_u) c_d^3 \big\{(m_b - m_s) c_d^3  s_u^2 +
    \big[(m_b - m_d) c_u^2  s_d + 2 (m_b - m_s) s_d s_u^2 \crn
    &&\hspace{0.675 cm}+\,      c_u (m_d-m_b) (V^{exp}_\mathrm{CKM})_{12}\big] c_d s_d +
    \big[(m_d + m_s-2 m_b) c_u  s_d + (m_b - m_s) (V^{exp}_\mathrm{CKM})_{12}\big] c_d^2 s_u \crn
    &&\hspace{0.675 cm}+\, 2  \big[(m_d + m_s-2 m_b) c_u  s_d + (m_b - m_s) (V^{exp}_\mathrm{CKM})_{12}\big] s_d^2 s_u\big\}, \\
&&\kappa_2=c_d^2 (m_t - m_u) \bigg\{(m_b - m_s) s_u^2 \big[(-m_c + m_t) s_d^4 + B^u_{31} s_u^4 (V^{exp}_\mathrm{CKM})_{13}\big]  \crn
    &&\hspace{0.675 cm}+\, B^u_{32} c_u^6 (m_b - m_s) (V^{exp}_\mathrm{CKM})_{23}
    -c_u^5 (m_b - m_s) s_u \big[B^u_{32} (V^{exp}_\mathrm{CKM})_{13} + B^u_{31} (V^{exp}_\mathrm{CKM})_{23}\big]  \crn
    &&\hspace{0.675 cm}-\, 2 c_u^3 (m_b - m_s) s_u^3 \big[B^u_{32} (V^{exp}_\mathrm{CKM})_{13} + B^u_{31} (V^{exp}_\mathrm{CKM})_{23}\big] \crn
    &&\hspace{0.675 cm}+\,   c_u^4 (m_b - m_s) s_u^2 \big[B^u_{31} (V^{exp}_\mathrm{CKM})_{13} + 2 B^u_{32} (V^{exp}_\mathrm{CKM})_{23}\big]  \crn
    &&\hspace{0.675 cm}+\, c_u \Big(2 m_b (m_c - m_t) s_d^3 (V^{exp}_\mathrm{CKM})_{12} + 2 m_d (m_t-m_c) s_d^3 (V^{exp}_\mathrm{CKM})_{12} \crn
    &&\hspace{0.675 cm}-\,      m_b s_u^5 \big[B^u_{32} (V^{exp}_\mathrm{CKM})_{13} + B^u_{31} (V^{exp}_\mathrm{CKM})_{23}\big] + m_s s_u^5 \big[B^u_{32} (V^{exp}_\mathrm{CKM})_{13} + B^u_{31} (V^{exp}_\mathrm{CKM})_{23}\big]\Big)  \crn
    &&\hspace{0.675 cm}+\,
    c_u^2 \Big\{m_c m_s - m_s m_t + 2 m_c m_d s_d^4 - 2 m_d m_t s_d^4 -
      2 B^u_{31} m_s s_u^4 (V^{exp}_\mathrm{CKM})_{13} - B^u_{32} m_s s_u^4 (V^{exp}_\mathrm{CKM})_{23}  \crn
    &&\hspace{0.675 cm}+\,   m_b \Big(m_t-m_c + 2 ( m_t-m_c) s_d^4 +s_u^4 \big[2 B^u_{31} (V^{exp}_\mathrm{CKM})_{13} + B^u_{32} (V^{exp}_\mathrm{CKM})_{23}\big]\Big)\Big\}\bigg\}, \\
    &&\kappa_3= (m_t - m_u)\big[c_u^2 s_d^5 + s_d s_u^2 - c_u s_d^4 (V^{exp}_\mathrm{CKM})_{12}\big]  + \big[c_u^3 (B^u_{31} c_u + B^u_{32} s_u) (1 + s_u^2)\big] s_d (V^{exp}_\mathrm{CKM})_{13} \crn
    &&\hspace{0.675 cm}+\, (m_u -m_t) s_u (V^{exp}_\mathrm{CKM})_{22} +  (V^{exp}_\mathrm{CKM})_{23} \big[B^u_{31} (c_u^5 + 2 c_u^3 s_u^2) + B^u_{32} s_u (1-2s_u^2 c_u^2)\big]s_d s_u\crn
    &&\hspace{0.675 cm}+\,  \big[B^u_{32} (V^{exp}_\mathrm{CKM})_{13} + B^u_{31} (V^{exp}_\mathrm{CKM})_{23}\big] c_u s_d s_u^5 +
 \big[B^u_{31} (V^{exp}_\mathrm{CKM})_{13} + 2 B^u_{32} (V^{exp}_\mathrm{CKM})_{23}\big] c_u^2 s_d s_u^4  \crn
    &&\hspace{0.675 cm}+\, (m_t - m_u) \big[c_u (V^{exp}_\mathrm{CKM})_{13} + s_u (V^{exp}_\mathrm{CKM})_{23}\big] (V^{exp}_\mathrm{CKM})_{32},
\eea
\vspace{-1.25 cm}
\newpage
\begin{center}
\textbf{Explicit expressions of $\epsilon^{1,2}_{d}$ and $\kappa_{l}\, (l=1\div6)$  (continued)}
\end{center}
\bea
&&\kappa_4=B^u_{32} s_d s_u^2 (V^{exp}_\mathrm{CKM})_{13} \big\{\big[2(m_c-m_t)m_s + (m_t - m_u)m_d +(m_t + m_u-2 m_c) m_b\big] s_u^2\crn
&&\hspace{0.675 cm}-\, \big[2 (m_u- m_t)m_d + (m_t-m_c)m_s  + (m_c + m_t - 2 m_u)m_b \big] c_u^2 \big\} \crn
    &&\hspace{0.675 cm}-\,  B^u_{31} s_d s_u^2 (V^{exp}_\mathrm{CKM})_{23} \big\{\big[2 (m_c-  m_t)m_s +  (m_t-  m_u)m_d +
         (m_t + m_u-2 m_c) m_b\big] c_u^2  \crn
    &&\hspace{0.675 cm}-\, \big[2 (m_u- m_t) m_d + (m_t -m_c) m_s+ (m_c + m_t - 2 m_u)  m_b\big] s_u^2\big\}
    -\,\Big\{c_u \big[ (m_t-m_c)m_s \crn
    &&\hspace{0.675 cm} +\, (m_t  - m_u) m_d  + (m_c + m_u- 2 m_t )m_b \big] (1 + s_u^2) s_d s_u \Big\} \Big[B^u_{31} (V^{exp}_\mathrm{CKM})_{13} - B^u_{32} (V^{exp}_\mathrm{CKM})_{23}\Big] \crn
    &&\hspace{0.675 cm}+\, c_u^4 s_d \big[-B^u_{32} (m_b - m_d) (m_t - m_u) (V^{exp}_\mathrm{CKM})_{13} +
     B^u_{31} (m_b - m_s)(m_c - m_t) (V^{exp}_\mathrm{CKM})_{23}\big] \crn
    &&\hspace{0.675 cm}+\, (m_b - m_s)(m_c - m_t) (m_t - m_u) (1 + s_u^2) \big[c_u (V^{exp}_\mathrm{CKM})_{23}-s_u (V^{exp}_\mathrm{CKM})_{13}\big] (V^{exp}_\mathrm{CKM})_{32},  \\
&&\kappa_5=m_s (m_t-m_c) \big[(m_t - m_u) s_d^3 (V^{exp}_\mathrm{CKM})_{12} - B^u_{32} s_u^5 (V^{exp}_\mathrm{CKM})_{13}\big] s_d \crn
    &&+ m_b \Big\{B^u_{31} (m_t - m_u) s_u^5 (V^{exp}_\mathrm{CKM})_{23} + (m_c - m_t) \big[(m_t - m_u) s_d^3 (V^{exp}_\mathrm{CKM})_{12} - B^u_{32} s_u^5 (V^{exp}_\mathrm{CKM})_{13}\big]\Big\} s_d \crn
    &&+ B^u_{31} m_d (m_u-m_t) s_d s_u^5 (V^{exp}_\mathrm{CKM})_{23}+(m_s-m_b)(m_c-m_t) (m_t - m_u) s_u^4 (V^{exp}_\mathrm{CKM})_{13} (V^{exp}_\mathrm{CKM})_{32}, \hspace{0.5 cm}\\
&&\kappa_6=\big[(m_d + m_s) m_t-m_c m_d\big] (m_t - m_u)  (1 - s_d^4) s_d s_u + \big[m_s m_t (m_u-m_t)\big] (V^{exp}_\mathrm{CKM})_{22} \crn
    &&\hspace{0.675 cm}-\, B^u_{31} \big\{\big[(m_d + m_s) m_t - m_d m_u\big] s_d s_u^5\big\} (V^{exp}_\mathrm{CKM})_{13}
    + s_u^4 (V^{exp}_\mathrm{CKM})_{23} \big\{m_s m_t (m_t - m_u) (V^{exp}_\mathrm{CKM})_{32} \crn
    &&\hspace{0.675 cm}+\, B^u_{32} \big[(m_d + m_s) m_t - m_d m_u\big] s_d s_u\big\}
    + m_c m_s \Big\{s_d s_u^5 \big[B^u_{31} (V^{exp}_\mathrm{CKM})_{13} - B^u_{32} (V^{exp}_\mathrm{CKM})_{23}\big] \crn
    &&\hspace{0.675 cm} + (m_t - m_u) \big[s_d^5 s_u -s_d s_u + (V^{exp}_\mathrm{CKM})_{22} - s_u^4 (V^{exp}_\mathrm{CKM})_{23} (V^{exp}_\mathrm{CKM})_{32}\big]\Big\}
    + m_b \Big\{(2 m_t - m_u) s_d s_u^5 \crn
 &&\hspace{0.675 cm}\times \big[B^u_{31} (V^{exp}_\mathrm{CKM})_{13} - B^u_{32} (V^{exp}_\mathrm{CKM})_{23}\big]+
    m_t (m_t - m_u) \big[2 s_d^5 s_u -2 s_d s_u - s_u^4 (V^{exp}_\mathrm{CKM})_{23} (V^{exp}_\mathrm{CKM})_{32}\crn
    &&\hspace{0.675 cm}+ (V^{exp}_\mathrm{CKM})_{22}\big]
    + m_c \big\{\big[(m_u-m_t) \big[2 s_d^5 s_u -2 s_d s_u + (V^{exp}_\mathrm{CKM})_{22} - s_u^4 (V^{exp}_\mathrm{CKM})_{23} (V^{exp}_\mathrm{CKM})_{32}\big] \crn
    &&\hspace{0.675 cm} + B^u_{32} (V^{exp}_\mathrm{CKM})_{23}-B^u_{31} (V^{exp}_\mathrm{CKM})_{13}\big] s_d s_u^5\big\}\Big\}. \eea
\newpage
\section{\label{Appentau01234} 
The dependence of $\tau_s\, (s=0\div 4)$ on $s_u$ and $s_d$}
\begin{figure}[ht]
\begin{center}
\vspace{-1.0 cm}
\hspace{-6.0 cm}
\includegraphics[width=0.815\textwidth]{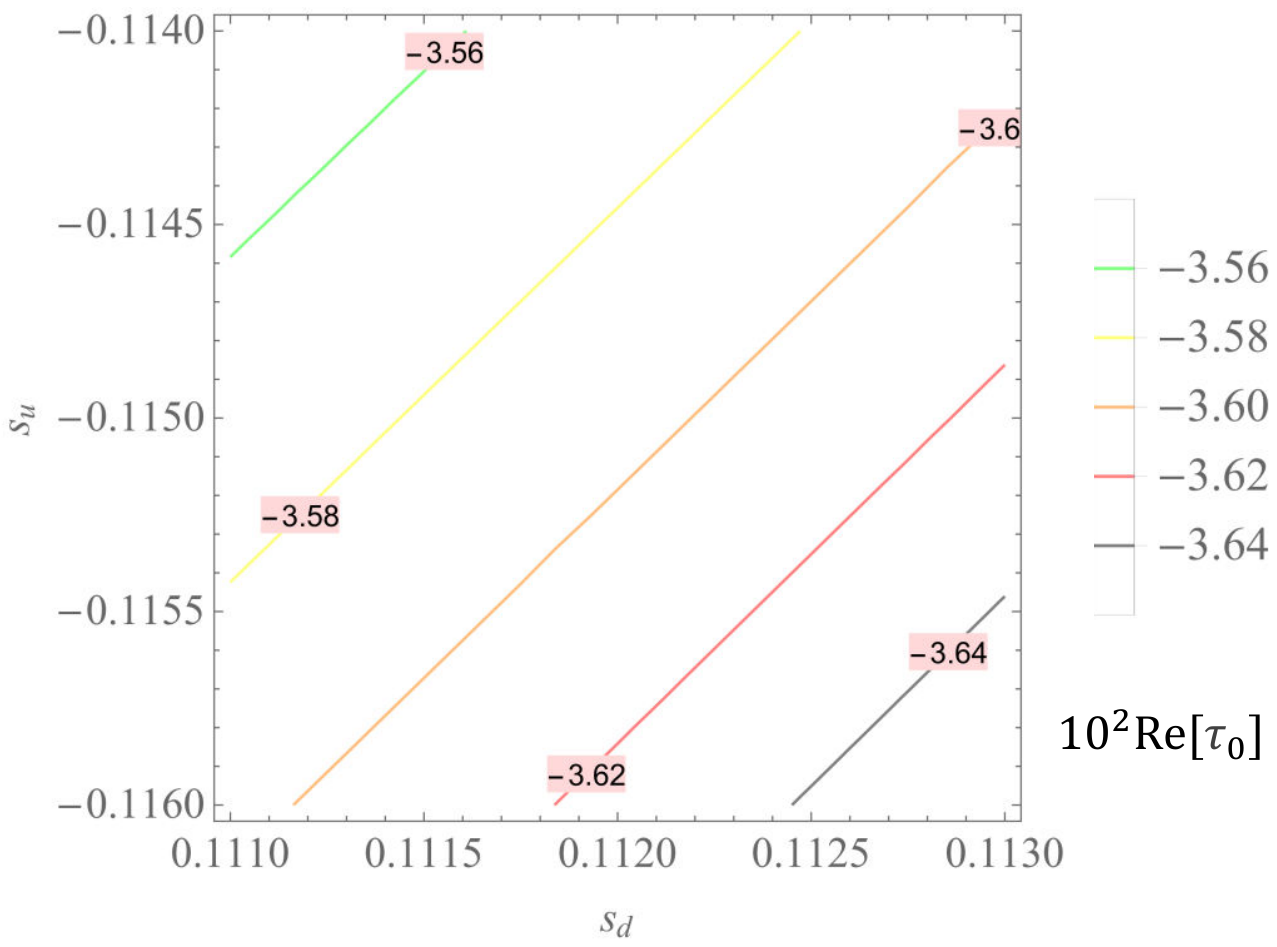}\hspace{-5.2 cm}
\vspace{-0.95 cm}
\includegraphics[width=0.815\textwidth]{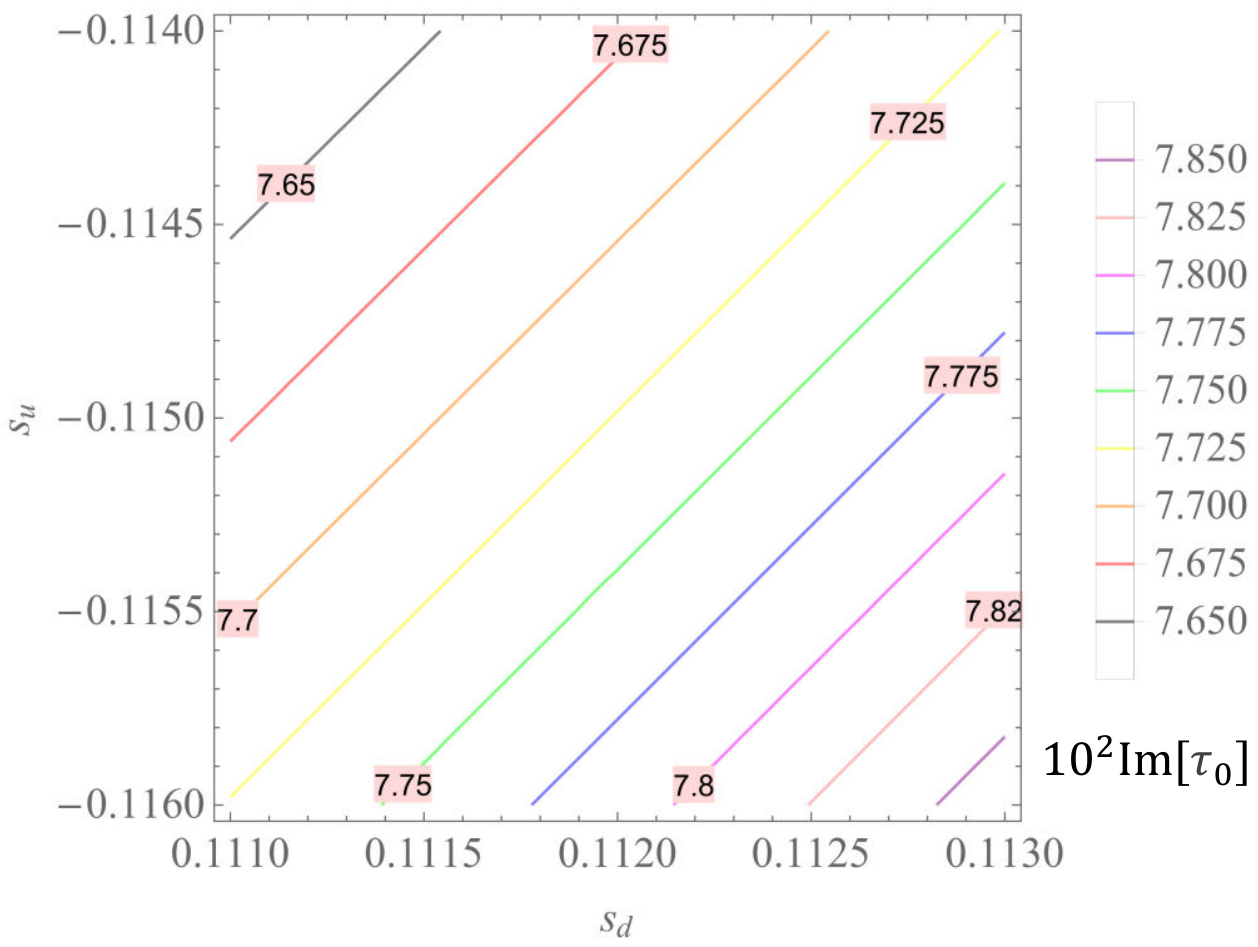}\hspace*{-5.25 cm}
\end{center}
\vspace{-9.75 cm}
\caption[(Colored lines) $\tau_0$ (in GeV, left panel) and $\tau_0$ (in GeV, right panel) versus $s_{u}$ and $s_{d}$ with $s_{u}\in (-0.116, -0.114)$ and $s_{d}\in (0.111, 0.113)$.]{(Colored lines) $\mathrm{Re} [\tau_0]$ (in GeV, left panel) and $\mathrm{Im} [\tau_0]$ (in GeV, right panel) versus $s_{u}$ and $s_{d}$ with $s_{u}\in (-0.116, -0.114)$ and $s_{d}\in (0.111, 0.113)$.}
\label{tau0F}
\vspace{-0.5 cm}
\end{figure}
\begin{figure}[ht]
\begin{center}
\vspace{-1.0 cm}
\hspace{-6.0 cm}
\includegraphics[width=0.815\textwidth]{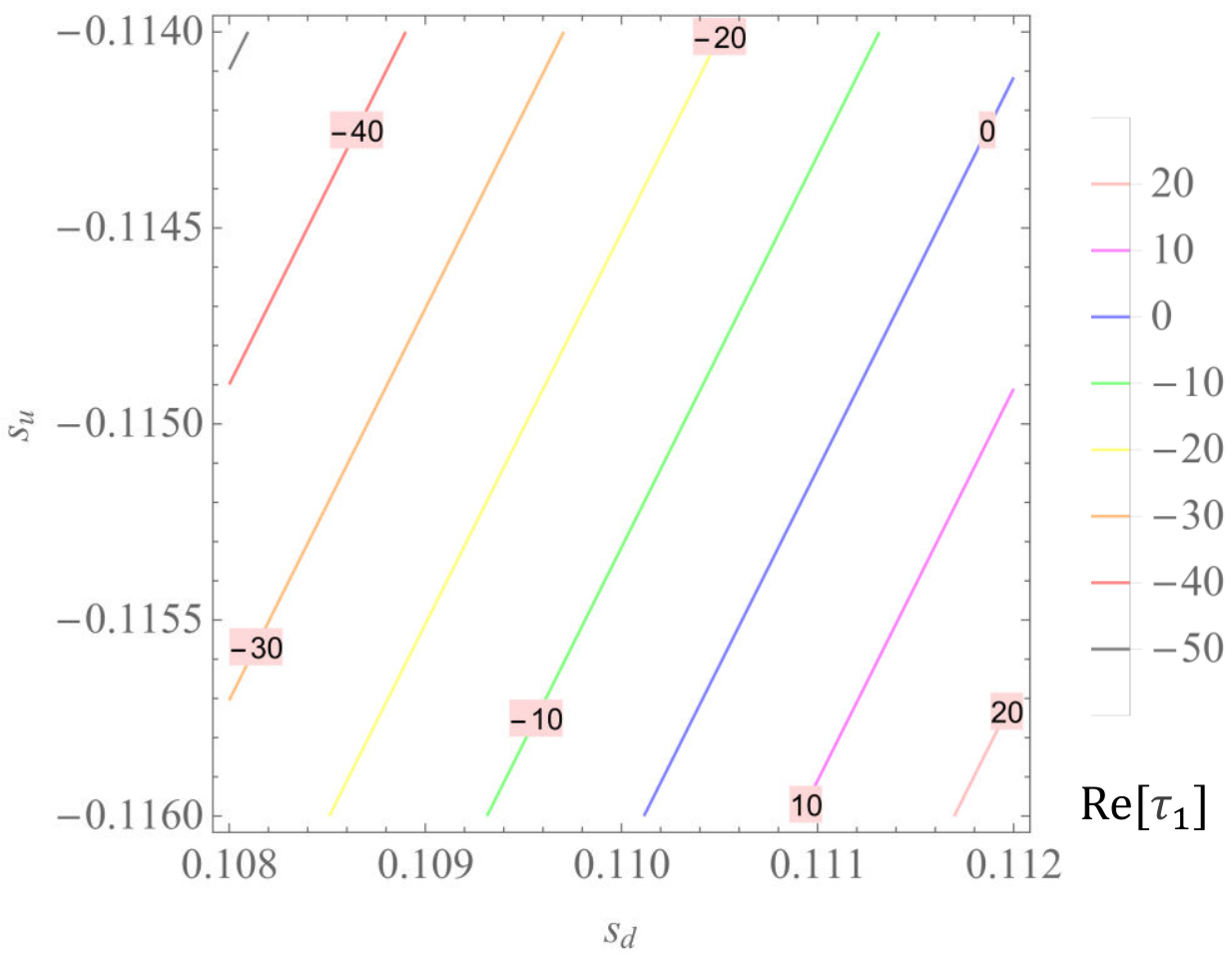}\hspace{-5.2 cm}
\vspace{-0.95 cm}
\includegraphics[width=0.815\textwidth]{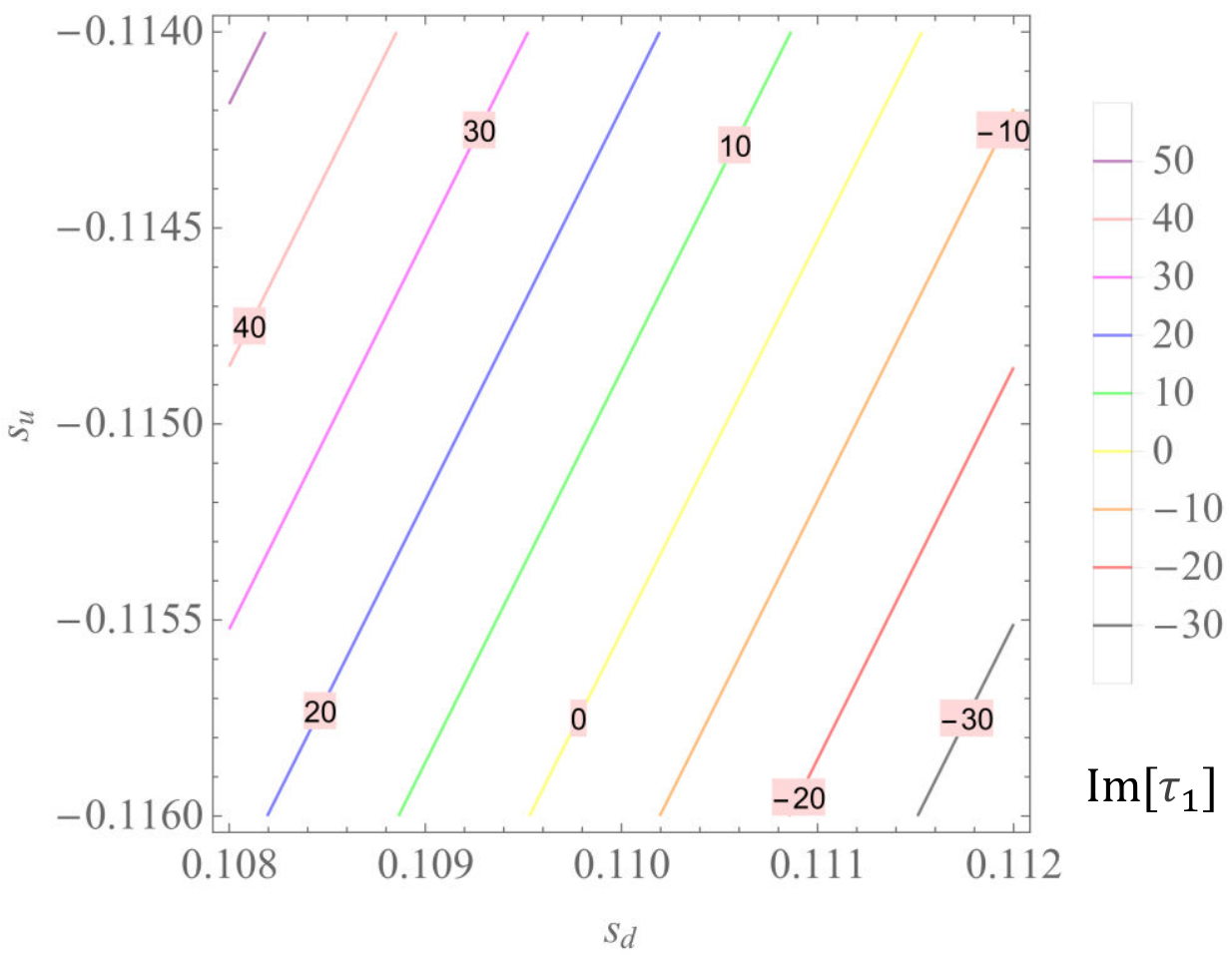}\hspace*{-5.25 cm}
\end{center}
\vspace{-9.75 cm}
\caption[(Colored lines) $\tau_1$ (in GeV, left panel) and $\tau_1$ (in GeV, right panel) versus $s_{u}$ and $s_{d}$ with $s_{u}\in (-0.116, -0.114)$ and $s_{d}\in (0.111, 0.113)$.]{(Colored lines) $\mathrm{Re} [\tau_1]$ (in GeV, left panel) and $\mathrm{Im} [\tau_1]$ (in GeV, right panel) versus $s_{u}$ and $s_{d}$ with $s_{u}\in (-0.116, -0.114)$ and $s_{d}\in (0.111, 0.113)$.}
\label{tau1F}
\vspace{-0.5 cm}
\end{figure}
\newpage
\vspace*{-1.25 cm}
\begin{center}
\textbf{Appendix \ref{Appentau01234}. The dependence of $\tau_s\, (s=0\div 4)$ on $s_u$ and $s_d$  (continued)}
\end{center}
\begin{figure}[ht]
\begin{center}
\vspace{-1.5 cm}
\hspace{-6.0 cm}
\includegraphics[width=0.815\textwidth]{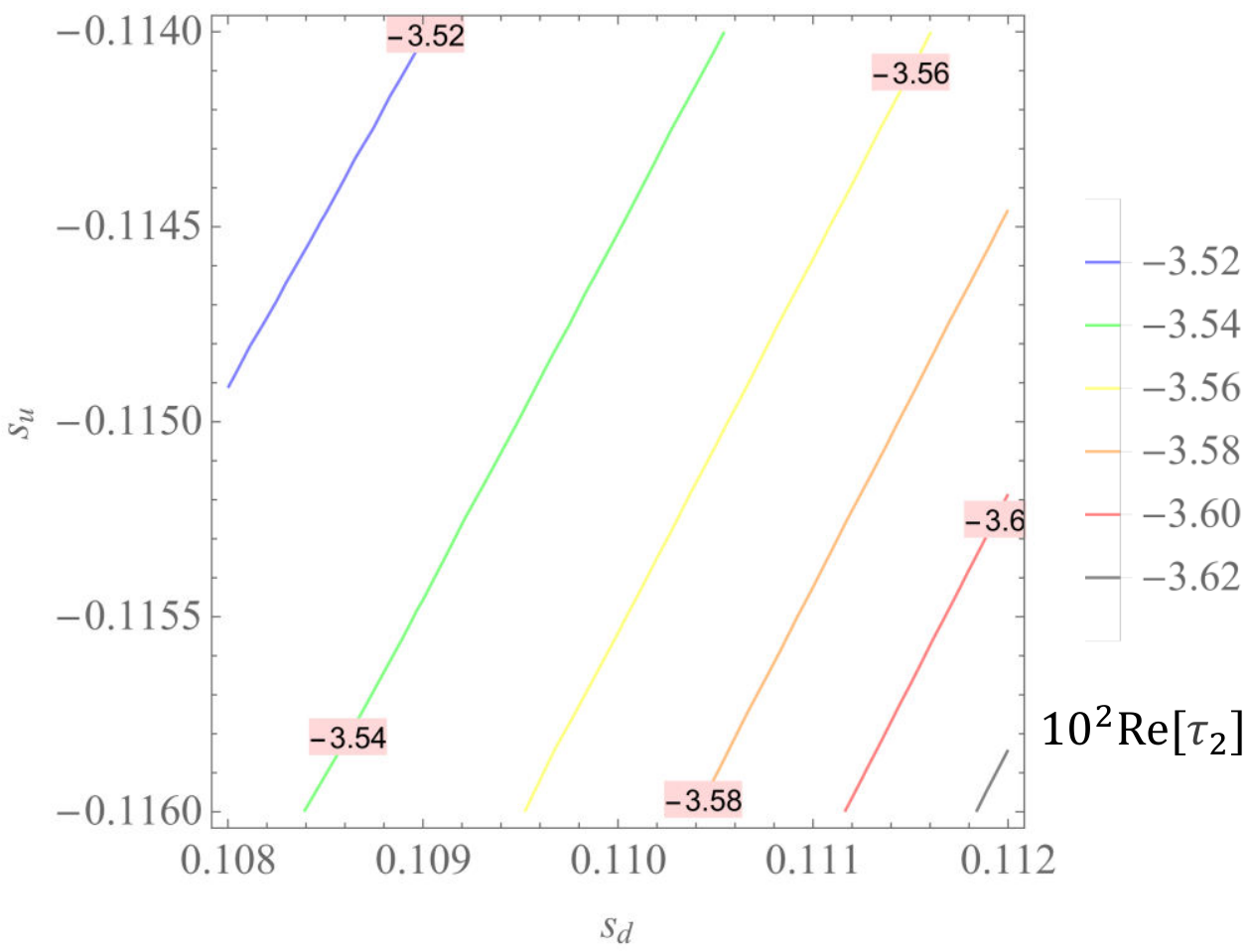}\hspace{-5.2 cm}
\vspace{-0.95 cm}
\includegraphics[width=0.815\textwidth]{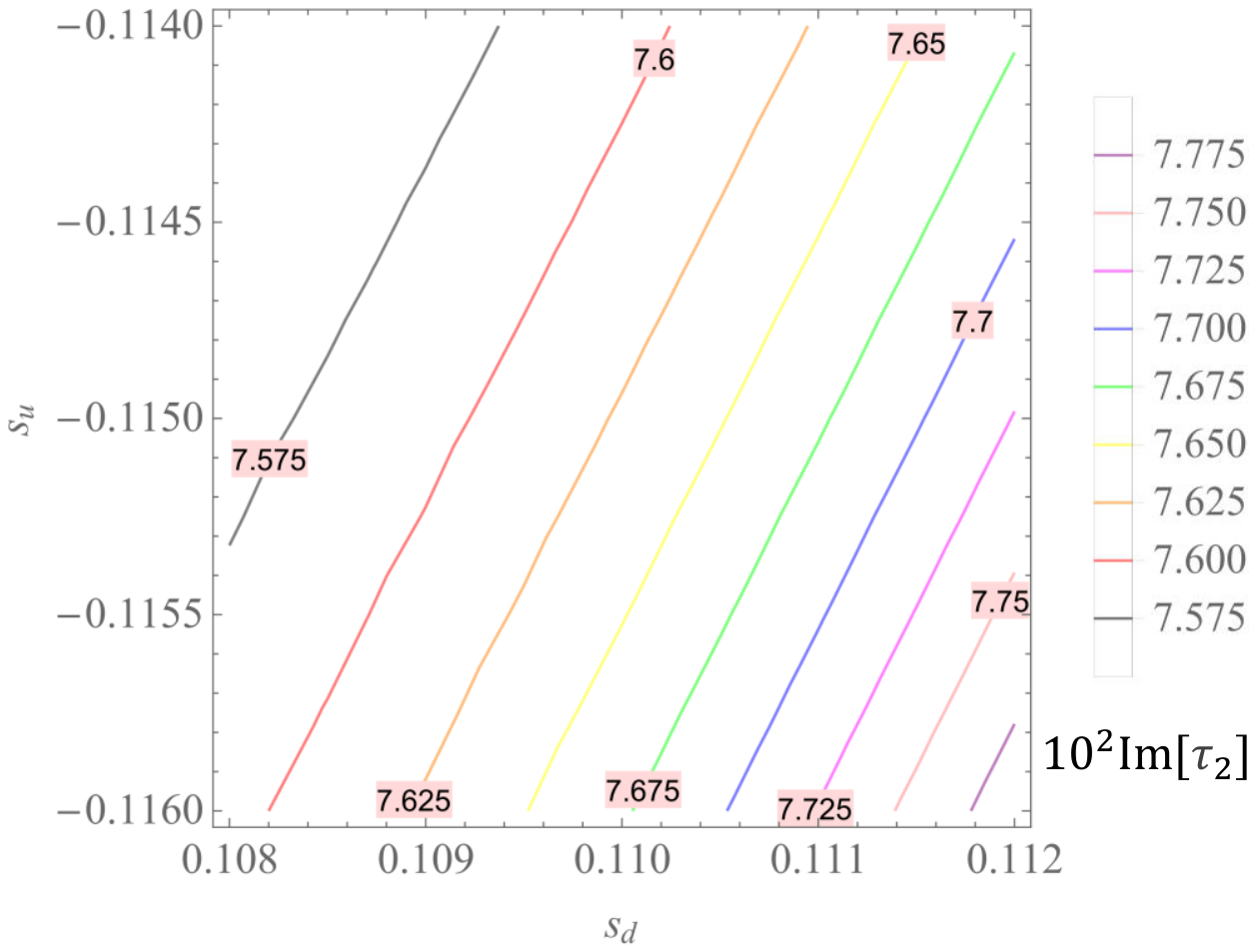}\hspace*{-5.25 cm}
\end{center}
\vspace{-10.0 cm}
\caption[(Colored lines) $\tau_2$ (in GeV, left panel) and $\tau_2$ (in GeV, right panel) versus $s_{u}$ and $s_{d}$ with $s_{u}\in (-0.116, -0.114)$ and $s_{d}\in (0.111, 0.113)$.]{(Colored lines) $\mathrm{Re} [\tau_2]$ (in GeV, left panel) and $\mathrm{Im} [\tau_2]$ (in GeV, right panel) versus $s_{u}$ and $s_{d}$ with $s_{u}\in (-0.116, -0.114)$ and $s_{d}\in (0.111, 0.113)$.}
\label{tau2F}
\vspace{-1.3 cm}
\end{figure}
\begin{figure}[ht]
\begin{center}
\vspace{-0.65 cm}
\hspace{-6.0 cm}
\includegraphics[width=0.815\textwidth]{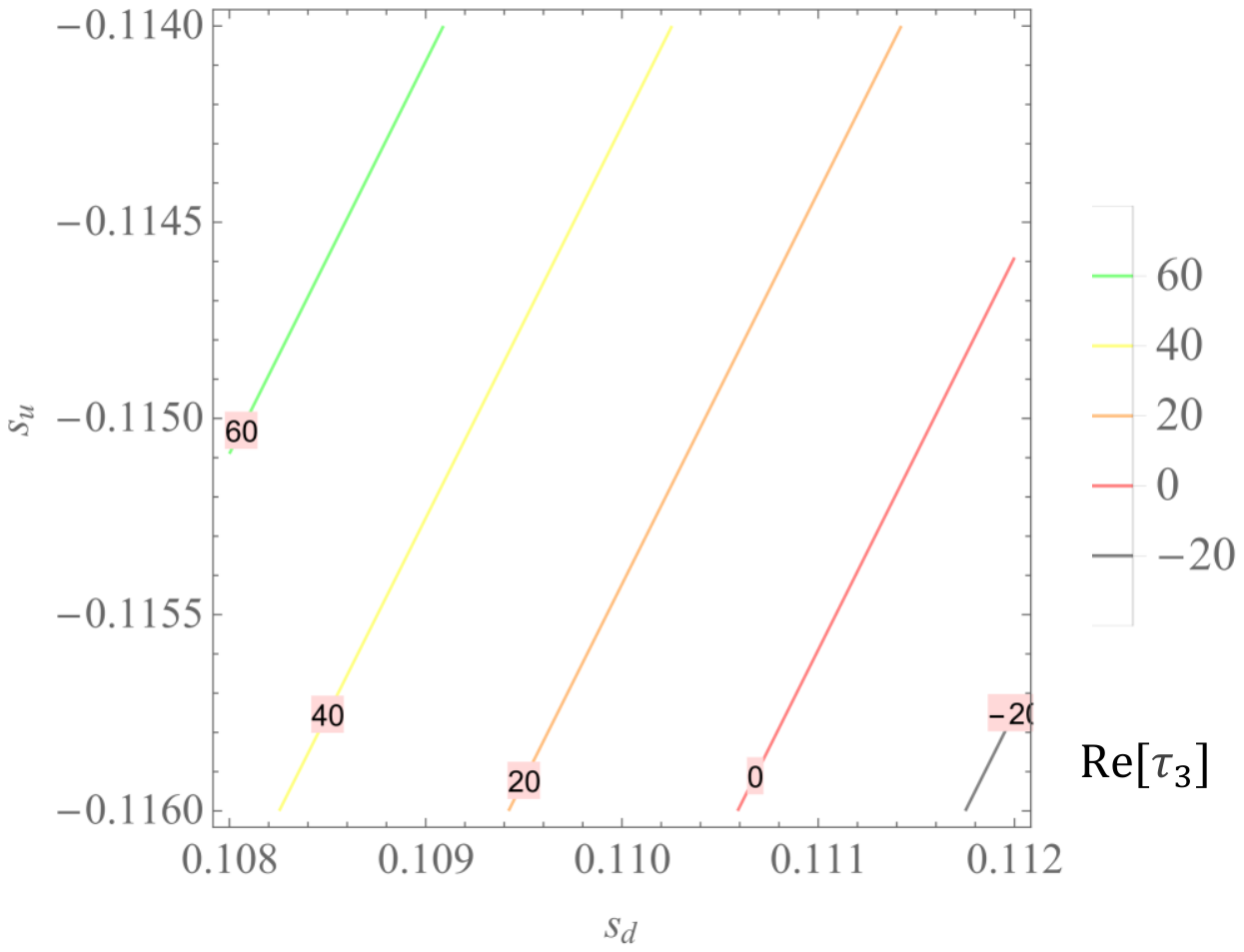}\hspace{-5.2 cm}
\vspace{-0.95 cm}
\includegraphics[width=0.815\textwidth]{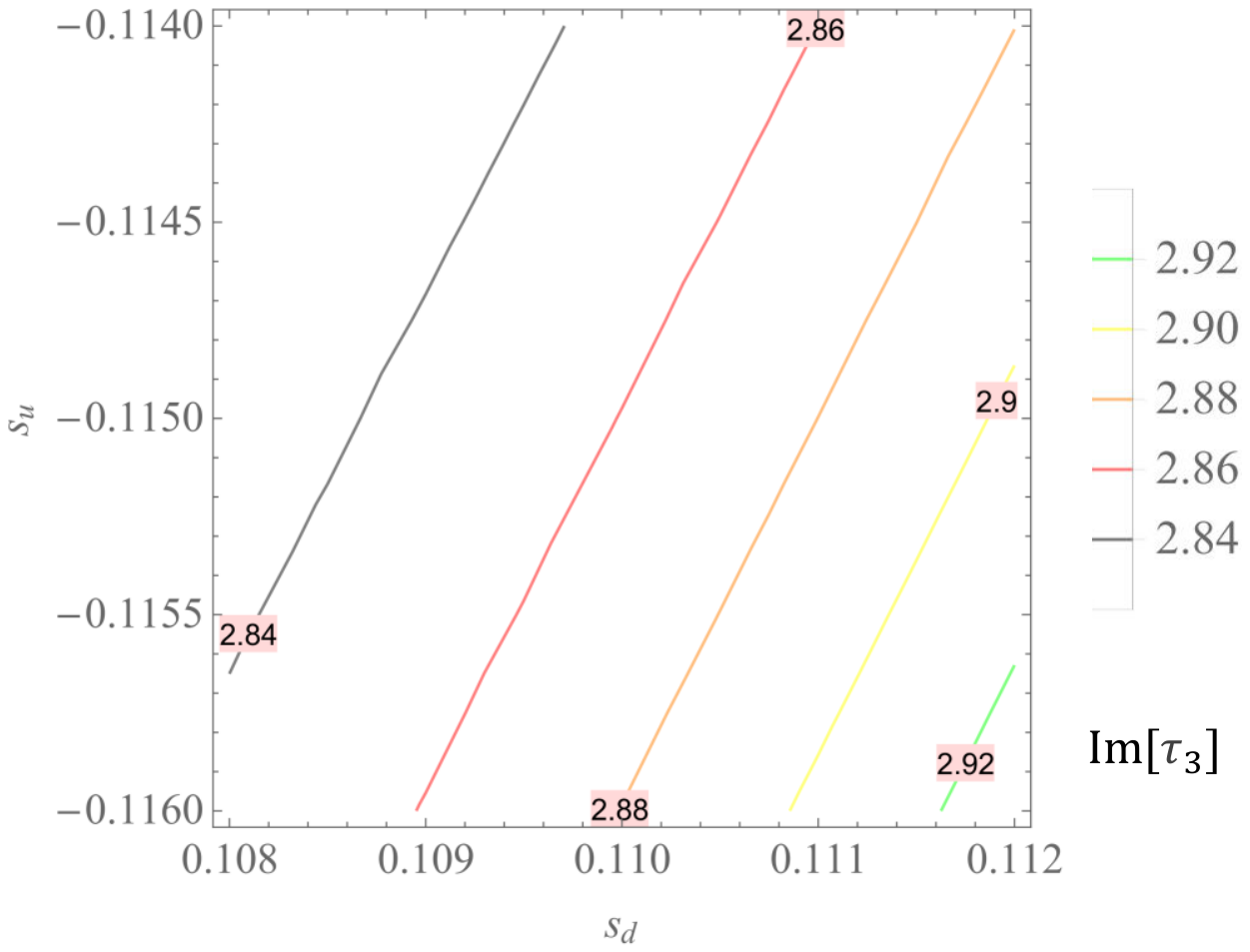}\hspace*{-5.25 cm}
\end{center}
\vspace{-10.0 cm}
\caption[(Colored lines) $\tau_3$ (in GeV, left panel) and $\tau_3$ (in GeV, right panel) versus $s_{u}$ and $s_{d}$ with $s_{u}\in (-0.116, -0.114)$ and $s_{d}\in (0.111, 0.113)$.]{(Colored lines) $\mathrm{Re} [\tau_3]$ (in GeV, left panel) and $\mathrm{Im} [\tau_3]$ (in GeV, right panel) versus $s_{u}$ and $s_{d}$ with $s_{u}\in (-0.116, -0.114)$ and $s_{d}\in (0.111, 0.113)$.}
\label{tau3F}
\vspace{-0.25 cm}
\end{figure}
\begin{figure}[ht]
\begin{center}
\vspace{-1.6 cm}
\hspace{-6.0 cm}
\includegraphics[width=0.815\textwidth]{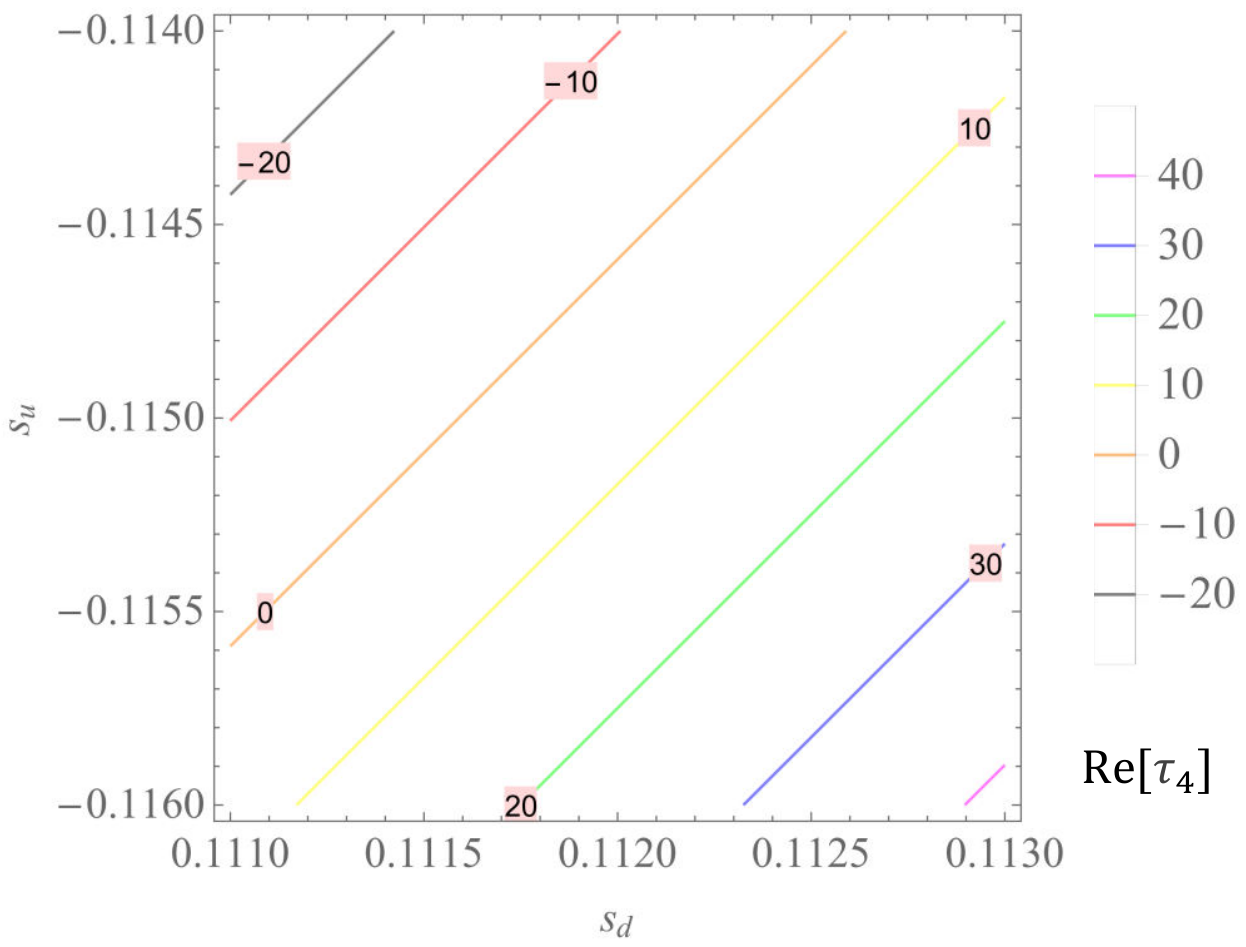}\hspace{-5.2 cm}
\vspace{-0.95 cm}
\includegraphics[width=0.815\textwidth]{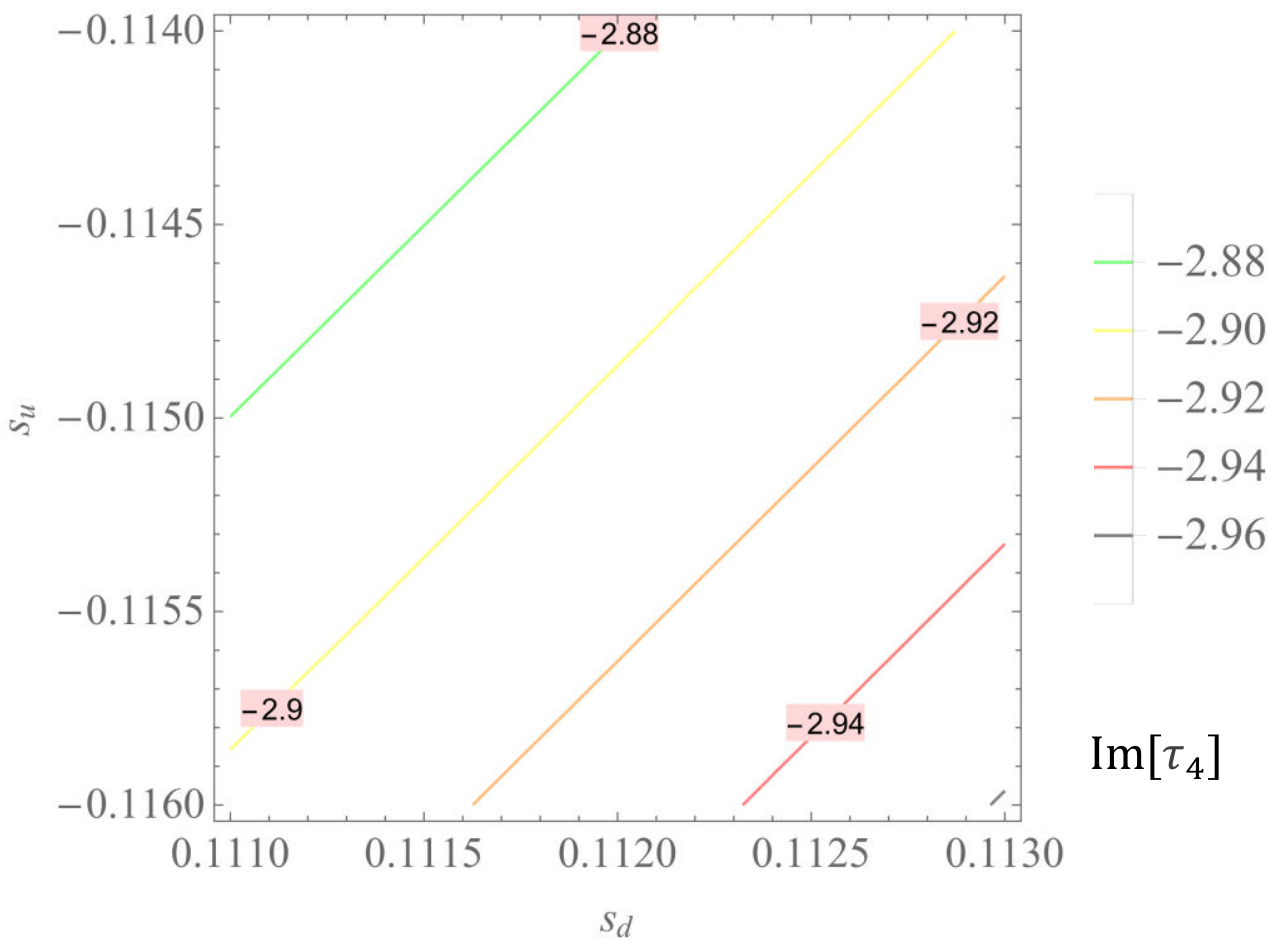}\hspace*{-5.25 cm}
\end{center}
\vspace{-10.0 cm}
\caption[(Colored lines) $\tau_4$ (in GeV, left panel) and $\tau_4$ (in GeV, right panel) versus $s_{u}$ and $s_{d}$ with $s_{u}\in (-0.116, -0.114)$ and $s_{d}\in (0.111, 0.113)$.]{(Colored lines) $\mathrm{Re} [\tau_4]$ (in GeV, left panel) and $\mathrm{Im} [\tau_4]$ (in GeV, right panel) versus $s_{u}$ and $s_{d}$ with $s_{u}\in (-0.116, -0.114)$ and $s_{d}\in (0.111, 0.113)$.}
\label{tau4F}
\vspace{-2.75 cm}
\end{figure}

\end{document}